\documentclass{jfm}
\usepackage{graphicx}
\usepackage{xcolor}
\usepackage{epstopdf, epsfig}
\usepackage{color}
\usepackage{tikz}
\usetikzlibrary[plotmarks]
\usepackage[caption=false]{subfig}
\usepackage[export]{adjustbox}
\usepackage{amsmath}
\usepackage{algorithm2e}
\usepackage{booktabs}

\shorttitle{A physics-infused Immersed Boundary Method using sequential Data Assimilation}
\shortauthor{M. M. Valero and M. Meldi}

\title{A physics-infused Immersed Boundary Method using online sequential Data Assimilation}

\author{Miguel M. Valero\aff{1}
  \corresp{\email{miguel.martinez\_valero@ensam.eu}}
 \and M. Meldi\aff{1}}

\affiliation{\aff{1}Univ. Lille, CNRS, ONERA, Arts et Métiers ParisTech, Centrale Lille, UMR 9014- LMFL- Laboratoire de Mécanique des fluides de Lille - Kampé de Feriet, F-59000 Lille, France}

\begin{document}

\newcommand{\blacklinedashed}{\raisebox{2pt}{\tikz{\draw[-,black,dashed,line width = 1pt](0,0) -- (5.1mm,0);}}}

\newcommand{\blacklinesolid}{\raisebox{2pt}{\tikz{\draw[-,black,solid,line width = 1pt](0,0) -- (5mm,0);}}}

\newcommand{\blacklinedotted}{\raisebox{2pt}{\tikz{\draw[-,black,dotted,line width = 1.5pt](0,0) -- (5mm,0);}}}

\newcommand{\blacklinemarker}{\raisebox{1pt}{\tikz{\draw[-,black,solid,line width = 0.7pt](0,0) -- (5mm,0); \filldraw[black] (2.5mm,0) circle (1.5pt);}}}

\definecolor{greySTRONG}{RGB}{127.5,127.5,127.5}
\definecolor{greySOFT}{RGB}{204,204,204}

\newcommand{\greylinesolidstrong}{\raisebox{2pt}{\tikz{\draw[-,greySTRONG,solid,line width = 1.5pt](0,0) -- (5mm,0)}}}

\newcommand{\greylinedashedstrong}{\raisebox{2pt}{\tikz{\draw[-,greySTRONG,dashed,line width = 1.5pt](0,0) -- (5mm,0)}}}

\newcommand{\greylinedottedstrong}{\raisebox{2pt}{\tikz{\draw[-,greySTRONG,dotted,line width = 1.5pt](0,0) -- (5mm,0)}}}

\newcommand{\greylinesolidsoft}{\raisebox{2pt}{\tikz{\draw[-,greySOFT,solid,line width = 1.5pt](0,0) -- (5mm,0)}}}

\newcommand{\greylinedashedsoft}{\raisebox{2pt}{\tikz{\draw[-,greySOFT,dashed,line width = 1.5pt](0,0) -- (5mm,0)}}}

\newcommand{\redline}{\raisebox{2pt}{\tikz{\draw[-,red,solid,line width = 1pt](0,0) -- (5mm,0);}}}

\newcommand{\blueline}{\raisebox{2pt}{\tikz{\draw[-.,blue,dashdotted,line width = 1pt](0,0) -- (5mm,0);}}}

\definecolor{green}{RGB}{0,150,0}
\newcommand{\greenline}{\raisebox{2pt}{\tikz{\draw[-,green,dotted,line width = 1.5pt](0,0) -- (5mm,0);}}}

\definecolor{magenta}{RGB}{255,0,255}
\newcommand{\magentaline}{\raisebox{2pt}{\tikz{\draw[-,magenta,dotted,line width = 0.7pt](0,0) -- (5mm,0);}}}

\maketitle

\begin{abstract}
A physics-infused strategy relying on the Ensemble Kalman Filter (EnKF) is here used to augment the accuracy of a continuous Immersed Boundary Method (IBM). The latter is a classical penalty method accounting for the presence of the immersed body via a volume source term which is included in the Navier--Stokes equations. The model coefficients of the penalization method, which are usually selected by the user, are optimized here using an EnKF data-driven strategy. The parametric inference is governed by the physical knowledge of local and global features of the flow, such as the no-slip condition and the shear stress at the wall. The C++ library CONES (Coupling OpenFOAM with Numerical EnvironmentS) developed by the team is used to perform an online investigation, coupling on-the-fly data from synthetic sensors with results from an ensemble of coarse-grained numerical simulations.  The analysis is performed for a classical test case, namely the turbulent channel flow with $Re_\tau = 550$. 
The comparison of the results with a high-fidelity Direct Numerical Simulation (DNS) shows that the data-driven procedure exhibits remarkable accuracy despite the relatively low grid resolution of the ensemble members. 
\end{abstract}


\section{Introduction}
\label{sec:introduction}

The development of reliable, efficient, and cost-effective numerical tools for the accurate prediction of multi-physics problems is a timely key challenge in Computational Fluid Dynamics (CFD). In fact, the development of predictive tools to investigate flow configurations, including several complex aspects (turbulence, compressibility effects, fluid-structure interaction, transport of active and passive scalars), must take into account requirements exhibiting goal rivalry. Future paradigms in the numerical development of flow solvers strive for increased accuracy of the predictive computational strategies while reducing the computational resources required. This latter criterion is not only associated with a global reduction of the cost and the accessibility of methods on different computational machines and architectures but also with the decrease of the carbon footprint related to calculations, in order to comply with progressively more strict regulations. Therefore, efficient numerical modelling in fluid mechanics is essential for progress in fundamental studies as well as industrial applications.  

Among the flow problems previously introduced, the accurate numerical representation of near-wall flow features for bodies immersed in turbulent flows needs key advancement. The prediction of numerous flow features of unstationary flows, such as aerodynamic forces, is driven by the precise representation of localized near-wall dynamics. This aspect is particularly relevant and, at the same time, challenging for the flow prediction around complex geometries. In this case, classical body-fitted approaches may have to deal with high deformation of the mesh elements, possibly leading to poor numerical prediction. Additionally, the simulation of moving bodies may require prohibitively expensive mesh updates. In the last decades, several numerical strategies have been proposed to handle these two problematic aspects. Among these proposals, the Immersed Boundary Method (IBM) \citep{Peskin1972_jcp,Peskin2002_an,Mittal2005_arfm,Kim2019_ijhff,Verzicco2022_arfm} has emerged as one of the most popular approaches. The IBM includes a spectrum of tools that operate on non-body-fitted grids i.e. the mesh is not tailored around the shape of the immersed body, which is used as an internal boundary condition for classical body-fitted approaches. For IBM, numerical strategies are developed instead to mimic the presence of the body. The techniques, which rely, for example, on the addition of source terms in the dynamic equation or on imposing discrete values at the centre of grid elements for some physical quantities investigated, are usually distinguished in continuous and discrete methods. Despite the significant differences between the numerous approaches presented in the literature, state-of-the-art methods can account for body movement and deformation with reduced computational cost. However, the complete resolution of near-wall features, in particular for high Reynolds regimes, is a challenging issue in IBM. In fact, owing to the regularity of the grid and the difficulties in applying arbitrary local directional stretching, a larger number of mesh elements is usually required in IBM to obtain similar accuracy of body-fitted tools \citep{Verzicco2022_arfm}.

Another difficulty for the numerical representation of near-wall turbulence, for both IBM and body-fitted methods, is linked to the performance of turbulence/subgrid-scale modelling in the proximity of the wall \citep{Pope2000_cambridge,Sagaut2006_springer,Wilcox2006_DCW}. Computational resources required for complete flow resolution at the wall may be prohibitive, and, at the same time, wall modelling can show a lack of accuracy in regions exhibiting flow recirculation or strong pressure gradients. In the last decades, several approaches coming from the Estimation Theory \citep{Simon2006_wiley} have been employed to complete turbulence closures and/or their associated wall functions with the aim of obtaining generalized predictive models. Among these approaches, data-driven approaches for Uncertainty Quantification and Propagation (UQ-UP) \citep{Xiao2019_pas}, Data Assimilation (DA) \citep{Asch2016_siam} and Machine Learning (ML) \citep{Duraisamy2019_arfm} have been extensively used to improve the accuracy of turbulence closures \citep{Gorle2013_pof,Edeling2014_jcp,Margheri2014_cf,Tracey2015_AIAA,Wu2018_prf,Srinivasan2019_prf,Volpiani2021_prf} and subgrid-scale models \citep{Meldi2011_pof,Vollant_jot,Meldi2018_ftc,Chandramouli2020_jcp,Lozano2020_arb,Mons2021_prf,Moldovan2022_jfm}. These works have identified theoretical and practical issues and proposed efficient parametric/structural corrections to such models. However, most of these analyses also indicate that a universal predictive model is arguably not achievable because of the high sensitivity to the physical features of the turbulent flows as well as to the numerous numerical details of the simulation process. In addition, data-driven methods require the availability and manipulation of data sets to perform their parametric optimization. The information required from such databases can become prohibitively large when \textit{data hungry} techniques such as deep learning are used. Considering that the optimized turbulence closures via data-driven techniques do not grant universal predictive features, one may wonder if running a classical high-accuracy simulation may be computationally less expensive and more accurate in a large number of cases. 

In the present work, a \textit{physics infused} data-driven strategy based on Data Assimilation is proposed to improve the accuracy of a classical IBM, namely the penalization model \citep{Angot1999}. This analysis aims to exploit available physical knowledge of the flow instead of blindly feeding the algorithms with available data. In this scenario, the physical information available about the flow configuration is mapped in a data set including local and global features, and it is made available over synthetic sensors, which are appositely created and placed. This information is used to enhance the performance of the model. Whenever possible, exploiting physical information is preferable to the usage of Big Data. First, the loading and manipulation of databases are alleviated, significantly reducing the computational costs. Second, a larger degree of flexibility about the number and positioning of the sensors can be easily achieved. This last aspect can be beneficial to enhance the rate of convergence of the underlying optimization processes, mitigating key problems such as over-constraining. In addition, the inclusion of physical criteria and information is not exclusive, and it can be easily integrated with available databases for the analysis of complex cases.

The test case chosen for the present investigation is the turbulent plane channel flow for $Re_\tau \approx 550$. This test case, which has been extensively studied in the literature \citep{Pope2000_cambridge}, is driven by the shear effects at the wall. In this framework, the physical knowledge of the near-wall dynamics observed for this test case will be infused into the DA algorithm to optimize the IBM. In particular, the Ensemble Kalman Filter (EnKF) \citep{Evensen2009_ieee,Asch2016_siam} will be used to infer the parametric behaviour of the penalization method mimicking the near-wall flow dynamics. As three-dimensional, scale-resolving simulations will be performed within the EnKF, the C++ platform CONES recently developed by the team \citep{Villanueva2023_arXiV} will be used to perform the DA strategy \textit{on-the-fly}. This means that the integration of the physical information will be performed while an ensemble of numerical simulations is running and producing instantaneous flow fields, without the need to stop and restart the calculations to perform the optimization process.

The article is structured as follows. In \S\ref{sec:proposedSec2}, the numerical equations are presented, together with the formulation of the IBMs used and the DA algorithm. In \S\ref{sec:proposedSec3}, the test case is introduced and discussed. Preliminary results are presented, and the setup of the DA experiment is detailed. In \S\ref{sec:proposedSec4}, the main results are discussed. The sensitivity of the data-driven method to changes in its hyperparametric description, as well as the underlying CFD model, is investigated. In \S\ref{sec:proposedSec5}, the conclusions are drawn, and future perspectives are discussed.

\section{Numerical Ingredients}\label{sec:proposedSec2}

\begin{figure}
        \centering
        
    \includegraphics[width=1\linewidth, trim = {0 0 0 3cm}]{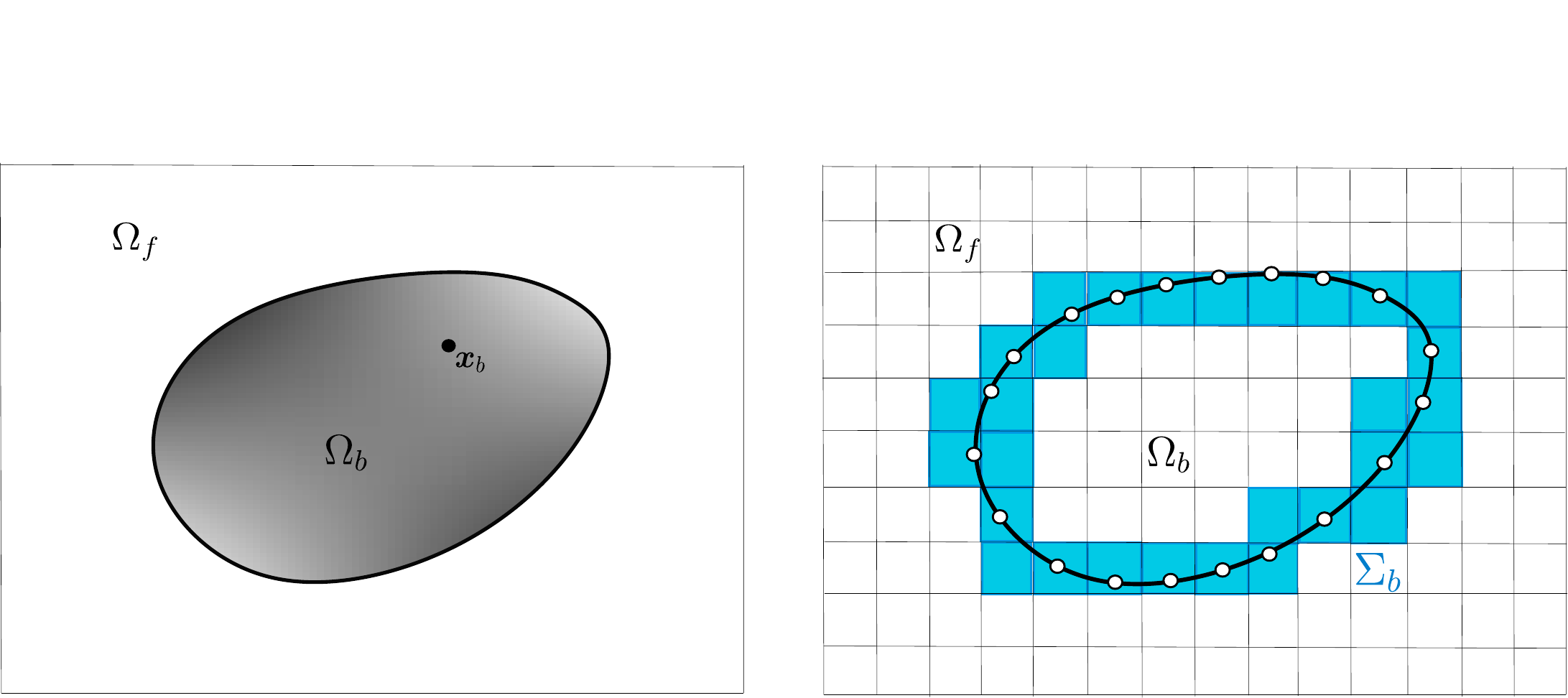} 
    \caption{Qualitative representation of a Cartesian grid used for IBM calculations. $\Omega_f$ and $\Omega_b$ represent the fluid and body domains, respectively. The blue area $\Sigma_b$ represents the body interface.}
    \label{fig:IBM_General}
\end{figure}

Since the original work by \citet{Peskin1972_jcp}, the term IBM has been used to include a large spectrum of methods that simulate immersed (or embedded) boundaries in viscous flows on grids that do not conform to these boundaries. A qualitative representation is shown in figure \ref{fig:IBM_General}. The usage of Cartesian grids permits formulating efficient spectral, finite differences, or finite volume approximations of the partial derivative equations used to represent the physical system. In the present work, we will focus on IBM strategies which rely on the usage of source terms to account for the presence of the immersed body.

\subsection{Dynamic equations and numerical solver}\label{sec:Solver}


An incompressible flow for a Newtonian fluid can be described by the classical 
 form of the Navier--Stokes equations:

\begin{eqnarray}
\bnabla \bcdot \boldsymbol{u} &=& 0 \label{eqn:NavierStokesM} \\
\frac{\partial \boldsymbol{u}}{\partial t} + \bnabla \bcdot (\boldsymbol{u}\boldsymbol{u}) &=& -\nabla p + \nu \nabla^2 \boldsymbol{u} + \boldsymbol{f}_P
    \label{eqn:NavierStokes}
\end{eqnarray}
 
$\boldsymbol{u}$, $p$ and $\nu$ are the velocity field, the pressure field (normalized over the density $\rho$), and the kinematic viscosity, respectively. The term $\boldsymbol{f}_P$ represents a normalized volume force, which could be, for example, the IBM source term used to account for the presence of the immersed body. 




The complete numerical resolution of all active dynamic scales governing the evolution of turbulence via Direct Numerical Simulation (DNS) requires prohibitive computational resources for $\Rey$ values commonly observed in realistic industrial flows. One popular approach used to mitigate this problem and reduce the computational resources required is the Large Eddy Simulation (LES) \citep{Sagaut2006_springer}, which relies on explicit/implicit filtering of the dynamic equations. This procedure excludes the direct calculation of small eddies, whose dynamic effects are modelled. This subgrid-scale closure usually relies on asymptotic theories based on the hypothesis of universal behaviour of the small eddies (i.e. they are statistically independent of the macroscopic features of the flow). Numerical discretization for LES is obtained starting from the filtered equations resolved for the reduced-order variables ($\tilde{\bcdot}$): 

\begin{eqnarray}
\bnabla \bcdot \boldsymbol{\tilde{u}} &=& 0 \label{eqn:NavierStokesFilteredM} \\
\frac{\partial \boldsymbol{\tilde{u}}}{\partial t} + \bnabla \bcdot (\boldsymbol{\tilde{u}}\boldsymbol{\tilde{u}}) &=& -\nabla \tilde{p} + \nu \nabla^2 \boldsymbol{\tilde{u}} - \bnabla \bcdot \boldsymbol{\tau}_{SGS} + \boldsymbol{\tilde{f}}_P
    \label{eqn:NavierStokesFiltered}
\end{eqnarray}


$\boldsymbol{\tau}_{SGS}$ is the subgrid stress tensor and its components are defined as $\tau_{SGS,ij} = \widetilde{u_i \, u_j} - \tilde{u}_i\tilde{u}_j$, where $i,j=x,y,z$ corresponds to the spatial directions of the frame of reference. The tensor $\boldsymbol{\tau}_{SGS}$ must be modelled to close the problem. Among the different proposals in the literature \citep{Sagaut2006_springer}, the \emph{SubGrid-Scale} (SGS) model proposed by \citet{Smagorinsky1963_mwr} is integrated into most of the CFD numerical solvers. This model is based on the Boussinesq turbulent viscosity hypothesis and assumes that $\Rey$ is sufficiently high so that scale separation and turbulence equilibrium are satisfied \citep{Katopodes2018_efm}. Hence, to close the problem, the Boussinesq hypothesis relies on an SGS viscosity $\nu_{SGS}$:

\begin{eqnarray}
    \tau_{SGS,ij} &=& \frac{1}{3} \delta_{ij} \tilde{S}_{kk} -2\nu_{SGS} \tilde{S}_{ij} \\
    \nu_{SGS} &=& \left(C_s \Delta \right)^2 \left(2 \tilde{S}_{ij}\tilde{S}_{ij} \right)^{1/2} \label{eq:filterWidth} \\
    \tilde{S}_{ij} &=& \frac{1}{2}\left(\frac{\partial{\tilde{u}_i}}{\partial{x_j}} + \frac{\partial{\tilde{u}_j}}{\partial{x_i}}\right)
\end{eqnarray}

$C_s \in [0.1, \, 0.2]$ is the so-called Smagorinsky coefficient, $\Delta$ is the filter width,
and $\tilde{S}_{ij}$ is the resolved rate-of-strain tensor. In this work, both DNS and LES will be used. 

Numerical simulations are performed using the C++ finite-volume open-source software \emph{OpenFOAM}. CFD solvers available on this platform, which are based on Finite Volume discretization, have been extensively used by the scientific community in recent years for both academic studies and industrial applications \citep{Tabor2010553,Meldi2012_pof,Selma2014241,Constant2017_cf}. The calculations have been performed using second-order centred schemes for spatial discretization, and a second-order backward scheme has been selected for the time discretization. The numerical system is resolved using a \emph{Pressure-Implicit with Splitting of Operators} (PISO) algorithm  \citep{ISSA198640,Ferziger1996_springer,Versteeg2007_pearson,Greenshields2022_springer}, which is detailed in \S\ref{appAinit} adding the velocity-dependent forcing term in the momentum equations. The simulation performed includes a set of DNS (i.e. numerical resolution of the dynamic system in equations \ref{eqn:NavierStokesM} - \ref{eqn:NavierStokes}) as well as two LES (equations \ref{eqn:NavierStokesFilteredM} - \ref{eqn:NavierStokesFiltered}) which have been closed using the Smagorinsky model. For the latter, two proposals available in the code for the calculation of the filter width $\Delta$ (see equation \ref{eq:filterWidth}) have been selected:

\begin{enumerate}
\item The \emph{cube-root volume delta} $\Delta_c^{CRV}$ method ties the value of $\Delta$ to the geometric local features of the mesh element. The expression used is $\Delta_c^{CRV} = a \,(V_c)^{1/3}$, where $V_c$ is the volume of the mesh element $c$, and $a$ is a parameter to be defined by the user. In this analysis, $a = 1$.
\item The \emph{van Driest damping function} $D=1 - \textrm{exp}\left(-y^+/b \right)$ is used to improve the accuracy of the SGS closure in the near-wall region. $y$ represents the wall-normal direction, and $y^+ = y / \delta_\nu$ is the adimensionalized wall distance over the viscous wall unit $\delta_\nu= \nu / u_\tau$. The friction velocity $u_\tau = \sqrt{\tau_w / \rho}$ is determined via the calculated shear stress at the wall $\tau_w$. In this case, the filter width is $\Delta_c^{vD} = \textrm{min} \left( D \kappa y/d, \Delta_c^{CRV} \right)$. The values of the empirical coefficients are $b = 26$, $\kappa = 0.41$ (\emph{von Karman} constant), and $d = 0.158$. In practice, $\Delta$ performs here as a \emph{cube-root volume delta} model far from the wall, while it exhibits significantly smaller values in the proximity of the body surface. 
\end{enumerate}

\subsection{Immersed Boundary Method}\label{sec:introIBM}


As discussed in \S\ref{sec:introduction}, the Immersed Boundary Method (IBM) includes a very wide range of applications that target the accurate representation of immersed bodies using non-body-fitted grids. In the present study, two proposals relying on the usage of source terms that are included in the dynamic equations will be introduced and used in the numerical simulations.

The first approach considered is the classical penalization method proposed by \cite{Angot1999}. This continuous method employs Darcy's law to obtain a closed expression for the source term $\boldsymbol{f}_P$ in the body interface $\Sigma_b$ and the solid region $\Omega_b$. To do so, a spatial dependent tensor $\mathsfbi{D}(\boldsymbol{x}, t)$ must be defined. The resulting penalty term is modelled as:

\begin{equation}
     \boldsymbol{f}_P = \left\{
        \begin{array}{ll}
        \boldsymbol{0} & \textrm{if } \boldsymbol{x} \in \Omega_f \\
        -\nu \, \mathsfbi{D} \left(\boldsymbol{u} - \boldsymbol{u_{ib}} \right) & \textrm{if } \boldsymbol{x} \in (\Sigma_b \cup \Omega_b)
        \end{array} \right.
    \label{eqn:forcePenDarcy}
\end{equation}



$\boldsymbol{u_{ib}}(\boldsymbol{x}, t)$ is a target velocity representing the immersed body's physical behaviour. If the body surface is not moving, then $\boldsymbol{u_{ib}}=\boldsymbol{0}$. The components $D_{ij}$ of the tensor $\mathsfbi{D}$ are usually controlled to obtain the best compromise between accuracy and stability of the numerical solver \citep{Verzicco2022_arfm}. The choice performed is particularly important in the proximity of the grid transition between the flow region $\Omega_f$ and the interface region $\Sigma_b$.

The second IBM approach considered in this study is a discrete penalty method already validated in \emph{OpenFOAM} by \citet{Constant2017_cf}. The method is based on the work by \cite{Uhlmann2005_jcp} and \citet{Pinelli2010_jcp}, but it is improved by including a \textit{Reproducing Kernel Particle Method} (RKPM) \citep{Liu1995_nmf}. A complete description of this method is reported in \S\ref{appB}. This discrete IBM relies on two complementary physical spaces. The \textit{Eulerian} domain is described by the grid used for calculation, while the \textit{Lagrangian} Markers are discrete points representing the immersed body's surface. The method determines the source term via a two-step procedure:
\begin{enumerate}
\item In the \textit{interpolation step}, the physical variables describing the flow, which are calculated on the Eulerian mesh, are interpolated on the Lagrangian Markers. The source term is then calculated on the Lagrangian space. Its structural form is very similar to the one seen for the penalization method in equation \ref{eqn:forcePenDarcy}:

\begin{equation}
\boldsymbol{F}_P = a_s \left( \boldsymbol{U^\star} - \boldsymbol{U_{ib}} \right)
\label{eqn:forceLagrangian}
\end{equation}

Capital letters are used to indicate the variables in the Lagrangian space. $\boldsymbol{U^\star}$ is the velocity field interpolated from the Eulerian grid while $a_s$ is a coefficient resulting from the discretization procedure. $\boldsymbol{U_{ib}}$ is homologous to $\boldsymbol{u_{ib}}$ in the Lagrangian framework.

\item After the source term is calculated on the Lagrangian space, it is projected on the Eulerian grid during the \textit{spreading step}. This procedure allows obtaining a closed expression for the source term to be integrated within the computational solver.
\end{enumerate}
The main difference between the two algorithms presented is that the penalization method is strictly local i.e. the forcing depends exclusively on the flow field predicted in the correspondent mesh element. On the other hand, the discrete method relies on an interpolation stencil to communicate between the Eulerian grid and the Lagrangian Markers. The user can select the size in terms of grid elements for this structure. Larger interpolation stencils connect a larger number of mesh elements with each Lagrangian Markers. This improves the stability of the numerical algorithms as well as the smoothness of the solution. However, larger computational stencils are also responsible for higher computational requirements. In addition, the size of the computational stencil (for the discrete method) and the width of the interface region $\Sigma_b$/value of the coefficients $D_{ij}$ (for the penalization model) can be responsible for diffused interfaces, which can affect the precision of the numerical results.

\subsection{Sequential Data Assimilation: The Ensemble Kalman Filter}\label{sec:introDA}


Data Assimilation (DA) techniques \citep{Daley1991_cup, Asch2016_siam} have emerged in recent times as a powerful tool to enhance the reliability of numerical simulations in Fluid Dynamics. These approaches enable the integration of high-accuracy observation (DNS or experiments, for example) into reduced-order numerical models to obtain improved predictions of the underlying physical system. DA approaches are usually grouped into two main families, namely variational methods and sequential methods. The former, which includes well-known methods such as 3DVar and 4DVar,  resolves the DA problem as an optimization task in which initial conditions and/or dynamic models are determined to minimize a prescribed cost function. These models are extremely accurate, but they may not converge for the analysis of multi-scale time-evolving physical processes \citep{Sirkes1997_mwr}. Applications in fluid mechanics mainly deal with stationary configurations \citep{Artana2012_jcp,Foures2014_jfm,Mons2021_jfm}. On the other hand, sequential methods mainly rely on Bayesian techniques to resolve the DA problem. One of the most powerful methods in sequential DA is the Ensemble Kalman Filter (EnKF) \citep{Evensen2009_ieee, Katzfuss2016_tas}. This technique has proved to be efficient in reconstructing turbulent flows in recent years, for both stationary and unstationary configurations \citep{Labahn2019_pci,Zhang2020_jcp, Mons2021_prf, Moldovan2022_jfm}.

The EnKF, which is the strategy selected for the present research work, is an advanced tool based on the Kalman Filter (KF) \citep{Kalman1960_jbs}. The KF assumes that a physical state $\boldsymbol{u}$ can be estimated from a linear discrete \textit{model} $\mathsfbi{M}$ and available \textit{observations} $\boldsymbol{y}$. A common assumption, in particular in fluid mechanics applications, is that the model $\mathsfbi{M}$ can create a quasi-continuous map of the physical phenomena investigated, but it provides a lower fidelity representation. On the other hand, the observation $\boldsymbol{y}$ is more accurate, but it is sparse in space and time. Both sources of information are affected by uncertainties, which are referred to as $v$ and $w$ for the model and the observation, respectively. These uncertainties are usually included in the system as random processes, and they can be approximated through unbiased Gaussian distributions: $v = \mathcal{N} (0, \mathsfbi{Q})$ and $w = \mathcal{N}(0, \mathsfbi{R})$. $\mathsfbi{Q}$ and $\mathsfbi{R}$ are time-dependent matrices representing the covariance matrix for the model and the observation. In the framework of this hypothesis, the process $\boldsymbol{u}$ can be accurately described by the two statistical moments of lower order (mean and variance) of its probability density function (pdf). This implies that the state estimation process performed by the DA approach is governed by the error covariance matrix $\mathsfbi{P} = \mathbb{E}[(\boldsymbol{u}-\mathbb{E}[\boldsymbol{u}])(\boldsymbol{u}-\mathbb{E}[\boldsymbol{u}])^T]$. The assimilation scheme is composed of two essential steps: 

\begin{enumerate}
    \item A \emph{forecast} $(f)$ step where the physical state $\boldsymbol{u}$ and its error covariance matrix $\mathsfbi{P}$ are advanced in time using the model. Considering a time advancement from the time step $k-1$ to $k$, this step is described by the following equations:
    \begin{eqnarray}
        \boldsymbol{u}_k^f &=&\mathsfbi{M}_{k:k-1} \boldsymbol{u}_{k-1} \label{eq:XforecastKF} \\
        \mathsfbi{P}_k^f &=&\mathsfbi{M}_{k:k-1} \mathsfbi{P}_{k-1} \mathsfbi{M}_{k:k-1}^T + \mathsfbi{Q}_k \label{eq:QforecastKF}
    \end{eqnarray}
    \item An \emph{analysis} $(a)$ phase where the physical state and the model are updated by the DA procedure. The analysis step is performed only if observation $\boldsymbol{y}$ is available at the time step $k$. This update is obtained as:

    \begin{eqnarray}
        \boldsymbol{u}_k^a &=&\boldsymbol{u}_k^f + \mathsfbi{K}_k \left( \boldsymbol{y}_k - \mathsfbi{H}_{k:k-1} \boldsymbol{u}_k^f \right) \\
        \mathsfbi{P}_k^a &=& \left(\boldsymbol{I} - \mathsfbi{K}_k \mathsfbi{H}_{k:k-1} \right) \mathsfbi{P}_k^f 
    \end{eqnarray}
    
\end{enumerate}

$\mathsfbi{H}_{k:k-1}$ is a mathematical operator which projects the state predicted by the model into the space of the observations. The term $\boldsymbol{y}_k - \mathsfbi{H}_{k:k-1} \boldsymbol{u}_k^f$ is referred to as \emph{innovation}, and it measures the discrepancy between model and observation. $\mathsfbi{K}_k$ is the \emph{Kalman gain}, which is a matrix computing the optimal correlation between the system's state prediction and the observed data, is obtained by minimizing the updated error covariance matrix $\mathsfbi{P}_k^a$: 
\begin{equation}
    \mathsfbi{K}_k =\mathsfbi{P}_k^f \mathsfbi{H}_{k:k-1}^T \left(\mathsfbi{H}_{k:k-1} \mathsfbi{P}_k^f \mathsfbi{H}_{k:k-1}^T + \mathsfbi{R}_k\right)^{-1}
\end{equation}
$\mathsfbi{K}_k$ reduces the uncertainty of the final state, taking into account the level of confidence for the model and the observation. One can see in equations \ref{eq:XforecastKF} and \ref{eq:QforecastKF} that the terms $\boldsymbol{u}_{k-1}$ and $\mathsfbi{P}_{k-1}$ are indicated without an affix. The reason is tied to the availability of observation at the time $k-1$ and, therefore, whether an analysis step is performed or not at the previous time step. In the case sensors are fixed in time, $\mathsfbi{H}_{k:k-1}$ is time-independent. This is the case for the present analysis, hence it will be indicated as $\mathsfbi{H}$ for the sake of conciseness.

While KF approaches are used in a number of fields in science, applications in CFD exhibit a number of problematic aspects. First, accurate state estimations are granted only in the framework of the constitutive hypotheses introduced i.e. linearity of the model and Gaussian behaviour of the uncertainties affecting the system. While modifications can be proposed to improve the performance of the KF outside these constraints \citep{Asch2016_siam,Carrassi2018_WIREs}, the resulting global accuracy is sub-optimal.  Second, the time advancement and update of the error covariance matrix $\mathsfbi{P}_k$ is prohibitively expensive for the number of degrees of freedom that need to be simulated in CFD of realistic flows.

These two problematic aspects are strongly mitigated, if not bypassed, by the EnKF. This method relies on an ensemble Monte-Carlo approximation of the error covariance matrix $\mathsfbi{P}_k$, which is calculated only during the analysis step. To this purpose, an ensemble of state vectors $\boldsymbol{u}_i$ ($i \in \left[1, m \right]$, where $m$ is the number of ensemble members) is used to propagate the uncertainty of the model during the forecast step:

\begin{equation}
    \boldsymbol{u}^f_{i,k} = \mathcal{M}_{i,k:k-1} \, \boldsymbol{u}_{i,k-1}
    \label{eqn:forecast}
\end{equation}

Considering that the time advancing of $\mathsfbi{P}_k$ is here suppressed, the ensemble members can be separately advanced in time. Therefore, the model $\mathcal{M}$ does not need to be linear anymore, as the EnKF only employs the physical solutions of the ensemble members calculated at each analysis step. 
When observation is available in time, the model realisations are combined to obtain an \emph{anomaly matrix} $\mathsfbi{X}$, which measures the deviation of the ensemble with respect to the mean $\boldsymbol{\overline{u}}$. Using the hypothesis of statistical independence of the realizations, an ensemble approximation $\mathsfbi{X}$ of the error covariance matrix can be obtained. Assuming that the analysis phase is performed at the time step $k$ as before, the determination of such matrix is obtained via the following equations:

\begin{eqnarray}
    \boldsymbol{\overline{u}}^f_k &=& \frac{\sum_i^m \boldsymbol{u}_{i,k}^f}{m} \\
    \left[\mathsfbi{X}_k^f \right]_i &=& \frac{\boldsymbol{u}^f_{i,k}-\boldsymbol{\overline{u}}^f_k}{\sqrt{m-1}} \label{eqn:anomalies_state} \\
    \mathsfbi{P}^f_k &=& \mathsfbi{X}^f_k \left(\mathsfbi{X}^f_k \right)^T
    \label{eqn:anomalies_state1}
\end{eqnarray}

The states $\boldsymbol{u}$ and $\boldsymbol{\overline{u}}$ are manipulated as vectors of matrix size $[n,1]$, where $n$ is the number of degrees of freedom characterizing the random process. For a CFD application for incompressible flows, the most classical choice is to consider the full velocity field discretized over the $n_{GE}$ grid elements. Therefore, $n=3*n_{GE}$. Similarly, the size of the matrices $\mathsfbi{X}$ and $\mathsfbi{P}$ is $[n,m]$ and $[n,n]$, respectively. In order to obtain a well-posed numerical system, an ensemble of $n_o$ observations is also obtained via a perturbation of the observation vector $\boldsymbol{y}_k$ whose size is $[n_o,1]$. This procedure allows obtaining an observation matrix $\mathsfbi{Y}_k$ of size $[n_o,m]$, whose $m$ columns are defined as $\mathsfbi{Y}_{i,k} = \boldsymbol{y}_k + \boldsymbol{e}_i, \, i=1,2,...,m$. The random noise $\boldsymbol{e}_i$ is described by a Gaussian probability function $\boldsymbol{e}_i \sim \mathcal{N} (0, \mathsfbi{R}_k)$. As for the KF, $\mathsfbi{R}_k$ represents the observation covariance matrix. 

Similarly, the EnKF requires the use of a non-linear sampling matrix $\mathcal{H} (\boldsymbol{u}^f_k)$ to project the model into the position of the observations. Analogously to the system's state $\boldsymbol{u}_k^f$ in equation \ref{eqn:anomalies_state}, it can be decomposed into a matrix $\mathsfbi{S}_k^f$ where each column represents a normalized anomaly, and the mean is defined as $\overline{\mathcal{H}(\boldsymbol{u}^f_k}) = \sum_i^m \mathcal{H}(\boldsymbol{u}_{i,k}^f) / m$:

\begin{equation}
    \left[\mathsfbi{S}_k^f \right]_i = \frac{\mathcal{H}(\boldsymbol{u}_{i,k}^f) - \overline{\mathcal{H}(\boldsymbol{u}^f_k)}}{\sqrt{m-1}}
    \label{eqn:anomalies_stateSampling}
\end{equation}

For an infinite ensemble size, $\lim_{m \to +\infty} \mathbb{E}[(\boldsymbol{e} - \mathbb{E}[\boldsymbol{e}]) (\boldsymbol{e} - \mathbb{E}[\boldsymbol{e}])^T] = \mathsfbi{R}$, which becomes time-independent, and the Kalman gain matrix $\mathsfbi{K}_k$ describing the most optimal correlation between the state and the observations is simplified to \citep{Hoteit2015_mwr, Carrassi2018_WIREs}:

\begin{equation}
    \mathsfbi{K}_k = \mathsfbi{X}_k^f \left(\mathsfbi{S}_k^f \right)^T \left[\mathsfbi{S}_k^f \left(\mathsfbi{S}_k^f \right)^T + \mathsfbi{R}_k \right]^{-1}
    \label{eqn:KalmanGain}
\end{equation}

All the elements presented constitute the essential ingredients to calculate the updated system's state $\boldsymbol{u}_k^a$, which is then used for the time advancement of the numerical model $\mathcal{M}$ in the following iteration $k+1$:

\begin{equation}
    \boldsymbol{u}_k^a = \boldsymbol{u}_k^f + \mathsfbi{K}_k \left( \mathsfbi{Y}_k - \mathcal{H}(\boldsymbol{u}_k^f)\right)
    \label{eqn:updatedState}
\end{equation}

The step-by-step implementation of the EnKF is detailed in the algorithm \ref{alg:EnKF}. The following state-of-the-art modifications are also included in the classical EnKF formulation:
\begin{itemize}
    \item The EnKF is naturally used to update the system's physical state, but it can also be used to update free coefficients $\theta$ governing the model $\mathcal{M}$. In the present work, this task is performed via an \textit{extended state} approach \citep{Asch2016_siam}. In practice, the parameters are included in the state vector $\boldsymbol{u}_i^\prime = [\boldsymbol{u}_i \, \theta_i]^T$ for each realisation, and they are updated with the physical field using equation \ref{eqn:updatedState}.
    \item Some \emph{deterministic inflation} \citep{Asch2016_siam} is included to increase the variability of the state predicted by the EnKF:

    \begin{equation}
        \boldsymbol{u}_{i,k}^a = \boldsymbol{\overline{u}}_k^a + \lambda_k \left(\boldsymbol{u}_{i,k}^a - \boldsymbol{\overline{u}}_k^a   \right)
        \label{eqn:inflation}
    \end{equation}

    where $(\lambda > 1)$. This procedure usually improves the accuracy of the updated system's state $\boldsymbol{u}_k^a$, since $\mathsfbi{P}_k$ may be underestimated due to the sampling errors deriving from the use of a limited amount of members $m$. Also, this procedure may also prevent a premature collapse of the system leading to the divergence of the filter.
    
    \item \emph{Covariance localization} of the Kalman gain $\mathsfbi{K}_k$ is used to gradually set to zero the correlation between the observations and the system's state with increasing distance. This choice controls the emergence of some undesired spurious fluctuations due to the underestimation of $\mathsfbi{P}_k$. Covariance localisation is usually performed premultiplying $\mathsfbi{K}_k$ with a used-defined matrix $\mathsfbi{L}_k$. This matrix, for which $[L]_{ij} \in [0, 1]$, is defined using an exponential decay in space so that the update of the model state tends to zero when far observation is considered. If the grid used by the model and the location of the sensors do not change in time, $\mathsfbi{L}$ is also time-independent. This is the case for the present investigation.
    
\end{itemize}

\begin{algorithm}
    \caption{Scheme of the EnKF used in the present study.}
    \label{alg:EnKF}
    \textbf{Input:} $\mathcal{M}$, $\mathsfbi{R}$, $\boldsymbol{y}_k$, and a prior/initial state system $\boldsymbol{u}_{i,0}$, where usually $\boldsymbol{u}_{i,0} \sim \mathcal{N}(\boldsymbol{\overline{u}}_0, \sigma^2)$ \\
    \For{$k = 1, 2,..., K$}{
        \For{$i = 1, 2, ..., m$}{
    \nl Advancement in time of the state vectors:\\
    \qquad $\boldsymbol{u}_{i,k}^f = \mathcal{M}\,\boldsymbol{u}_{i,k-1}$ \\
    \nl Generation of an observation matrix from the confidence level given to the observation data:\\
    \qquad$\mathsfbi{Y}_{i,k} = \boldsymbol{y}_k + \boldsymbol{e}_i$, with $\boldsymbol{e}_i \sim \mathcal{N}(0,\mathsfbi{R})$\\
    \nl Estimation of the ensemble means (system state and projection matrix):\\
    \qquad$\boldsymbol{\overline{u}}_k^f = \frac{1}{m}\sum_{i = 1}^{m}\boldsymbol{u}_{i,k}^f$,\,
    $\overline{\mathcal{H}(\boldsymbol{u}^f_k}) = \frac{1}{m}\sum_{i = 1}^m \mathcal{H}(\boldsymbol{u}_{i,k}^f)$ \\
    \nl Computation of the anomaly matrices (system state and projection matrix):\\
    \qquad$\mathsfbi{X}_k = \frac{\boldsymbol{u}_{i,k}-\boldsymbol{\overline{u}}_k}{\sqrt{m-1}}$,\,
    $\mathsfbi{S}_k = \frac{\mathcal{H}(\boldsymbol{u}_{i,k}^f) - \overline{\mathcal{H}(\boldsymbol{u}^f_k)}}{\sqrt{m-1}}$ \\
    \nl Calculation of the Kalman gain (with covariance localization):\\
    \qquad$\mathsfbi{K}_k = \mathsfbi{L} \odot \mathsfbi{X}_k^f(\mathsfbi{S}_k)^T \left[\mathsfbi{S}_k(\mathsfbi{S}_k)^T + \mathsfbi{R}\right]^{-1}$\\
    \nl Update of the state matrix:\\
    \qquad$\boldsymbol{u}_{i, k}^a = \boldsymbol{u}_{i,k}^f + \mathsfbi{K}_k \left(\mathsfbi{Y}_{i,k}- \mathcal{H}(\boldsymbol{u}_{i,k}^f) \right)$ \\
    \nl Application of (deterministic) covariance inflation: \\
    \qquad$ \boldsymbol{u}_{i,k}^a = \boldsymbol{\overline{u}}_k^a + \lambda_k \left(\boldsymbol{u}_{i,k}^a - \boldsymbol{\overline{u}}_k^a   \right)$
    }
    }
\end{algorithm}

\section{Test case: turbulent plane channel flow, $\boldsymbol{Re_{\tau}=550}$}\label{sec:proposedSec3}


The turbulent plane channel flow is a fundamental benchmark test case in CFD, and open databases are available online, providing high-accuracy DNS data for reference. On the one hand, the simple geometry of this test case allows the exclusion of a number of physically complex aspects in the optimization process, such as, for example, the separation of the boundary layer. On the other hand, because of the velocity fluctuations observed in the proximity of the wall, traditional IBMs usually fail to provide an accurate flow prediction. For all of these reasons, this test case is a suitable candidate for an ambiguous assessment for IBM augmentation via DA.   

The friction Reynolds number considered is $\Rey_\tau=u_\tau h / \nu = 550$. The friction velocity was previously introduced, and it is defined as $u_\tau= \sqrt{\tau_w / \rho}$ where $\tau_w$ is the wall shear stress, while $h$ is the half-height of the channel. The coordinate system is set so that $x$ is the streamwise direction, $y$ is the wall-normal direction, and $z$ is the spanwise direction. The two walls are positioned at $y=0, \, 2h$, respectively. Periodic boundary conditions are applied to the side walls, and the mass flow rate is conserved in time using a source term for the dynamic equations already implemented in \textit{OpenFOAM}. Results are usually presented in non-dimensional form (suffix $+$) using the friction velocity $u_\tau$ and the viscous wall unit $\delta_\nu = \nu / u_\tau$ for normalization. A second non-dimensional form used in the present analysis (suffix $\star$) relies on $u_\tau$ calculated by a reference simulation (R-DNS-BF) introduced in \S\ref{sec:preliminaryAnalysis}. In addition, statistical moments are also averaged over the two directions $x$ and $z$, for which statistical homogeneity is observed. 

\subsection{Reference simulation and preliminary results}\label{sec:preliminaryAnalysis}

\begin{figure}
    \begin{tabular}{ccc}
    \includegraphics[width=0.32\linewidth]{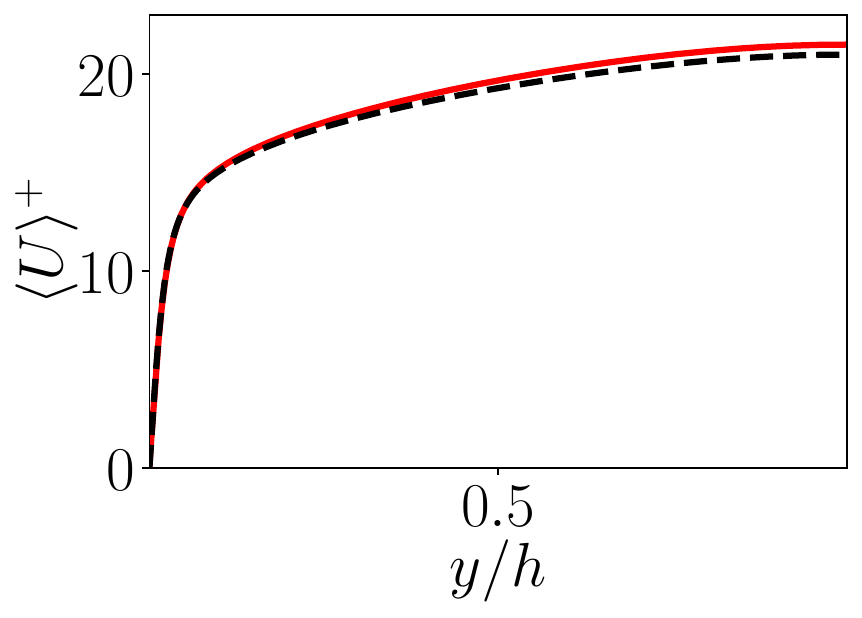} & \includegraphics[width=0.32\linewidth]{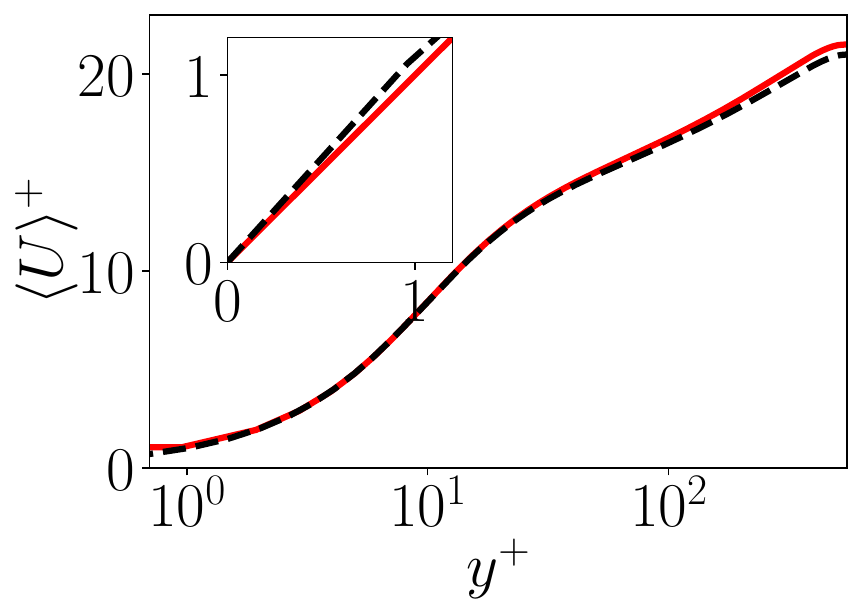} &
    \includegraphics[width=0.32\linewidth]{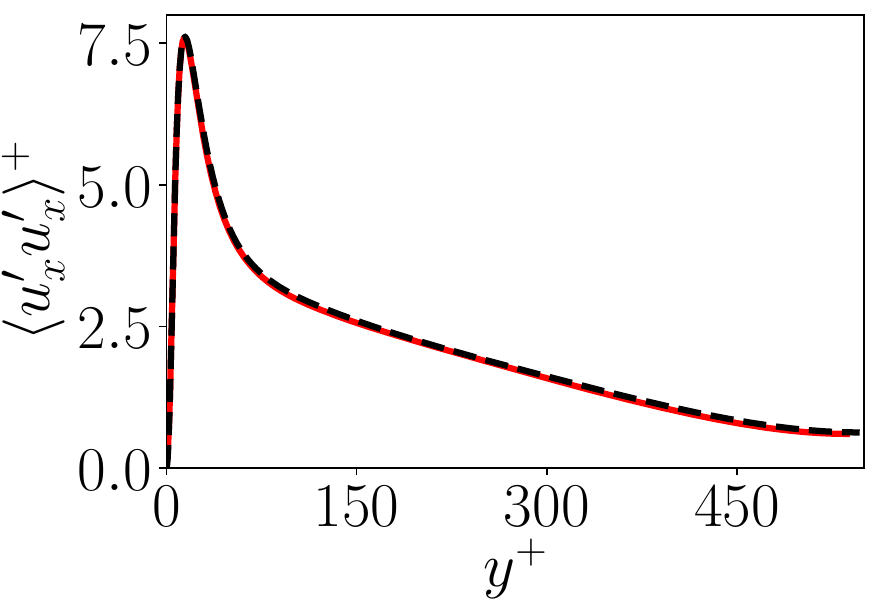} \\
    \textit{(a)} & \textit{(b)} & \textit{(c)} \\
    \includegraphics[width=0.32\linewidth]{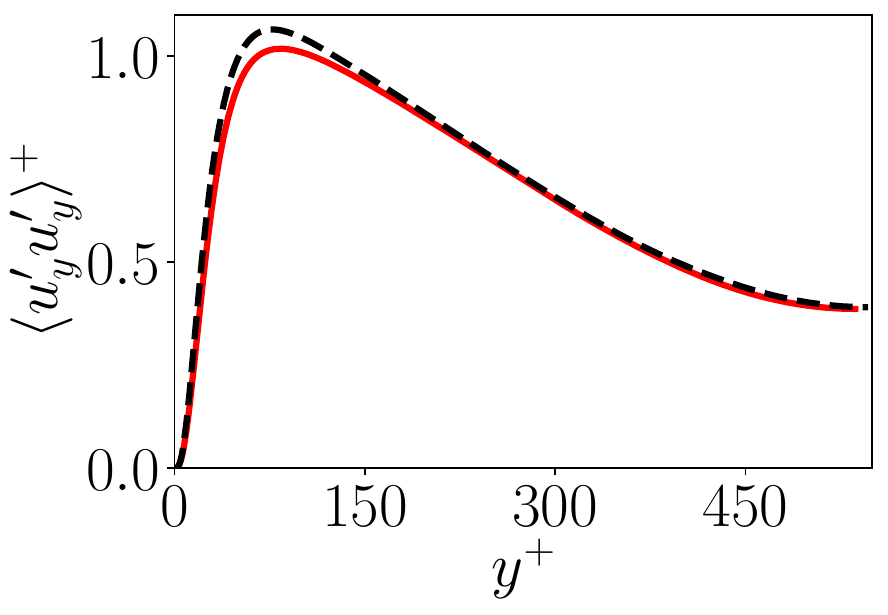} & \includegraphics[width=0.32\linewidth]{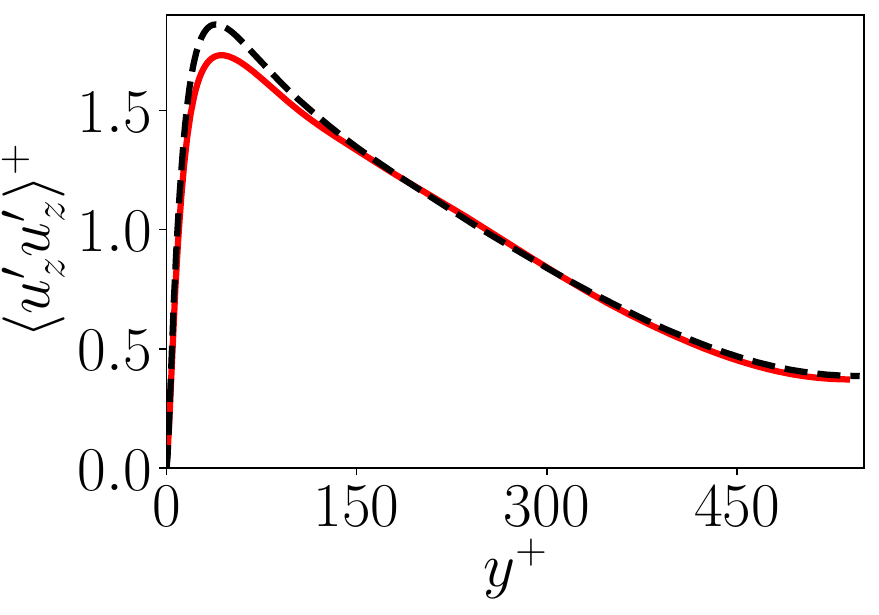} &
    \includegraphics[width=0.32\linewidth]{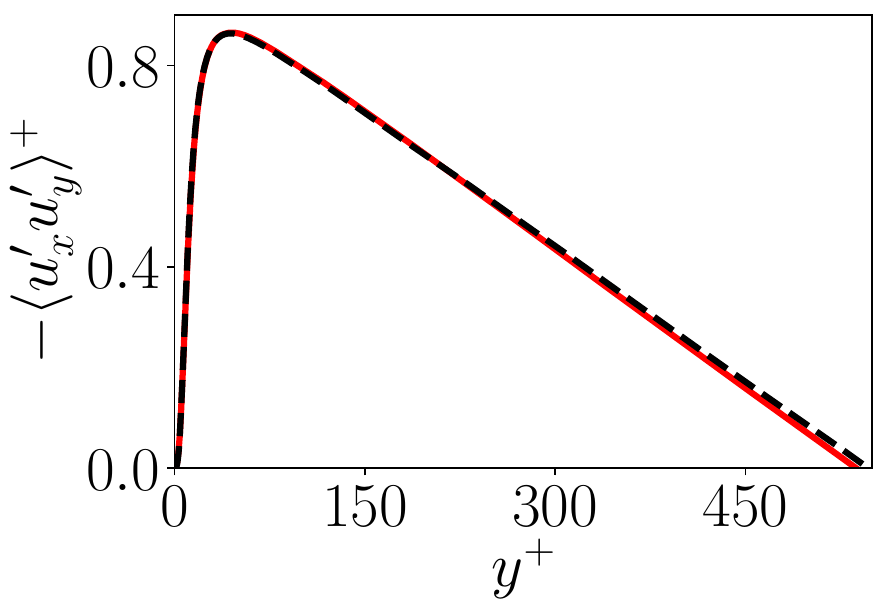} \\
    \textit{(d)} & \textit{(e)} & \textit{(f)}
    \end{tabular}
    \caption{Comparison of the main statistical moments of the velocity field. Results are shown for (\protect\redline) the simulation R-DNS-BF and (\protect\blacklinedashed) the database by \cite{DelAlamo2003_pof}.}
    \label{fig:Hoyas}
\end{figure}
\begin{figure}
        \centering\begin{tabular}{cc}
    \includegraphics[width=0.5\linewidth, trim = {0 -2.1cm 0 0}, clip]{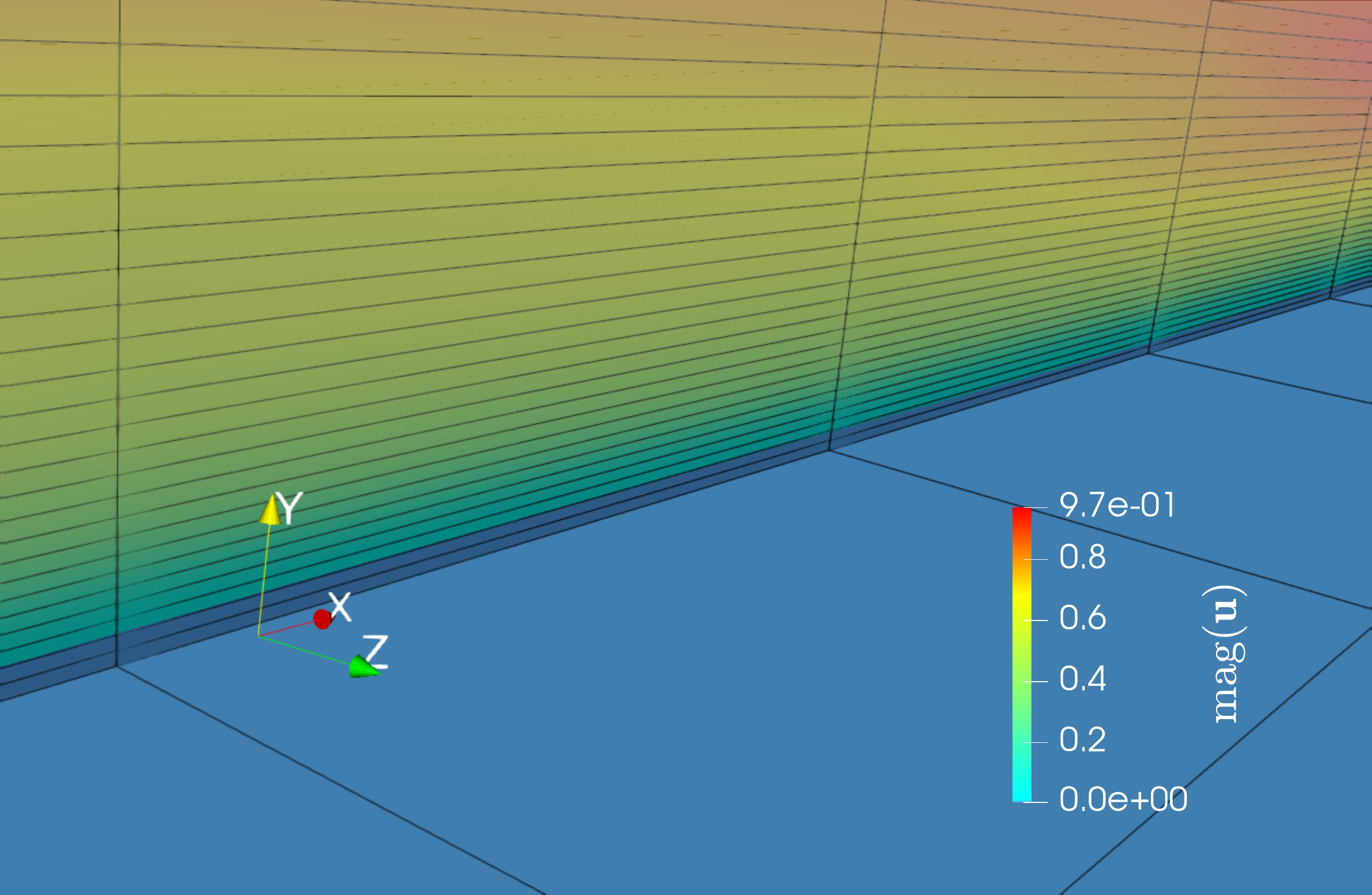} & \includegraphics[width=0.445\linewidth]{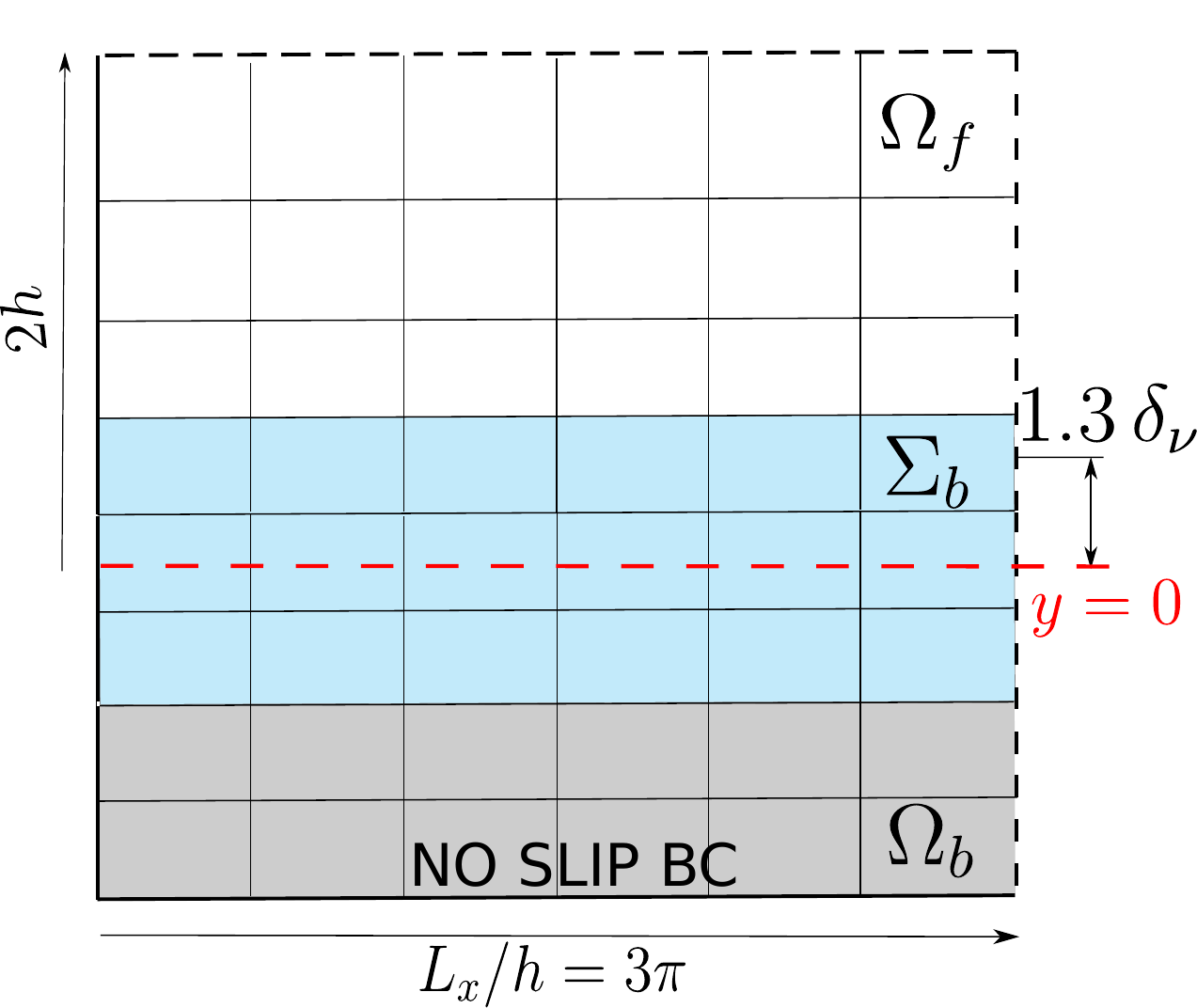} \\
    \textit{(a)} & \textit{(b)} \\
        \end{tabular}
    \caption{\textit{(a)} General view of the grid employed for the plane channel with IBM. \textit{(b)} Cartesian grid employed when using the IBM for the analysis of the turbulent plane channel. $\Omega_f$ and $\Omega_b$ represent the fluid and body domains, respectively (the latter displayed in grey). The blue area $\Sigma_b$ (three cell layers in the $y$ direction) constitutes the body interface.}
    \label{fig:IBM_Mesh}
\end{figure}

A database of classical numerical simulations has been performed to obtain preliminary results. Details are reported in the first six rows of table \ref{tab:summary}. The physical domain is discretized using a uniform distribution of the elements in the $x$ and in the $z$ direction. A geometric distribution is used in the vertical direction, increasing the size of the grid while moving away from the wall. Therefore, $\Delta y^\star_{min}$ represents the grid size at the wall, and $\Delta y^\star_{max}$ is the size at the centre of the channel. This analysis has been performed to obtain a suitable initial \textit{prior} state to be optimized with DA. The database includes several simulations obtained with different techniques:

\begin{itemize}
\item One body-fitted DNS (R-DNS-BF), which represents the reference simulation that will be used to validate the results. Statistical moments for this simulation have been obtained for $300$ advective times $t_A = h / U_c$, where $U_c$ is the averaged velocity at the centerline for $y=h$. The averaged velocity profile and the components of the Reynolds stress tensor compare well with results by \cite{DelAlamo2003_pof} and \cite{Hoyas2008_pof} for similar $\Rey_\tau$ as shown in figure \ref{fig:Hoyas}. The minor discrepancies observed could be due to a small difference in $u_\tau$ (around $2.55 \%$), to the rate of convergence of the statistical moments, to the different discretisation strategies or to the different distribution of the mesh elements in the wall-normal direction.
\item One body-fitted DNS (DNS-BF), which is run on a smaller domain and uses a lower grid resolution when compared with the simulation R-DNS-BF. Details of the grid are provided in table \ref{tab:summary}. One can see that a factor of $4$ coarsens the resolution in the $x$ and $z$ directions. In the vertical direction, the number of mesh elements is also reduced by a factor of $4$, and the ratio coefficient of the geometric distribution of the elements is different. This last choice has been performed to obtain a similar resolution in the wall region when compared with the refined grid used for the simulation R-DNS-BF.
\item Two LES, which are run on the same grid as DNS-BF. The first simulation, BF-LES, is performed using Smagorinsky's model as subgrid-scale closure. The second one, BF-LES-VD, includes \emph{van Driest}'s correction at the wall for the subgrid-scale model.
\item Two IBM runs, which are performed on a grid very similar to the one used for DNS-BF shown in figure \ref{fig:IBM_Mesh}. The differences emerge in the proximity of the wall. For the IBMs, the wall mesh element has its centre in $y=0$ (or $y=2h$ at the top), and three additional layers of mesh elements of the same size are included in the solid region, considering one of them in the interface $\Sigma_b$. These layers are placed to ensure the numerical stability of the algorithm. The two calculations are referred to as DNS-IBM-CF for continuous forcing (penalization) and DNS-IBM-DF for discrete forcing. For both simulations, the bottom and top walls are not moving, so $\boldsymbol{u_{ib}} = \boldsymbol{0}$ and $\boldsymbol{U_{ib}} = \boldsymbol{0}$, respectively. The volume force $\boldsymbol{f}_P$ is non-zero at the interface region $\Sigma_b$ (light blue region in figure \ref{fig:IBM_Mesh}\textit{(b)}), which consists of the three closest mesh elements in the $y$ direction to the immersed walls ($y=0, \, 2h$). The interface region $\Sigma_b$ has been chosen to be the same for the two methods in order to provide a rigorous comparison. The Lagrangian Markers for the discrete method are positioned at the centre of each mesh element for $y=0$, and the computational stencil is made by $3 \times 3 \times 3$ elements. For the penalized method, $\boldsymbol{f}_P$ is also included in the solid domain $\Omega_b$, which is in this case represented by a layer of two cells in the $y$ direction (grey in figure \ref{fig:IBM_Mesh}\textit{(b)}). 

\end{itemize}

Results obtained with the simulations of the database are now compared. The aim of the present analysis is to evaluate the accuracy of the calculations performed with the coarse grids, as well as to assess the efficacy of SGS modelling and IBM in this case. Comparisons will be performed against the simulation R-DNS-BF, which is considered to be the \textit{true} physical state.

First, the predicted friction velocity $u_\tau$ is analyzed. The values obtained by the different simulations, which are reported in table \ref{tab:summary}, show that most of the simulations significantly under-predict $u_\tau$. The only exception is represented by the simulation LES-BF i.e. the LES calculation for which the classical Smagorinsky model is used. In this case, discrepancies with the reference are of the order of $7.29\%$. However, one may argue that this apparently acceptable prediction is actually the result of compensation between different error sources, such as explicit filtering, the Smagorinsky model and the discretization schemes. Applications of LES to the turbulent plane channel exhibit very high sensitivity to these interactions \citep{Meyers2007_pof}. The lack of global accuracy for the simulation LES-BF is shown by the analysis of the physical flow fields. 

\begin{figure}
    \begin{tabular}{ccc}
    \includegraphics[width=0.32\linewidth]{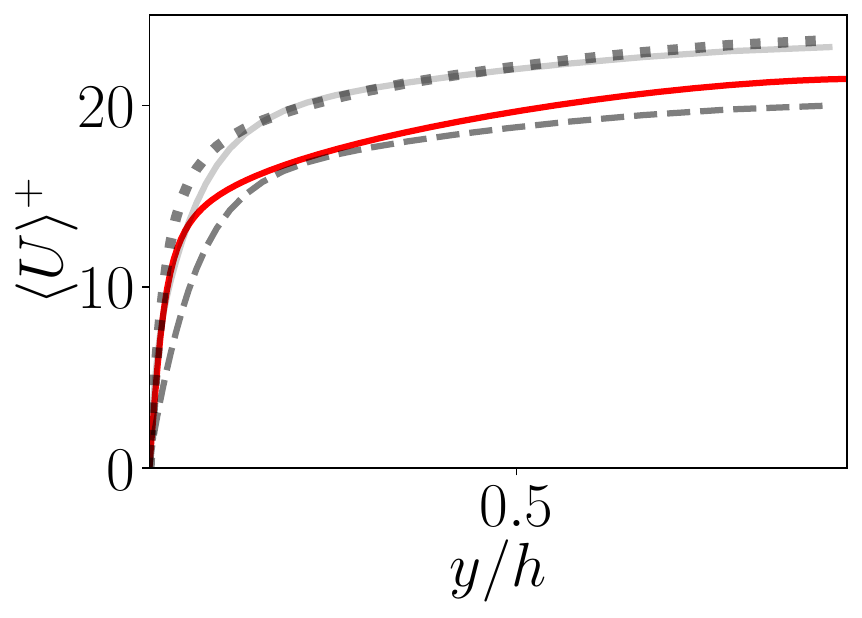} & \includegraphics[width=0.32\linewidth]{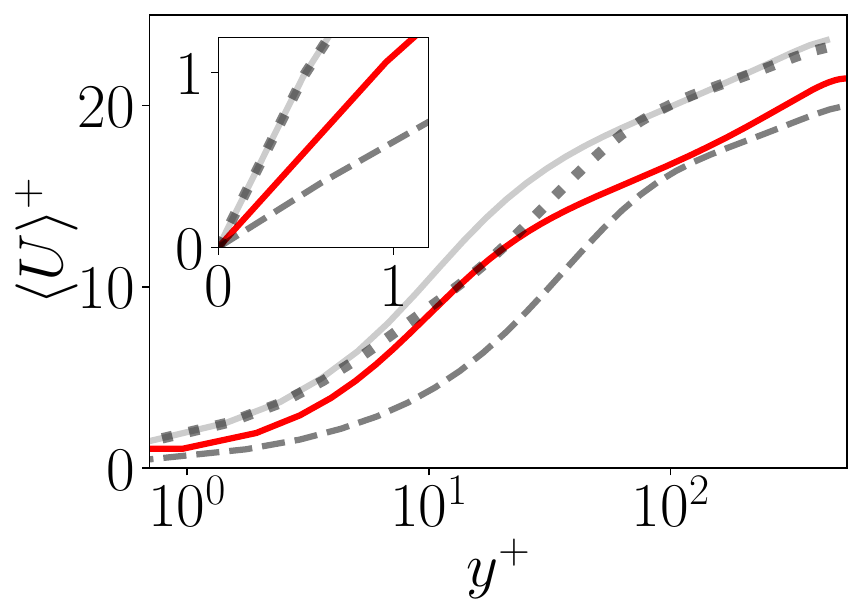} &
    \includegraphics[width=0.32\linewidth]{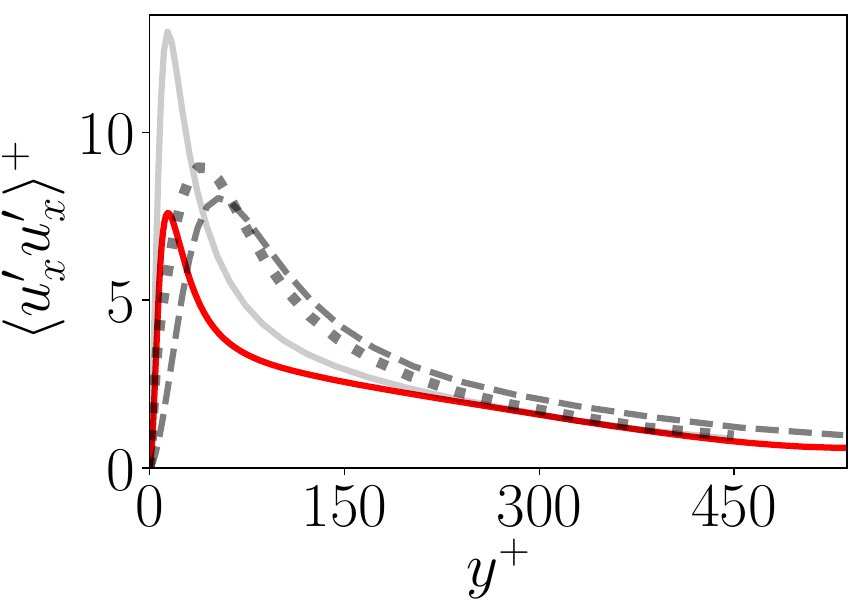} \\
    \textit{(a)} & \textit{(b)} & \textit{(c)} \\
    \includegraphics[width=0.32\linewidth]{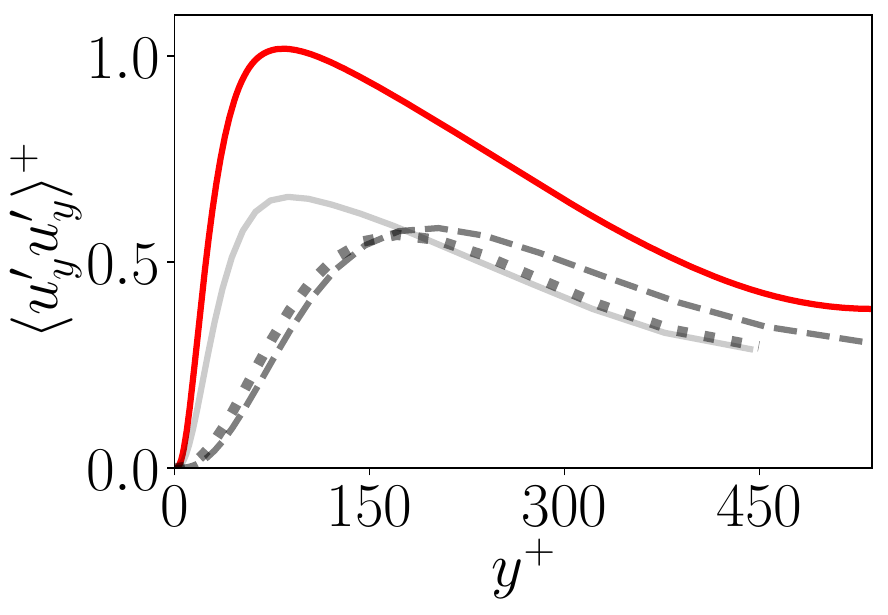} & \includegraphics[width=0.32\linewidth]{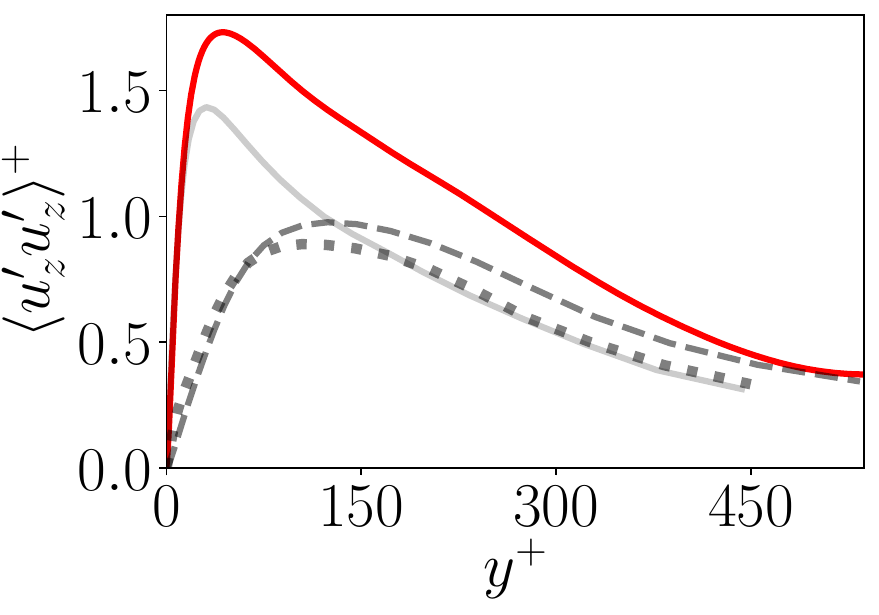} &
    \includegraphics[width=0.32\linewidth]{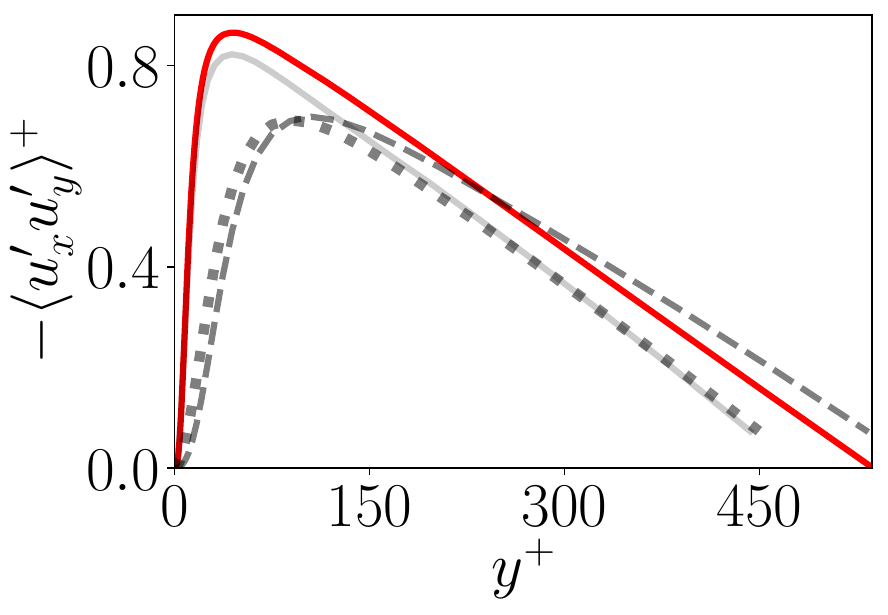} \\
    \textit{(d)} & \textit{(e)} & \textit{(f)}
    \end{tabular}
    \caption{Comparison of the main statistical moments of the velocity field. Results are shown for simulations (\protect\greylinesolidsoft) DNS-BF, (\protect\greylinedashedstrong) BF-LES, (\protect\greylinedottedstrong) BF-LES-VD, and (\protect\redline) R-DNS-BF.}
    \label{fig:bodyFitted}
\end{figure}

Figure \ref{fig:bodyFitted} shows the evolution of the main statistical quantities of the flow, which include the mean streamwise velocity $\langle U \rangle ^+ = \langle u_x \rangle / u_\tau $ and the components of the Reynolds stress tensor $\langle u_i ^\prime u_j ^\prime \rangle ^+ = \langle u_i ^\prime u_j ^\prime \rangle / u_\tau^2$, for the body-fitted simulations. 
Significant discrepancies are overall observed in figure \ref{fig:bodyFitted}\textit{(a)} and \textit{(b)} for the normalized mean streamwise velocity $\langle U \rangle ^+$. The simulation LES-BF complies well with the DNS data far from the wall, thanks to the predicted value for $u_\tau$. However, the velocity profile in the proximity of the wall is erroneous in this case. Simulations DNS-BF and LES-VD-BF perform reasonably better close to the wall for $y^+ < 10$ and $y^+ < 30$, respectively. However, the accuracy is significantly degraded in the outer layer, owing to the under-prediction of $u_\tau$. A high discrepancy is observed as well for the components of the resolved Reynolds stress tensor, which are shown in figure  \ref{fig:bodyFitted}\textit{(c)} to \textit{(f)}. One can see that the LES do not provide an accurate prediction for the position of the peak, which is significantly far from the wall for all the components. On the other hand, the simulation DNS-BF provides an accurate estimation of this feature. Differences are observed for the predicted magnitude of the different components. One can see in figure \ref{fig:bodyFitted}\textit{(c)} how both the LES and the calculation DNS-BF significantly over-predict the component $\langle u^{\prime}_x u^{\prime}_x \rangle^+ $. A general under prediction is instead observed for the components $\langle u^{\prime}_y u^{\prime}_y \rangle^+ $ and $\langle u^{\prime}_z u^{\prime}_z \rangle^+ $. A global reduction of the magnitude would be expected, considering that the lack of grid resolution is expected to dampen the velocity fluctuations. However, it also affects the accuracy in the prediction of their spatial gradients, which govern the dissipation of each component of the Reynolds stress tensor. Therefore, the complex results observed are mainly due to the choices performed in terms of mesh resolution. The component $\langle u^{\prime}_x u^{\prime}_y \rangle^+ $ shown in figure \ref{fig:bodyFitted}\textit{(f)}, which is tied to turbulence production effects, is the one for which the smaller discrepancy is globally observed. One can see that in every case, the magnitude of the components of the Reynolds stress tensor is higher for the DNS-BF calculation when compared with the two LES. The reason is associated with the dissipative effect of the Smagorinsky model used to close the equations.    

\begin{figure}
    \begin{tabular}{ccc}
    \includegraphics[width=0.32\linewidth]{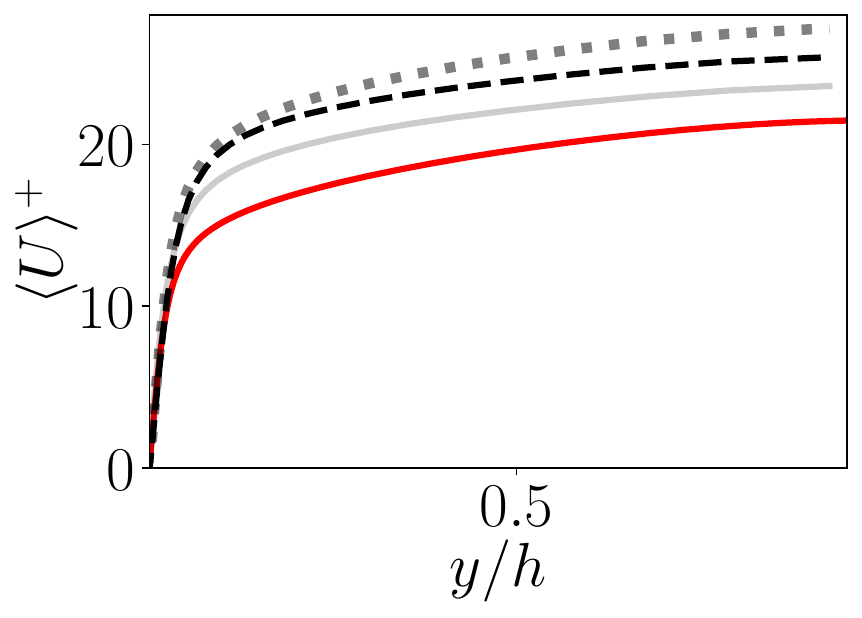} & \includegraphics[width=0.32\linewidth]{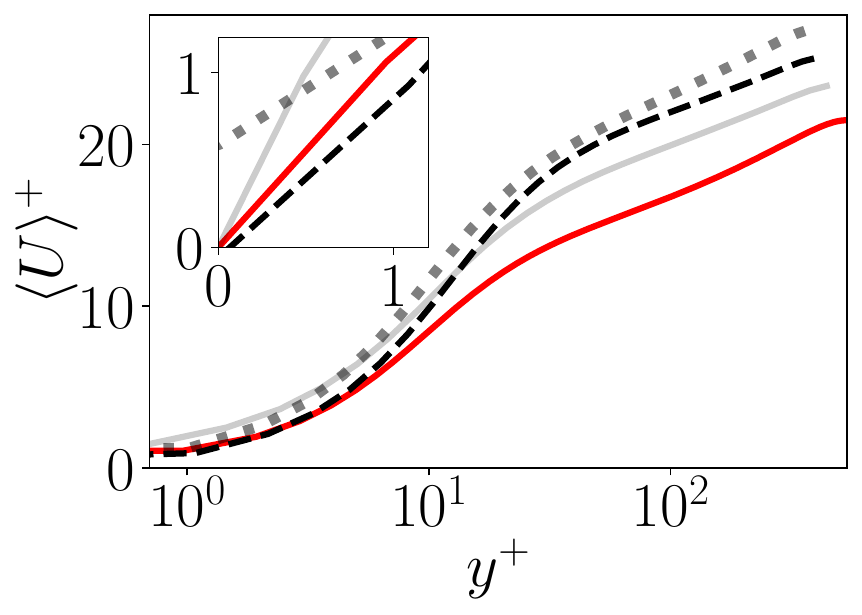} &
    \includegraphics[width=0.32\linewidth]{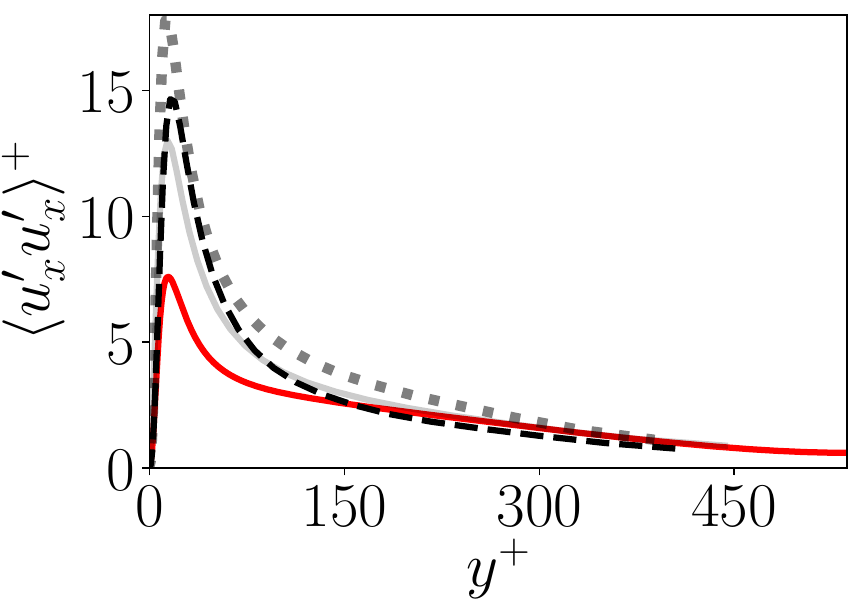} \\
    \textit{(a)} & \textit{(b)} & \textit{(c)} \\
    \includegraphics[width=0.32\linewidth]{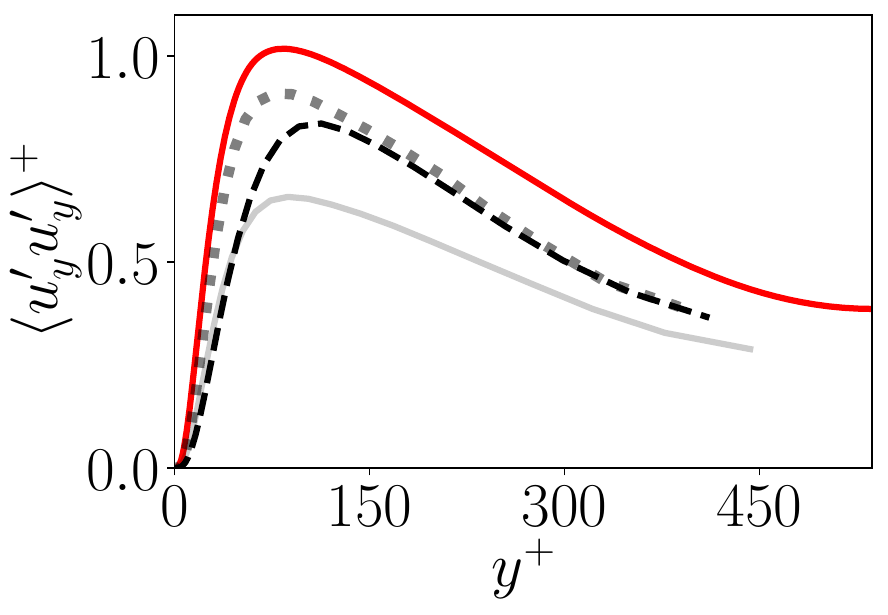} & \includegraphics[width=0.32\linewidth, height=3cm]{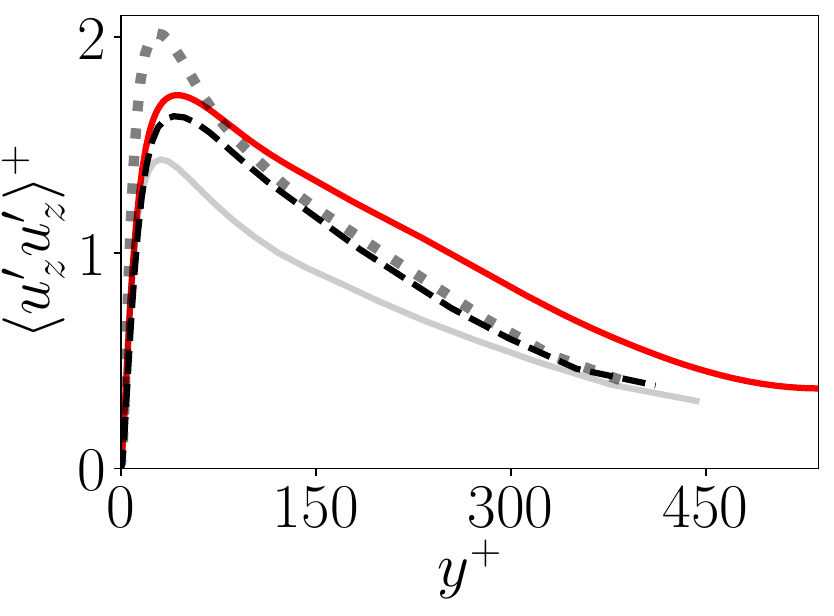} &
    \includegraphics[width=0.32\linewidth]{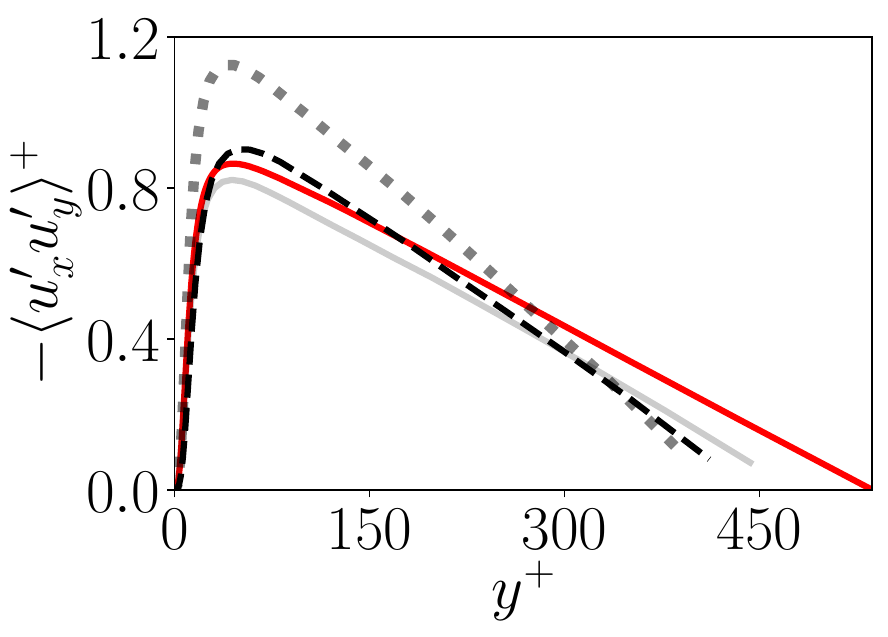} \\
    \textit{(d)} & \textit{(e)} & \textit{(f)}
    \end{tabular}
    \caption{Comparison of the main statistical moments of the velocity field. Results are shown for simulations (\protect\greylinesolidsoft) DNS-BF, (\protect\greylinedottedstrong) DNS-IBM-CF, (\protect\blacklinedashed) DNS-IBM-DF, and (\protect\redline) R-DNS-BF.}
    \label{fig:IBM}
\end{figure}

The simulations run with classical IBM using penalization (DNS-IBM-CF) and discrete (DNS-IBM-DF) approaches are now investigated. Comparisons are performed using the reference run R-DNS-BF as well as the body-fitted calculation DNS-BF, whose grid is almost identical to the one used in the IBM runs. For the continuous IBM method, the tensor $\mathsfbi{D} = D \; \mathsfbi{I}$, where $D=10^5 \, m^{-2}$ for the mesh elements in $\Sigma_b \cup \Omega_b$ and $D=0$ elsewhere. The analysis considers the mean velocity as well as the components of the Reynolds stress tensor, as previously done for the body-fitted simulations. Results are shown in figure \ref{fig:IBM}. One can see that the physical quantities calculated by both IBM strategies are similar, and they are close to findings obtained with the coarse-grained simulation DNS-BF. Looking in detail at the near-wall behaviour of the velocity profile in figure \ref{fig:IBM}\textit{(b)} one can see that simulation DNS-IBM-CF is not able to successfully impose $\boldsymbol{u_{ib}}=\boldsymbol{0}$ at the wall, nor to provide an accurate estimation of the velocity gradient close to the wall.

In summary, the present findings indicate that all the preliminary simulations fail to provide an accurate estimation of the physical flow features, using a coarse grid. One can also add that body-fitted LES provided unreliable results due to the non-linear interactions between different error sources. This uncertainty is arguably going to be magnified by the interactions between subgrid-scale modelling and IBM. For this reason, the DA analyses will be performed using CFD runs without SGS models. This decision excludes a complexity to be considered in the DA study, namely the interaction between the data-driven IBM model derived by DA and the SGS model itself.

\begin{table}
  \begin{center}
\def~{\hphantom{0}}
  \begin{tabular}{lcccccccccc}
       & $N_x \times N_y \times N_z$ & $\Delta x^\star$ &  $\Delta z^\star$ & $\Delta y^\star_{\mathrm{min}}$ & $\Delta y^\star_{\mathrm{max}} $ &
       $L_x/h$ & $L_z/h$ & $u_\tau $ & $\Delta t \, (t_A)$\\
       & & & & & & & & & & \\
       R-DNS-BF & $1\,024 \times 256 \times 512$ & $9.9$ & $6.6$ & $1$ & $11.2$ & $6\pi$ & $2\pi$ & $0.0480$ & $0.004$ \\
       DNS-BF & $128 \times 58 \times 64$ & $39.5$ & $26.3$ & $1.08$ & $52$ & $3\pi$ & $\pi$ & $0.0428$ & $0.02$  \\
       LES-BF& $128 \times 58 \times 64$ & $39.5$ & $26.3$ & $1.08$ & $52$ & $3\pi$ & $\pi$ & $0.0515$ & $0.02$ \\
       LES-VD-BF& $128 \times 58 \times 64$ & $39.5$ & $26.3$ & $1.08$ & $52$ & $3\pi$ & $\pi$ & $0.0434$ & $0.02$ \\
       DNS-IBM-CF & $128 \times 64 \times 64$ & $39.5$ & $26.3$ & $1.3$ & $52$ & $3\pi$ & $\pi$ & $0.0375$ & $0.02$  \\
       DNS-IBM-DF & $128 \times 64 \times 64$ & $39.5$ & $26.3$ & $1.3$ & $52$ & $3\pi$ & $\pi$ & $0.0397$ & $0.02$  \\
       & & & & & & & & & & \\
       DNS-IBM-DA & $128 \times 64 \times 64$ & $39.5$ & $26.3$ & $1.3$ & $52$ & $3\pi$ & $\pi$ & $0.0487$ & $0.02$ 
  \end{tabular}
  \caption{Grid resolution, domain size and time step $\Delta t$ used for the generation of the database of simulations.}
  \label{tab:summary}
  \end{center}
\end{table}

\subsection{Data Assimilation experiment}\label{sec:DA experiments}


A DA strategy is proposed here to infuse physical knowledge within a reduced-order IBM numerical solver. The aim of the analysis is i) to improve the accuracy of such a solver and ii) to do so with a limited increase in computational resources required. Discussion in \S\ref{sec:introDA} indicated how the DA methods rely on a \textit{model}, providing a quasi-continuous representation of the physical variables in the domain of investigation, as well as some local \textit{observation}. These elements are now discussed in detail.

The \textit{model} here considered is the numerical solver and test case used for the preliminary simulation DNS-IBM-CF. Such a model was not able to provide accurate estimations of the statistical moments. In addition, a no-slip condition at the wall was not obtained. The number of ensemble realizations, which allows us to explore the spaces associated with the parametric description and the physical solution, is set to $m=40$. The value chosen for $m$ is based on recommendations provided in the literature from analyses combining the EnKF with CFD solvers \citep{Mons2021_prf, Moldovan2021_jcp}. The data-driven strategy will be used to dynamically enhance the physical state at each analysis phase as well as to infer the local value of the diagonal components of the tensor $\mathsfbi{D}$. Therefore, the full parametric space to be optimized consists of the three components $D_{xx}$, $D_{yy}$ and $D_{zz}$ for $128 \times 64 \times 64 = 524 \, 288$ grid elements, for a total of $1\, 572 \, 864$ degrees of freedom to be determined in the parametric space investigated. The following strategies are applied to reduce the complexity of the problem:
\begin{enumerate}
    \item Coefficients for mesh elements $\in \Omega_f$ are automatically set to zero. This implies that, out of the $64$ layers in the $y$ direction, only $5$ at the bottom and $5$ at the top are considered. In addition, statistical symmetry around the plane $y=h$ is used to consider only $5$ layers. While symmetry is not observed for the instantaneous flow, viscous phenomena in the proximity of the wall reduce the intensity of velocity fluctuations, mitigating the effect of this approximation.
    \item Similarly, statistical invariance due to homogeneity in the $x$ and $z$ is used to neglect the dependency of the coefficients to those directions i.e. $D_{ii}(\boldsymbol{x},t)=D_{ii}(y,t)$ for $i=x,y,z$.
\end{enumerate}

Thus, the optimization task is reduced to a space of $15$ degrees of freedom, which is the determination of $3$ constants for $5$ grid layers in the $y$ direction. A second essential aspect is the \textit{prior} which is used as the initial condition for the model. The prior, which includes the physical condition imposed at the initial time as well as the free parameters prescribed, plays an essential role in the rate of convergence towards the optimized solution by DA. In this case, the initial physical state imposed for each ensemble member is obtained from interpolation of the simulation R-DNS-BF for $t=0$. In addition, the velocity field for the mesh elements in the subviscous layer has been set to $\boldsymbol{u}=[u_\tau y / \delta_\nu , \, 0 , \, 0]$. The values for the components $D_{ii}$ used in the simulation DNS-IBM-CF are considered to be the prior state for the parametric description. For each simulation of the ensemble, these values are perturbed using a Gaussian distribution $D_{ii} = \mathcal{N}(10^5, 2.5 \cdot 10^7) \, m^{-2}$ for each of the five grid layers in the $y$ direction where the source term is non-zero. These conservative choices for physical state and parametric description have been performed to unambiguously identify the sensitivity of the solution to the parametric variation, speeding up the optimization procedure by EnKF.   

\begin{figure}
    \centering
    \includegraphics[max size={\textwidth}, trim = {0 1.5cm 0 4cm}, clip]{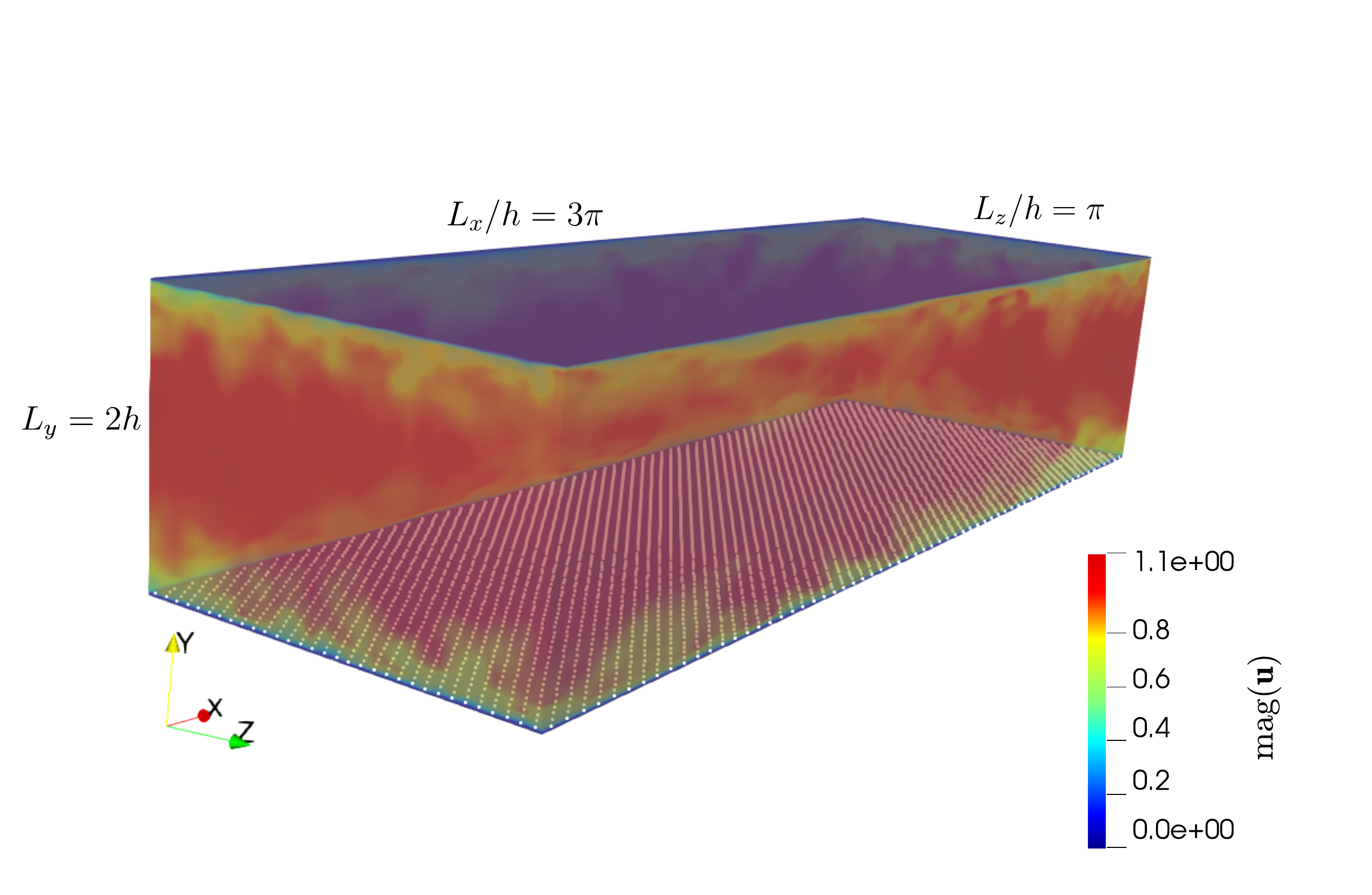} 
    \caption{Position at the bottom wall of the sensors providing constraints for the streamwise direction $(u_x = 0)$ when applying the EnKF.}
    \label{fig:position_constraints}
\end{figure}

The \textit{observation} used in the present analysis is now described. The \textit{physics infused} strategy proposed utilises physical knowledge of the flow in the form of data to be integrated into the data-driven method. The information employed deals with the physical behaviour of the flow in the proximity of the wall when the Reynolds number $Re_\tau$ is known. Thus, the non-slip condition is first applied on a number of sensors, which are distributed over the physical domain for $y=0$ and $y=2h$. A qualitative representation is shown in figure \ref{fig:position_constraints}. A total of $4\,096$ sensors are used for which the condition $u_x=0$ is imposed as surrogate observation. No constraint is imposed for $u_y$ and $u_z$. This choice complies with the intrinsic limitations of the EnKF, for which a matrix inversion of the size of the observed data $[n_o, \, n_o]$ must be performed \citep{Asch2016_siam}. Also, thanks to the solenoidal features of the resolved numerical schemes, the inferred field for $u_x$ also affects the other velocity components without the need to over-constrain the system to be optimized. A second physical information is infused. Once the value of $\Rey_\tau$ is known, $u_\tau$ is also fixed. This implies that the mean velocity gradient at the wall is known considering its relation with $\tau_w$ and $u_\tau$:

\begin{equation}
\tau_w = \rho \nu \, \left( \frac{\partial \langle u_x \rangle}{\partial y} \right) = \rho u_\tau^2 = \frac{1}{2} \rho C_f U_c^2
\end{equation}

In this case, the data inferred is the friction coefficient $C_f$ calculated using the friction velocity $u_\tau$ and the mean streamwise centerline velocity $U_c$ obtained via the simulation R-DNS-BF. The confidence in the physics-infused information is now discussed. First of all, the uncertainty affecting each sensor is supposed to be statistically independent, so that the matrix $\mathsfbi{R}$ in equation \ref{eqn:KalmanGain} can be reasonably approximated to a diagonal matrix. The standard deviation has been set to $0.05$ for the rows associated with sensors observing the velocity at the wall. For the sensor measuring the friction coefficient, the level of confidence is set to $0.05 \,{C_f}$. This value is very close to the variance of the time evolution of $C_f(t)$ observed for the simulation R-DNS-BF.

\begin{figure}
    \centering
    \includegraphics[max size={\textwidth},
    trim = {1cm 0 0 0}, clip]{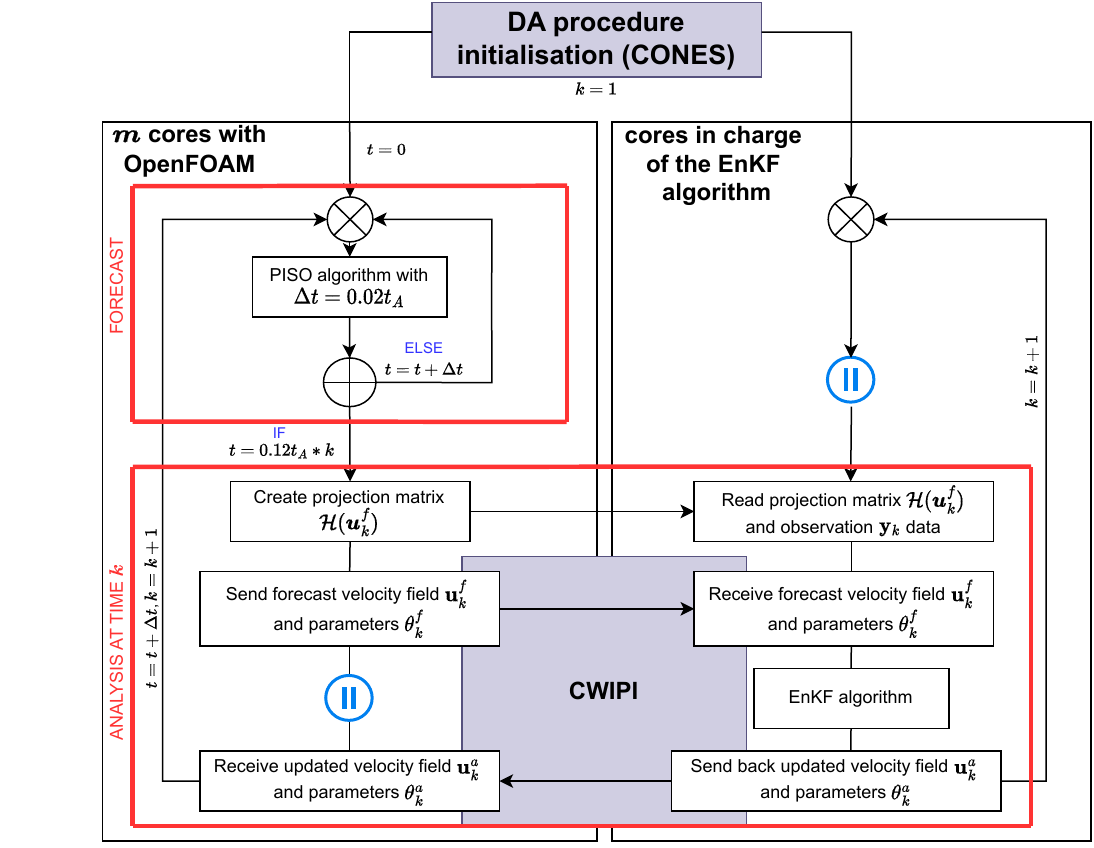} 
    \caption{Scheme of CONES application for sequential DA techniques.}
    \label{fig:EnKF_procedure}
\end{figure}

The DA procedure combining the \textit{model} and the \textit{observation} is now detailed. The sequential features of the EnKF are fully exploited, which means that the DA state estimation and optimization are performed on the fly during the run. For simulations of this size, non-intrusive approaches are too expensive in terms of computational resources required. In fact, simulations of the ensemble must be interrupted before every analysis phase and restarted after. This operation time, which sums up the writing and reading of files for the physical solution, may increase the computational costs by several orders of magnitude when compared with the forecast of the solution. For this reason, the numerical simulations of the ensemble are coupled online with the DA code using the application \emph{Coupling OpenFOAM with Numerical EnvironmentS} (CONES), recently developed by the research group \citep{Villanueva2023_arXiV}. CONES parallelizes the problem using an open-source coupler called CWIPI \citep{Reflox2011_al}, which identifies two modules (simulations in OpenFOAM and the library with the EnKF algorithm) and establishes HPC communications among them according to the user's specifications. CPU cores assigned for the module of OpenFOAM are occupied when carrying out the forecast step and become inactive during the analysis phases, whereas those destined for the correction step work oppositely. Thus, this algorithm completely excludes the costly operations of interrupting/restarting the simulations of the ensemble. In this work, a DA analysis phase is performed every six time steps of the numerical forecast. Considering that the time step $\Delta t = 0.02 t_A$, this implies that data are assimilated every $0.12 t_A$. This relatively high frequency of assimilation with respect to the characteristic physical scales of the flow naturally excludes the risk of lack of convergence of the DA procedure \citep{Meldi2018_ftc}. As detailed in \S\ref{appAinit}, the pressure is updated from the system's state estimated by the EnKF by means of a Poisson equation. This additional step grants conservativity of the discretized Navier--Stokes equations. A visualization of the EnKF procedure is shown in figure \ref{fig:EnKF_procedure}.

Finally, covariance inflation and localization are discussed. These state-of-the-art procedures aim to improve the accuracy and robustness of the EnKF, as well as reduce the computational costs associated \citep{Asch2016_siam}. For the former, deterministic inflation is applied by using equation \ref{eqn:inflation} with a constant $\lambda = 1.05$ during initial evolution phases for $t \in [0, 5\,t_A]$. This choice increases the variability of the ensemble and prevents the collapse of the parametric description of the system during the first analysis phases. Localization is applied taking into account that observation/physical constraints are located in the proximity of the wall. A matrix $\mathsfbi{L}$ premultiplying the Kalman Gain in equation \ref{eqn:KalmanGain} is generated where all the coefficients are zero with the exception of those referring to the mesh elements located in the subviscous layer $(y^+ < 5)$ and outside the physical space. These coefficients are determined using a decay exponential function $(L_{ij} = e^{-\Delta r_{ij}^2 / \eta})$, where $\Delta r_{ij}$ is the distance between the sensor providing observation $i$ and mesh element $j$. The parameter $\eta$ is tuned to avoid discontinuities in the DA state estimation update moving from the subviscous layer to the outer wall regions. In addition, the EnKF algorithm is clipped in the near-wall region, and only the velocity field $\boldsymbol{u}$ of the mesh elements located here is updated. This corresponds to a total number of $57\,344$ mesh elements (seven cell layers in the wall-normal direction), which makes $n = 172\,032$ degrees of freedom without accounting for the fifteen parameters of the tensor $\mathsfbi{D}$.

\section{Results}\label{sec:proposedSec4}

The physics-infused procedure developed relies on on-the-fly state estimation and parametric optimization of an ensemble of numerical simulations. These runs predict the instantaneous turbulent physical field. Results obtained by this procedure are first compared with the reference database presented in \S\ref{sec:preliminaryAnalysis}. A sensitivity analysis of the hyperparameters governing the performance of the numerical model and of the DA strategy is also investigated.  

\subsection{Comparison of DA results with reference simulations}\label{sec:researchWork1}

It is here reminded that the DA procedure based on the EnKF is used to optimize the parametric behaviour of an advanced IBM penalization model, as well as to perform an update of the flow field. These operations are performed sequentially, operating directly over the predicted instantaneous flow field.

\begin{figure}
\centering
\begin{tabular}{cc}
    \includegraphics[width=0.5\linewidth, trim = {8cm 4cm 4cm 1cm}, clip]{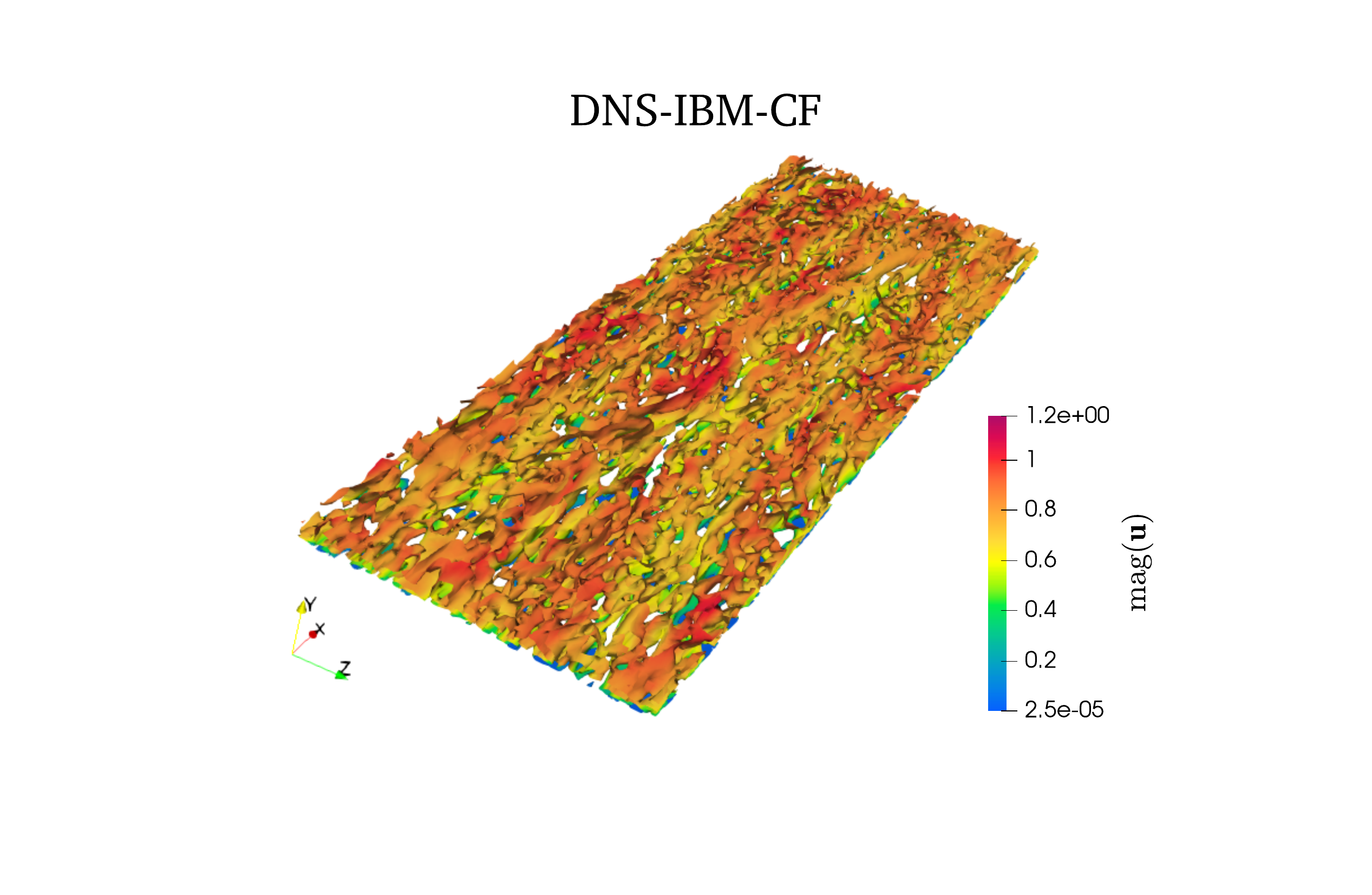} & \includegraphics[width=0.5\linewidth, trim = {8cm 4cm 4cm 1cm}, clip]{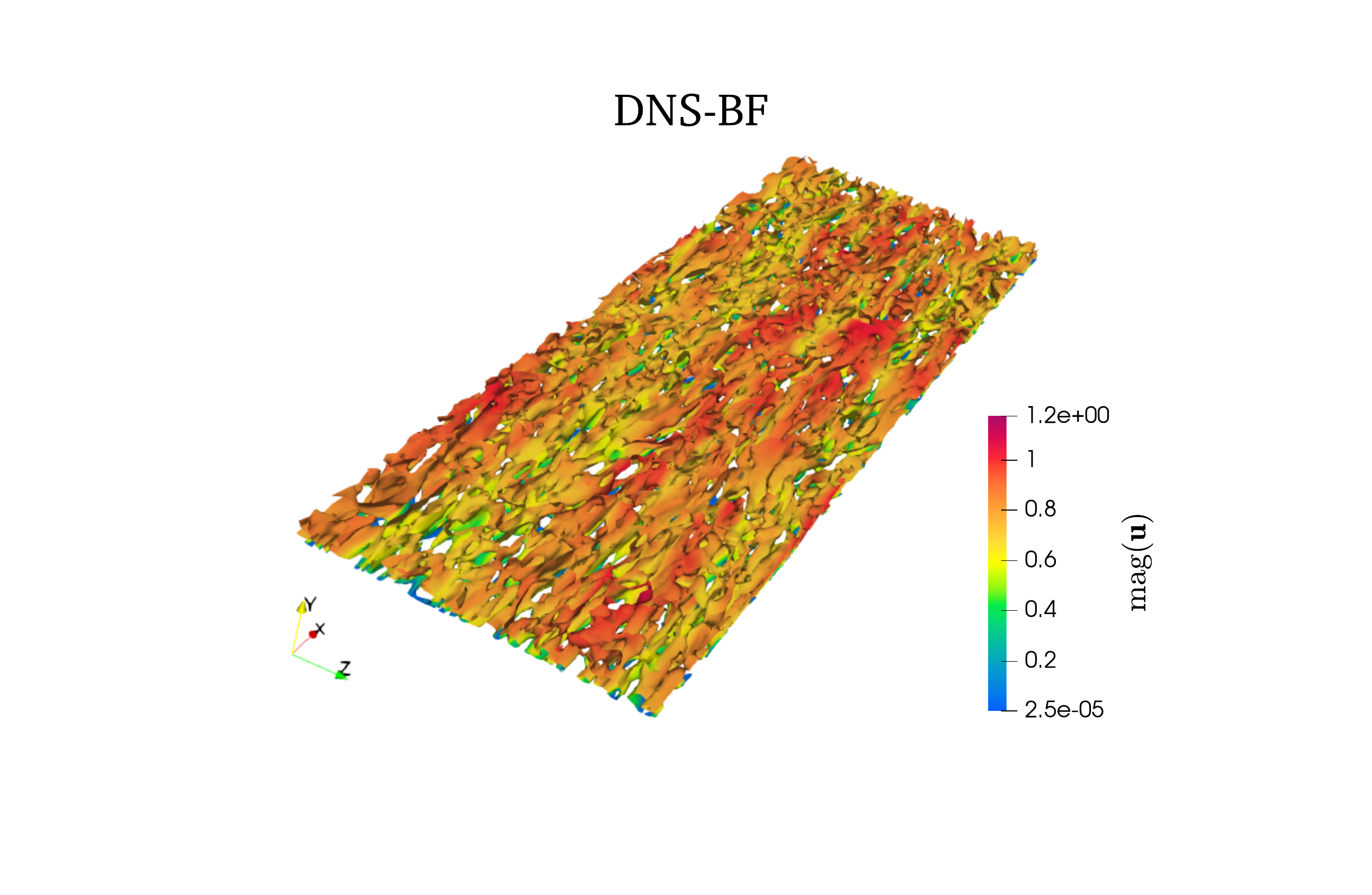} \\
    \textit{(a)} & \textit{(b)} \\
    \includegraphics[width=0.5\linewidth, trim = {8cm 4cm 4cm 1cm}, clip]{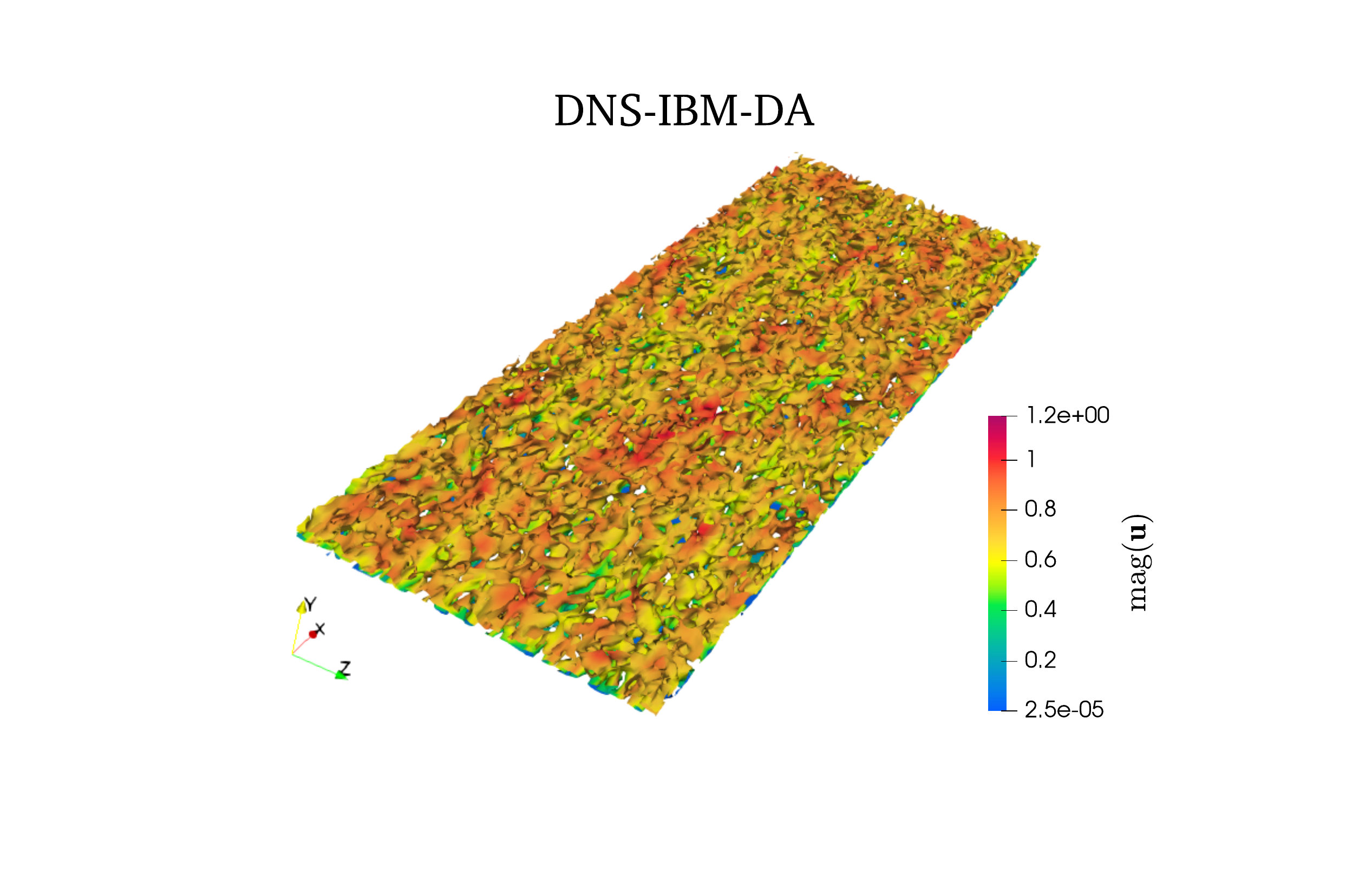} & \includegraphics[width=0.5\linewidth, trim = {8cm 4cm 4cm 1cm}, clip]{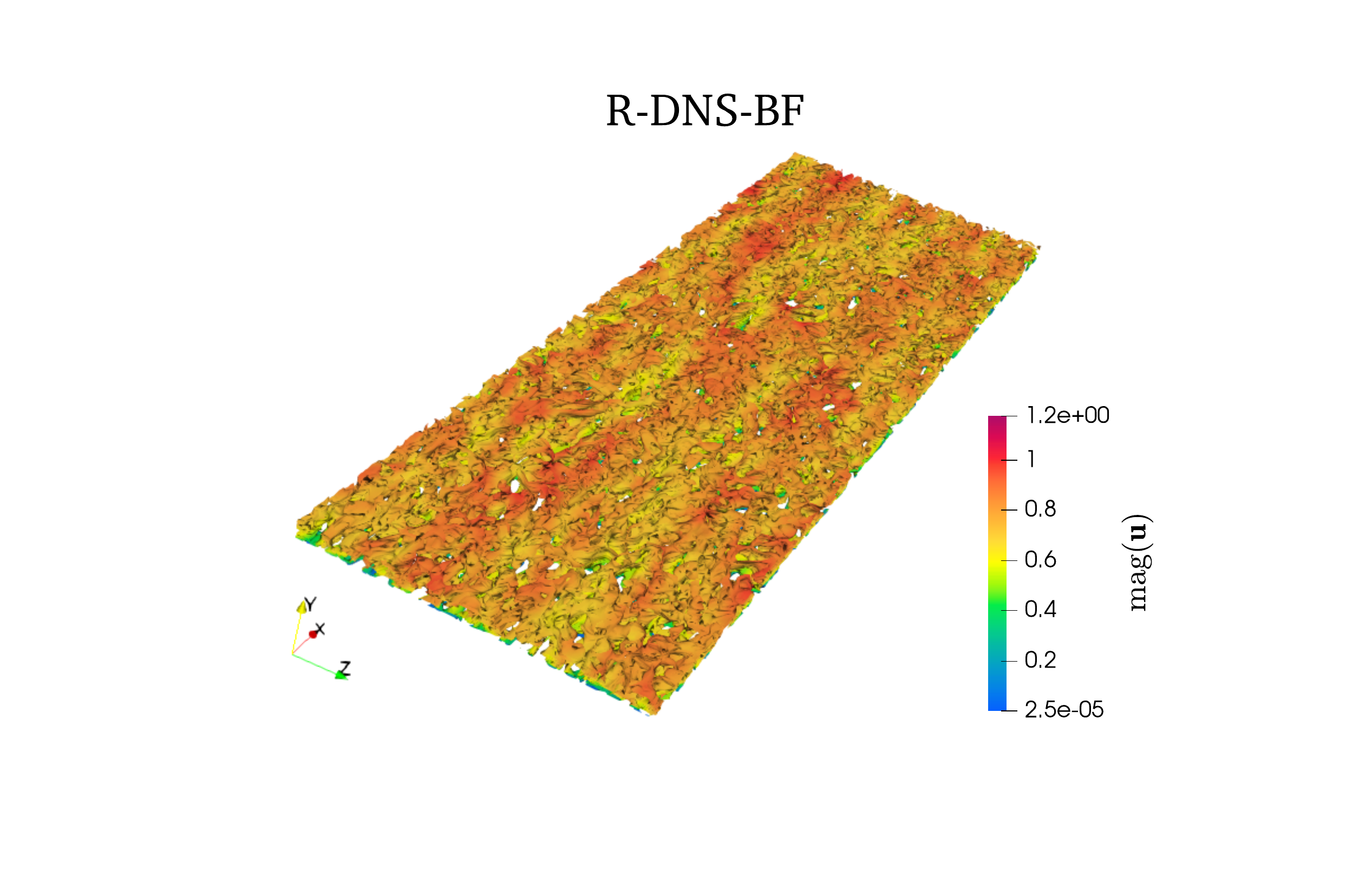} \\
    \textit{(c)} & \textit{(d)} 
    \end{tabular} 
\caption{Isocontours of Q-criterion calculated for $y/h = 0.18$ ($y^\star \approx 100$). Results are shown for simulations (a) DNS-IBM-CF, (b) DNS-IBM-DF, (c) DNS-IBM-DA, and (d) R-DNS-BF.}
\label{fig:Qcriterion}
\end{figure}

First of all, data in table \ref{tab:summary} seem to indicate that the infusion of the local no-slip constraint and the global knowledge of the friction coefficient $C_f$ is able to significantly improve the prediction of the friction velocity $u_\tau$. In fact, classical IBM calculations (DNS-IBM-DF) obtain a significant underprediction of this latter quantity for a discrepancy of around $17\%$. The physics-infused procedure obtains an overprediction of around $1.5\%$, thus a discrepancy of almost 12 times less than the classical simulations. Therefore, one can expect that the global accuracy of the flow is significantly increased, considering the essential role of $u_\tau$ in the establishment of the physical features observed for this test case. A first qualitative comparison of the flow features is shown in figure \ref{fig:Qcriterion} for the isocontours calculated via the Q-criterion. One can see that, despite the lack of grid resolution, the DA simulation (DNS-IBM-DA) appears to be able to capture fine structures that exhibit a better agreement with the reference DNS (R-DNS-BF). This result appears to imply that the instantaneous flow update performed by the EnKF provides a significant improvement in accuracy when compared with the classical IBM approaches, which are run on grids of identical resolution.

\begin{figure}
    \centering
\begin{tabular}{ccc}
    \includegraphics[width=0.32\linewidth]{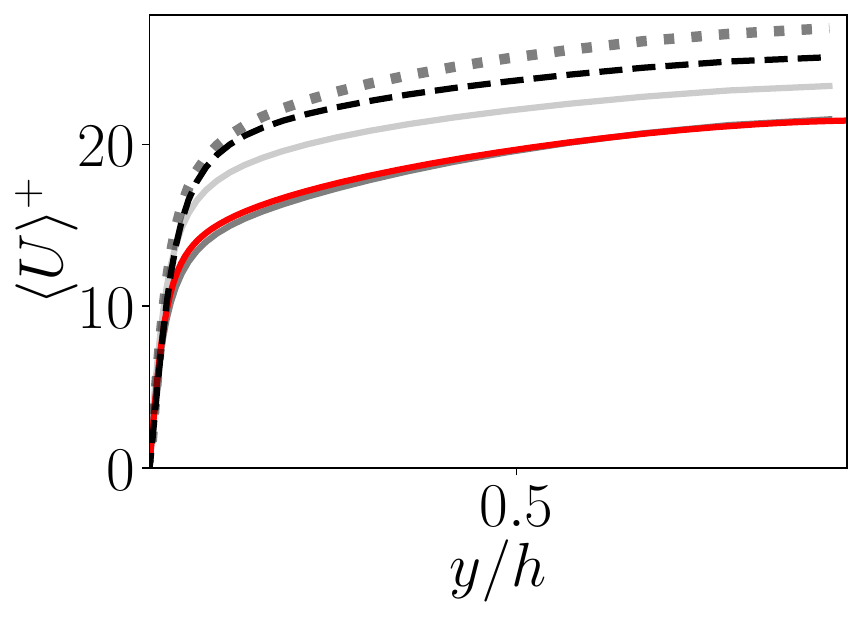} & \includegraphics[width=0.32\linewidth]{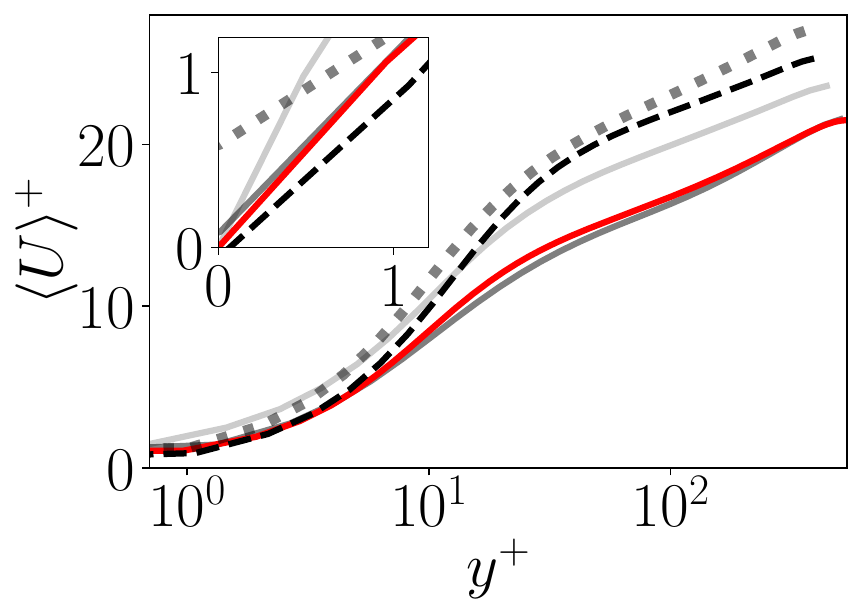} &
    \includegraphics[width=0.32\linewidth]{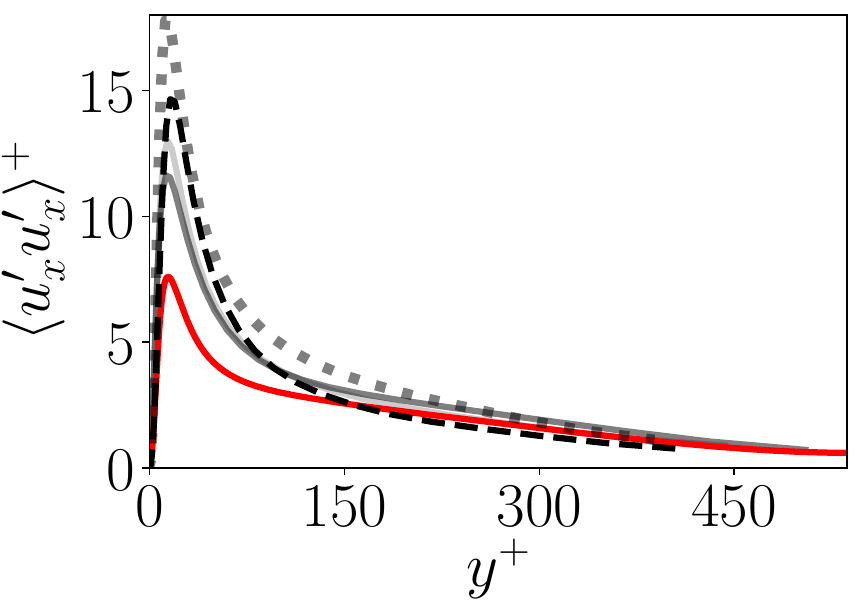} \\
    \textit{(a)} & \textit{(b)} & \textit{(c)} \\
    \includegraphics[width=0.32\linewidth]{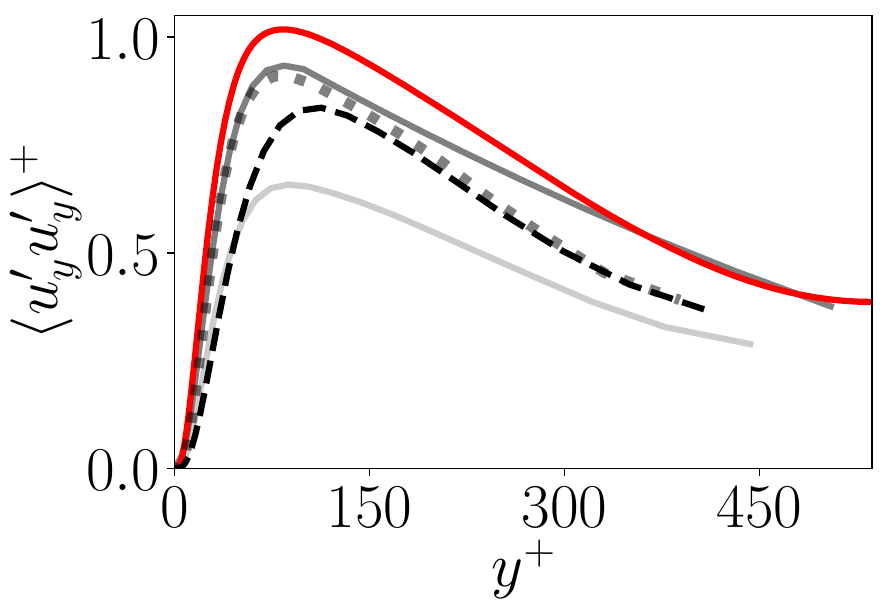} & \includegraphics[width=0.32\linewidth, height=3cm]{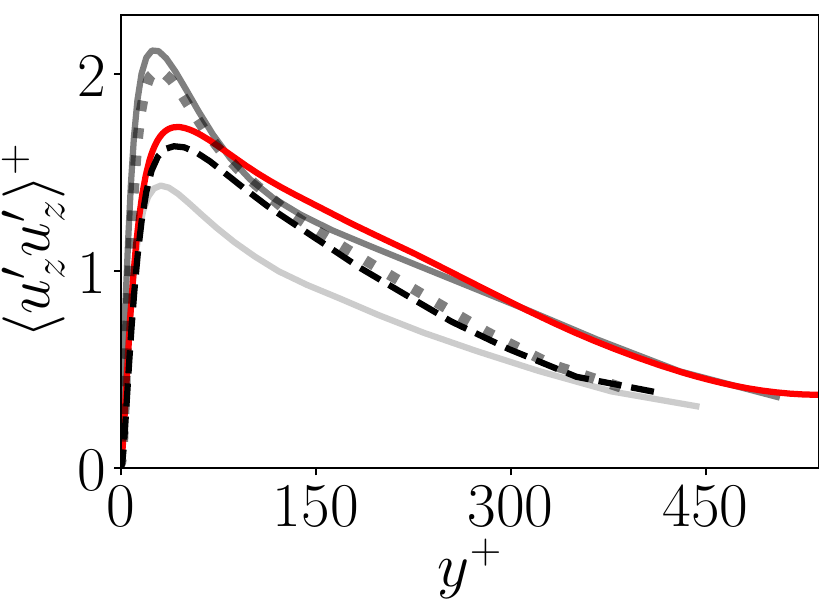} &
    \includegraphics[width=0.32\linewidth]{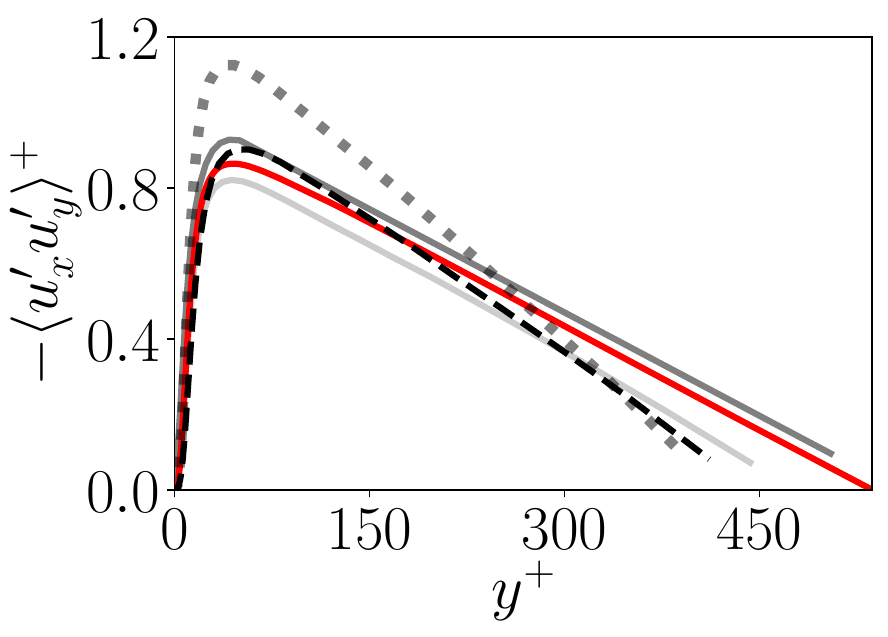} \\
    \textit{(d)} & \textit{(e)} & \textit{(f)}
    \end{tabular}
    \caption{Comparison of the main statistical moments of the velocity field. Results are shown for simulations  (\protect\greylinesolidstrong) DNS-IBM-DA, (\protect\greylinesolidsoft) DNS-BF, (\protect\greylinedottedstrong) DNS-IBM-CF, (\protect\blacklinedashed) DNS-IBM-DF, and (\protect\redline) R-DNS-BF.}
    \label{fig:meanProfilesDA}
\end{figure}

Similar conclusions can be drawn comparing the statistical moments obtained by the simulations, which are shown in figure \ref{fig:meanProfilesDA}. The normalized mean velocity profile $\langle U \rangle^+$ shown in figure \ref{fig:meanProfilesDA}\textit{(a)} and \textit{(b)} is significantly closer to the DNS reference both in the inner and the outer layers. In addition, one can see that the no-slip condition at the wall is well obtained by the DA simulation, within the confidence level prescribed for the observation, since $\langle U \rangle_{y=0, 2h}^+ = 0.097$. For the DNS-IBM-CF, $\langle U \rangle_{y=0, 2h}^+=0.557$, which involves an enhancement of $83\%$. Improvements using the DA method can also be observed for the Reynolds stress tensor components, in particular when compared with the continuous IBM model. These improvements, however, vary depending on the components. High accuracy is obtained for the prediction of $\langle u_x^\prime u_y^\prime \rangle^+$, which is shown in figure \ref{fig:meanProfilesDA}\textit{(f)}, while a small degradation is observed for $\langle u_z^\prime u_z^\prime \rangle^+$ in figure \ref{fig:meanProfilesDA}\textit{(e)}. 

\begin{figure}
    \centering
    \begin{tabular}{ccc}
    \includegraphics[width=0.32\linewidth]{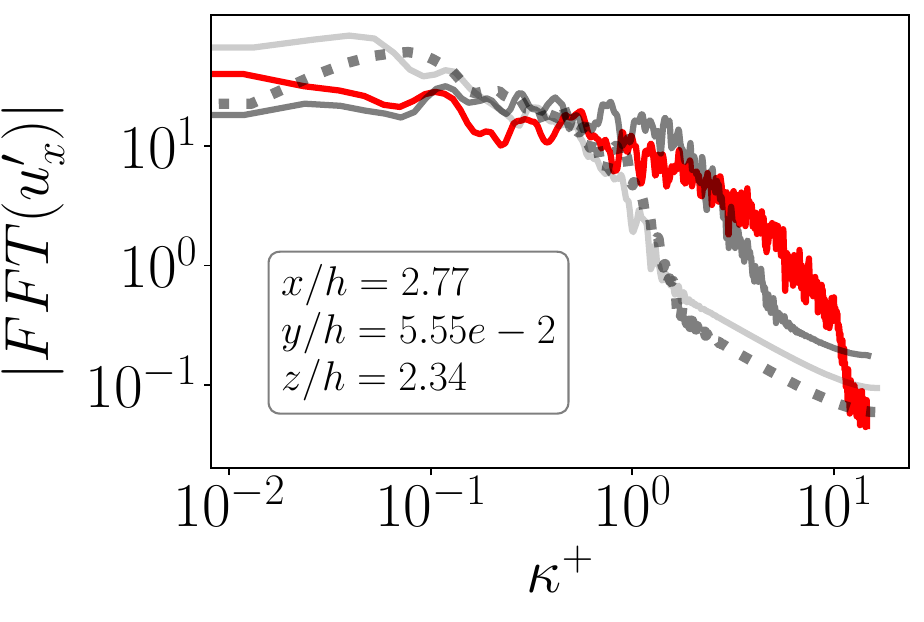} & \includegraphics[width=0.32\linewidth]{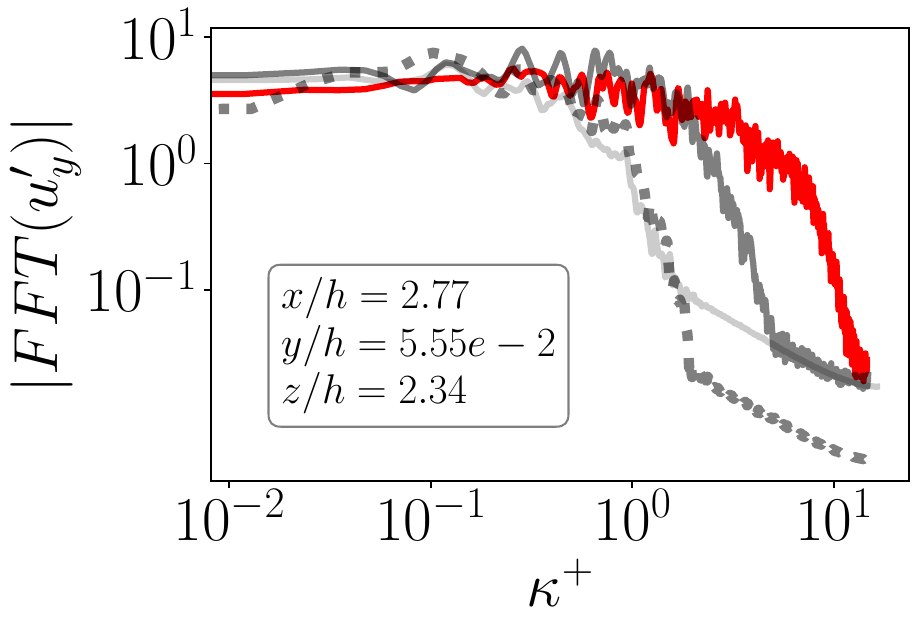} &
    \includegraphics[width=0.32\linewidth]{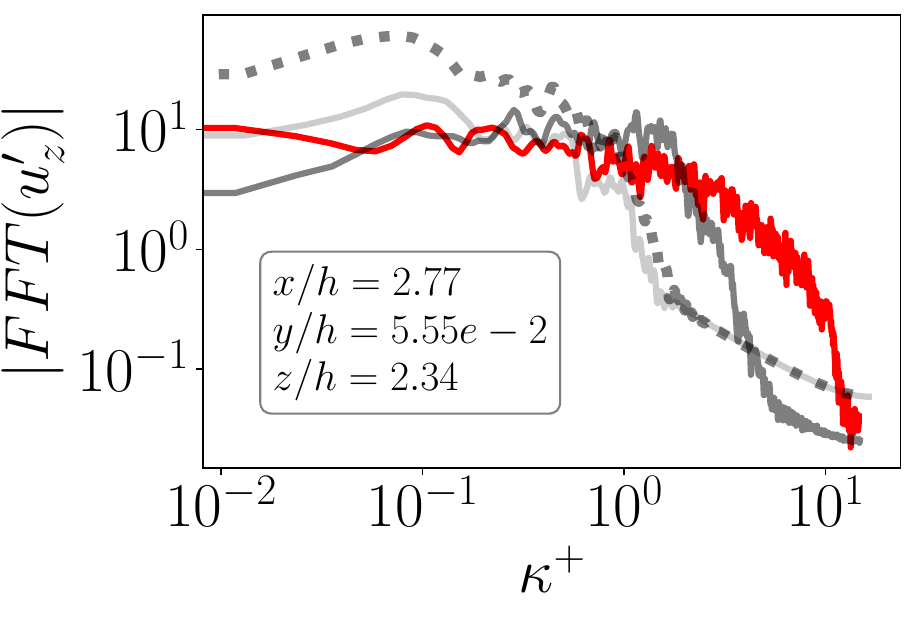} \\
    \textit{(a)} & \textit{(b)} & \textit{(c)} \\
    \includegraphics[width=0.32\linewidth]{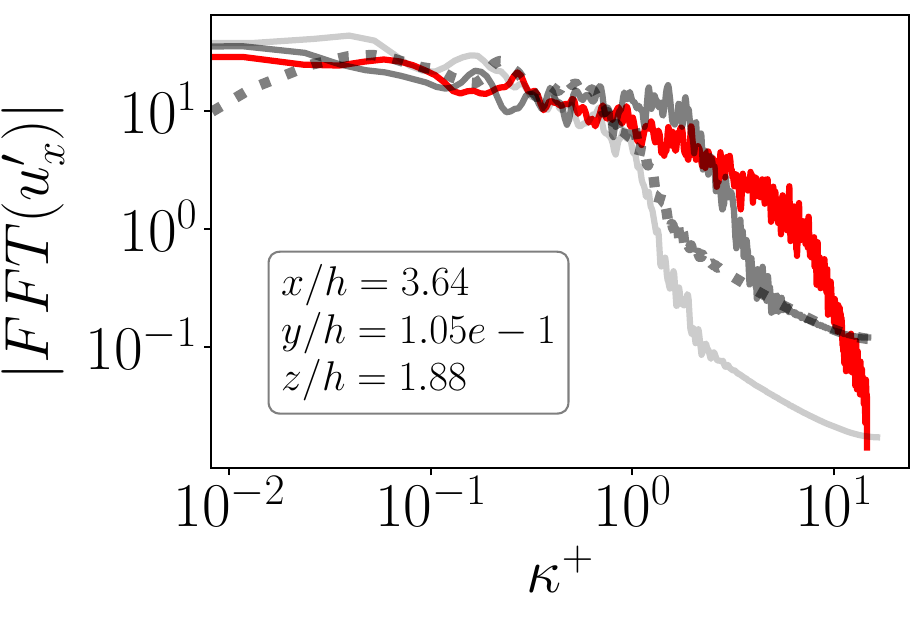} & \includegraphics[width=0.32\linewidth]{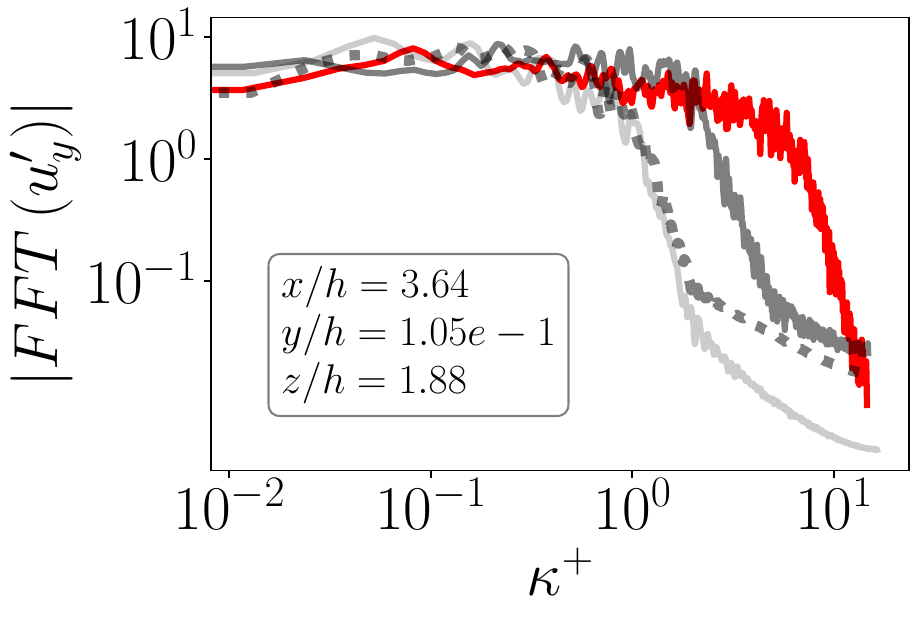} &
    \includegraphics[width=0.32\linewidth]{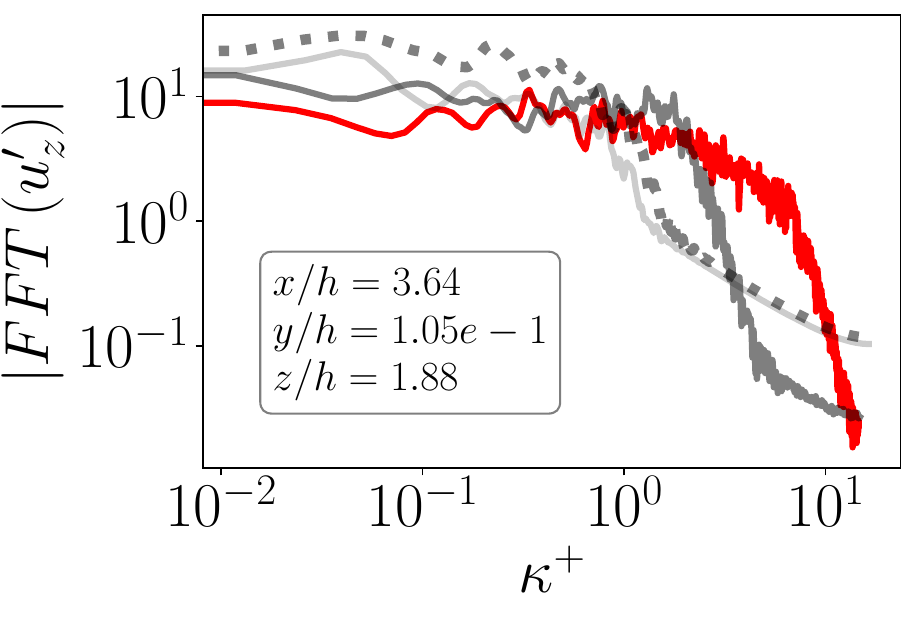} \\
    \textit{(d)} & \textit{(e)} & \textit{(f)}
    \end{tabular}
    \caption{$FFT$ calculated by sampling the fluctuating velocity field $\boldsymbol{u}^\prime$ located at (first row) $y^\star \approx 30$ and  (second row) $y^\star \approx 56$. Results are shown for the simulations (\protect\greylinesolidstrong) IBM-DNS-DA, (\protect\greylinedottedstrong) IBM-DNS-CF, (\protect\greylinesolidsoft) DNS-BF, and (\protect\redline) R-DNS-BF.}
    \label{fig:Spectra}
\end{figure}

The analysis is completed with the comparison of the time spectra shown in figure \ref{fig:Spectra}. The spectra are calculated using the 1-D discrete \emph{Fast Fourier Transform} [$FFT$; \cite{Press2017_cambridge}] to the time-series of the fluctuating velocity $\boldsymbol{u}^\prime (t)$ sampled at the locations $y^\star = 30, \, 56$:

\begin{equation}
    FFT(k) = \sum_{k = 0}^{N - 1} e^{-2\pi j \frac{k t}{N}} \boldsymbol{u}^\prime(t)
    \label{eqn:FFT}
\end{equation}

$N = 7\,500$ is the total number of samples. The sampling time is $\Delta t = 0.04\, t_A$ over a period $T = 100\,t_A$. 
Using Taylor's hypothesis for frozen turbulence \citep{Taylor1938_prsl}, results in figure \ref{fig:Spectra} are shown for wavenumbers $\kappa= 2\pi f / U_c$, where $f$ is the temporal frequency of the $FFT$ transform. In addition, $\kappa$ is adimensionalized with respect to the kinematic viscosity $\nu$ and the friction velocity $u_\tau$ so that $\kappa^+ = \kappa \nu / u_\tau$. To smooth out the curves, a first-order Butterworth low-pass filter \citep{Butterworth_ew} is used to eliminate the undesired noise. 
One can see that the coarse-grid simulations (IBM-DNS-CF and DNS-BF) produce similar spectra, and the fluctuation energy starts to decay for relatively low wavenumbers. This observation is related to two concurring phenomena, namely the lack of grid resolution and the poor accuracy in estimating the wall shear stress. On the other hand, DA results get significantly closer to the reference DNS results, indicating that the accurate prediction of $u_\tau$ provides a beneficial effect on the representation of the velocity fluctuating field. One could also observe that the gap between the reference DNS and the DA run is around half-octave i.e. proportional to the difference in resolution of the grids used for calculations. This confirms that the usage of the EnKF was able to efficiently compensate for the modelling error and that the discrepancies observed are intrinsically tied to the grid resolution.

\subsection{Sensitivity of the physics-infused strategy to mesh refinement and placement of sensors}\label{sec:researchWork2}

\begin{figure}
    \centering
\begin{tabular}{cc}

    \multicolumn{2}{c}{
    \includegraphics[width=0.48\linewidth]{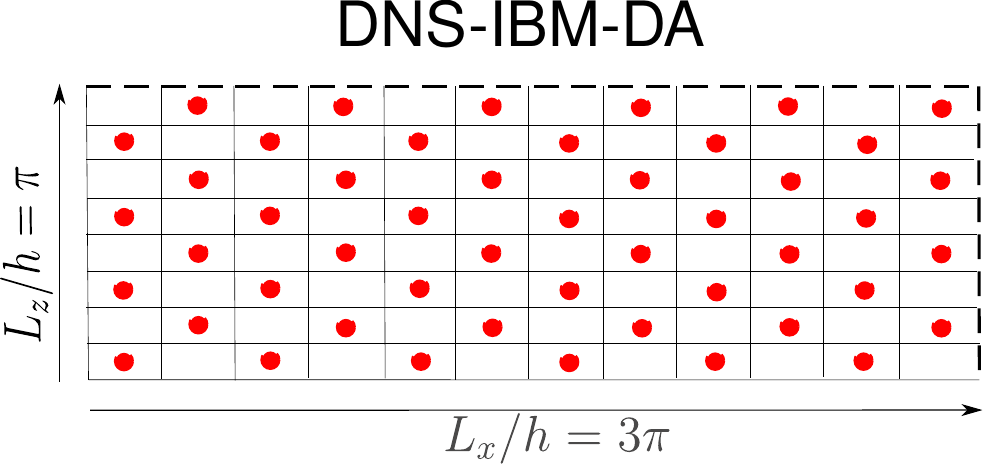}
} \\
    \multicolumn{2}{c}{
    \textit{(a)}
    } \\
    \includegraphics[width=0.48\linewidth]{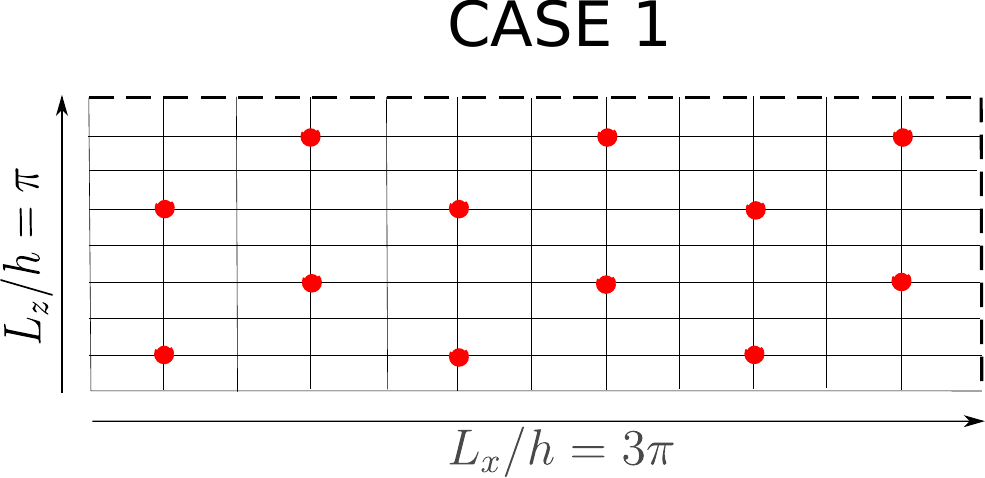} & \includegraphics[width=0.48\linewidth]{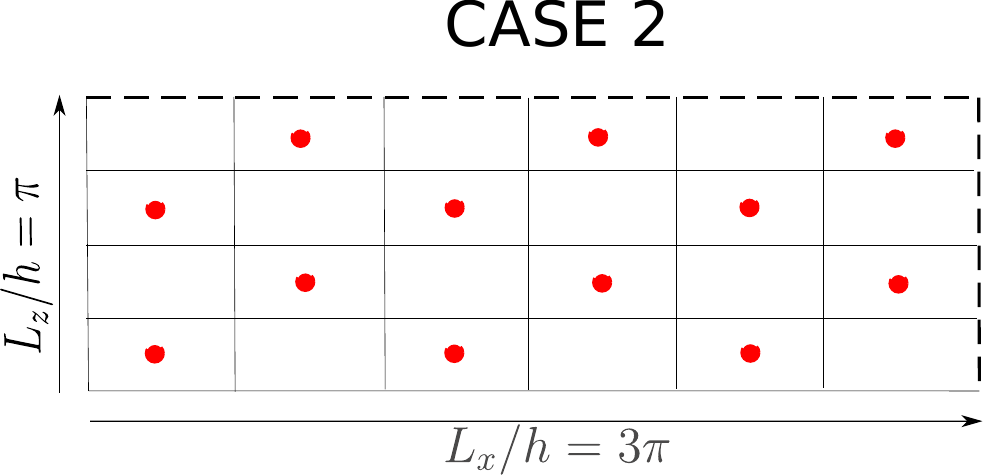} \\
    \textit{(b)} & \textit{(c)} \\
    & \\ \includegraphics[width=0.48\linewidth]{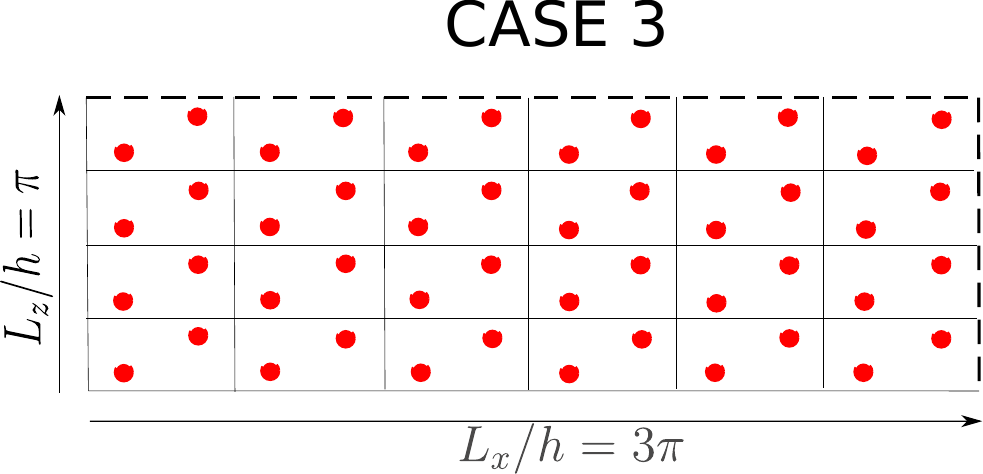} & \includegraphics[width=0.48\linewidth]{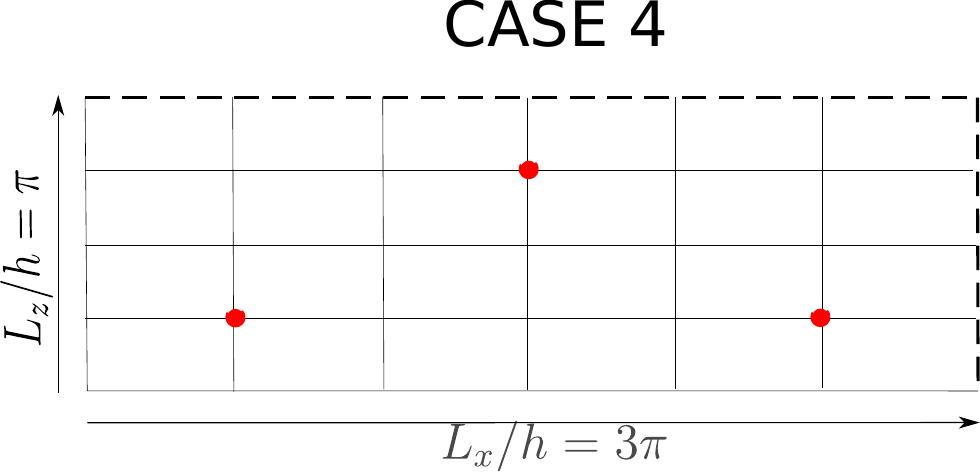} \\
    \textit{(d)} & \textit{(e)}
    \end{tabular}
    \caption{Location of the sensors where $u_x=0$ is infused during the DA procedure. Snapshots are shown for $y=0$.}
    \label{fig:mesh_constraints2}
\end{figure}

The analysis of the results performed in \S\ref{sec:researchWork1} highlighted how the infusion of physical information in the numerical process can improve the quality of the results even when simulations are performed using coarse grids. In the present section, the sensitivity of the DA strategy is tested against variations in the key elements that constitute the global methodology. More precisely, two aspects are considered. The first one is represented by variations in the grid resolution for the model. The second one consists of a different distribution and density of the sensors providing observation. To this purpose, four additional DA runs are performed varying the indicated parameters. Features of such simulations are reported in table \ref{tab:summary2}. The simulation DNS-IBM-DA from the previous section is now taken as the reference for the rest of the data-driven simulations. As one can see, case $1$ uses the same mesh as the reference DA run, while cases $2$, $3$, and $4$ are performed using a modified grid. In these cases, the resolution in the streamwise direction $x$ and in the spanwise direction $z$ is two times coarser, while a higher resolution is employed in the normal direction $y$. For the latter, a smaller expansion ratio between consecutive mesh elements $r_y = \Delta y_i / \Delta y_{i-1}$ is used. This parameter, which is equal to $r_y=1.171$ for the grid employed in the DNS-IBM-DA run, is reduced to $r_y=1.060$ in these studies. This new grid is composed of $N = 262\,144$ elements. The second aspect that is considered is the number of sensors employed to locally inform the model to respect the constraint $u_x=0$ for $y=0$. One can see in figure \ref{fig:mesh_constraints2} that different densities of the sensors, as well as different distributions, have been analysed. In particular, one can see that probes have been positioned both at the centre of the mesh element as well as at their interface. In addition, in case $3$, the problematic aspect of multiple sensors within one mesh element has been investigated.

The prior state for the new DA runs has been chosen using the same criteria previously presented for the DNS-IBM-DA case, and the optimization targets the values of the diagonal elements of the tensor $\mathsfbi{D}$. The prior state for the parameters is, for these cases, the solution obtained by the DNS-IBM-DA run. This choice, which has been performed to obtain a faster and more robust convergence of the optimisation procedure, also allowed for a lower deterministic inflation level to be applied. In this case, a constant value of $\lambda = 1.01$ is chosen for all parameters throughout the time range $t \in [0, 300\,t_A]$. The level has been selected empirically to obtain the best trade-off between fast convergence and stability of the optimization procedure.

\begin{figure}
    \centering
    \begin{tabular}{ccc}
    \includegraphics[width=0.32\linewidth]{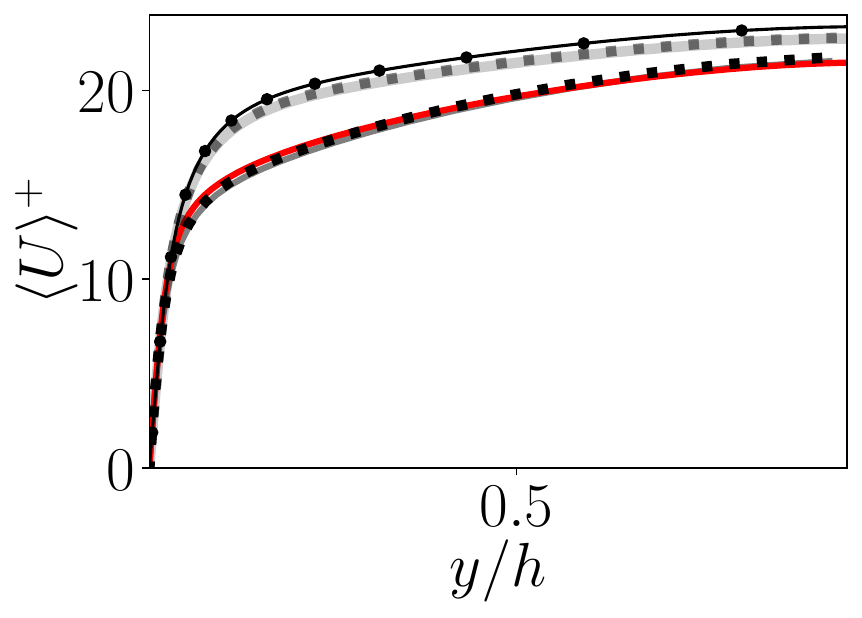} & \includegraphics[width=0.32\linewidth]{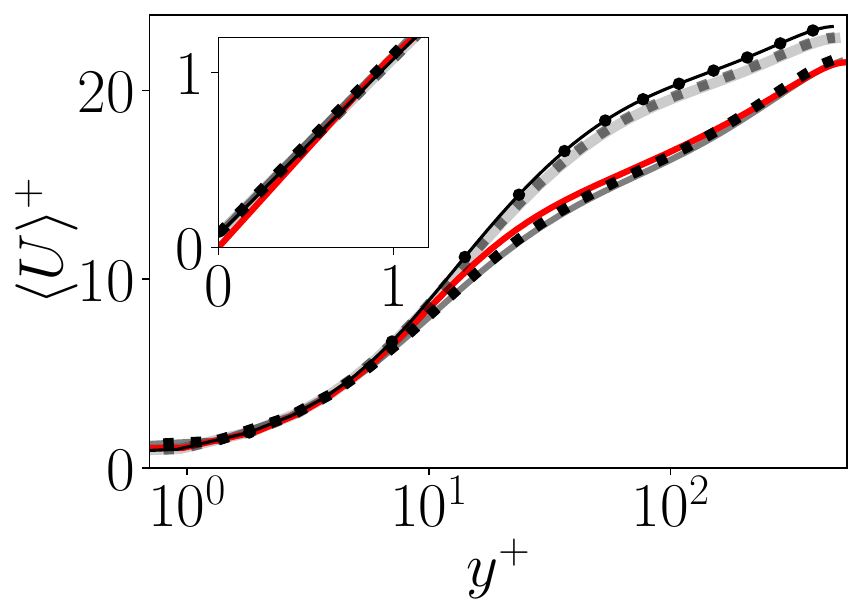} &
    \includegraphics[width=0.32\linewidth]{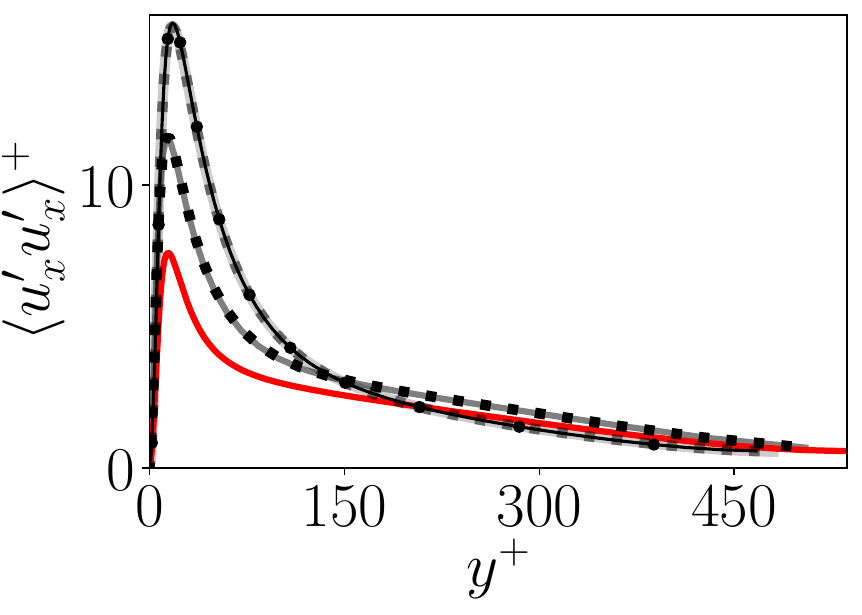} \\
    \textit{(a)} & \textit{(b)} & \textit{(c)} \\
    \includegraphics[width=0.32\linewidth]{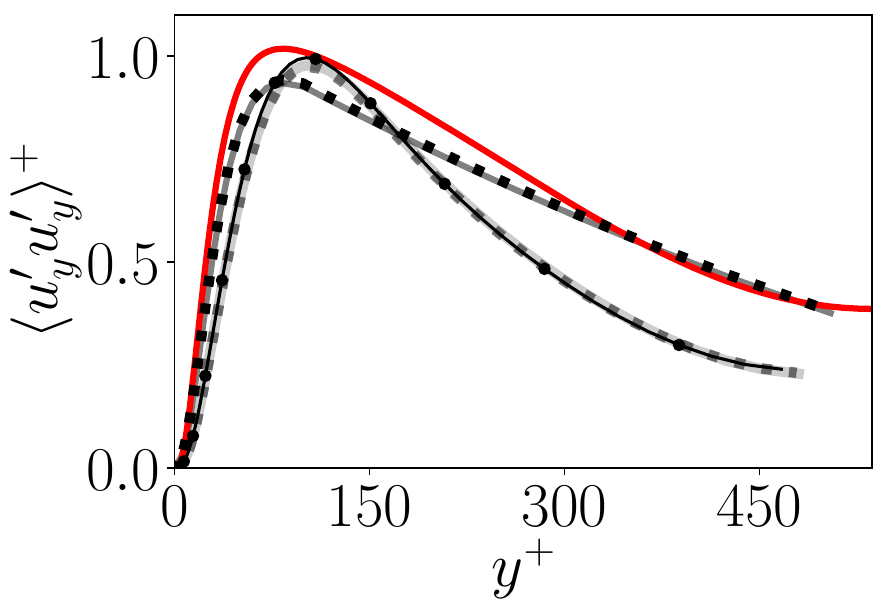} & \includegraphics[width=0.32\linewidth, height=3cm]{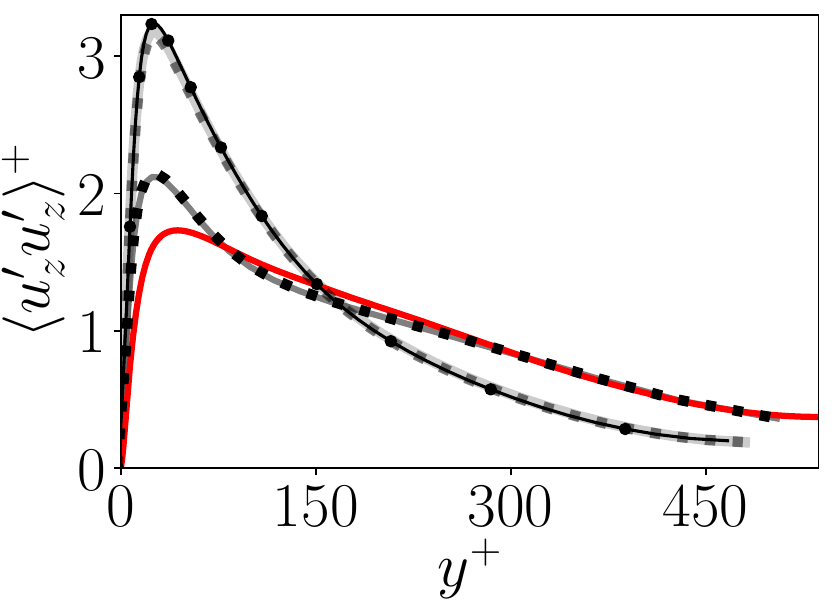} &
    \includegraphics[width=0.32\linewidth]{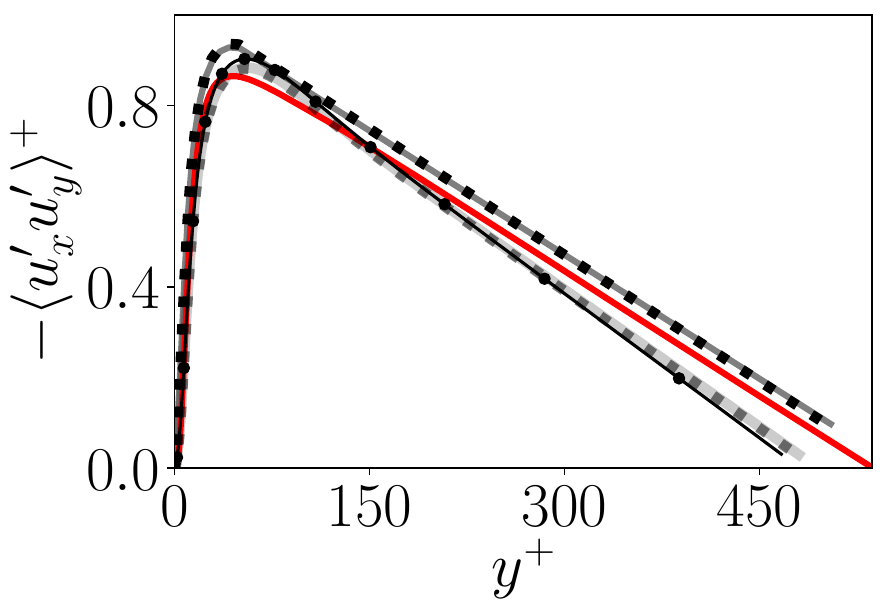} \\
    \textit{(d)} & \textit{(e)} & \textit{(f)}
    \end{tabular}
    \caption{Comparison of the main statistical moments of the velocity field for the data-driven cases. Results are shown for simulations (\protect\greylinesolidstrong) DNS-IBM-DA, (\protect\blacklinedotted) case 1, (\protect\greylinesolidsoft) case 2, (\protect\blacklinemarker) case 3, (\protect\greylinedottedstrong) case 4, and (\protect\redline) R-DNS-BF.}
    \label{fig:MeanProfiles2}
\end{figure}

Figure \ref{fig:MeanProfiles2} illustrates the main statistical moments of the flow. One can see that the choice of a different grid affects the prediction of the physical quantities investigated, while a weak sensitivity to the position and number of sensors is observed. In particular, results for case 1 are very similar to those obtained by the run DNS-IBM-DA, despite the smaller number of sensors which are located at the cell edges. A quantification of the differences between these two runs is provided by the error in the prediction of the friction coefficient, calculated as:
\begin{equation}
\Delta_{\overline{C_f}} = \frac{|\langle C_{f,\textrm{DA}}\rangle - \langle C_{f, \textrm{R-DNS-BF}}\rangle|}{\langle C_{f, \textrm{R-DNS-BF}} \rangle}
\end{equation}
The results, which are reported in table \ref{tab:summary2}, indicate a discrepancy of around $3\%$ for the reference DA run and $1\%$ for case 1. The analysis is now extended to the three DA runs performed over the modified grid. Results obtained for the statistical moment of the velocity field in figure \ref{fig:MeanProfiles2} indicate that degradation of the performance of the DA algorithm is observed. This observation is expected, because the IBM model is, in this case, less accurate owing to the lower grid resolution. However, it has been verified that the DA runs obtained using this grid are sensibly better than classical simulations on the same mesh, confirming the improvement in accuracy of the DA strategy whatever numerical model is used for the forecast. One can see that results from cases 2 and 4 are almost identical, also in terms of measured discrepancy $\Delta_{\overline{C_f}}$. On the other hand, results for case 3 are less precise, with an increase of $\Delta_{\overline{C_f}}$ when compared to the other two DA realizations on the same grid. A direct conclusion that can be drawn from this analysis is that, while features of the numerical model used for DA are important, also the number and position of sensors can play a role in the global accuracy of the results. More precisely, when sensors are positioned too close one to another, the uncertainties affecting the measurements can be responsible for convergence issues in the optimization process. In addition, the hypothesis of uncorrelated uncertainty of the observation (i.e. $\mathsfbi{R}$ being a diagonal matrix) becomes questionable when sensors are close. On the other hand, if sensors are far from the other in terms of characteristic scales of the flow, the global lack of information may preclude a satisfying state estimation. It must also be stressed that in this case, only one piece of information, namely the streamwise velocity, has been observed for each sensor. The global picture is arguably going to be more complex when multiple physical quantities are provided at each location.



The analysis of second-order statistical moments in figure \ref{fig:MeanProfiles2}\textit{(c)} to \textit{(f)} shows that no differences are observed among the simulations with the same grid. This may imply that the behaviour of such statistics is less sensitive to variations of the friction coefficient/friction velocity in the range here investigated. One can also see that, even if every DA run improves the performance of the underlying IBM, only the realizations with the first grid refinement (DNS-IBM-DA and case 1) are actually able to comply with the indicated level of confidence ($5\%$) for $\overline{C_f}$.

\begin{table}
  \begin{center}
\def~{\hphantom{0}}
  \begin{tabular}{lccccccccc}
       & $N_x \times N_y \times N_z$ & $\Delta x^\star$ &  $\Delta z^\star$ & $\Delta y^\star_{min}$ & $\Delta y^\star_{max} $ &
       $L_x/h$ & $L_z/h$ & $u_x = 0$ & $\Delta_{\overline{C_f}}$\\
       & & & & & & & & \\
       R-DNS-BF & $1\,024 \times 256 \times 512$ & $9.9$ & $6.6$ & $1$ & $11.2$ & $6\pi$ & $2\pi$ & $-$ & $-$\\
       & & & & & & & & \\
       DNS-IBM-DA & $128 \times 64 \times 64$ & $39.5$ & $26.3$ & $1.3$ & $52$ & $3\pi$ & $\pi$ & $4\,096$ & $0.033$ \\
       case 1 & $128 \times 64 \times 64$ & $39.5$ & $26.3$ & $1.3$ & $52$ & $3\pi$ & $\pi$ & $1\,024$ & $0.012$ \\
       case 2 & $64 \times 128 \times 32$ & $79.0$ & $52.7$ & $1$ & $30$ & $3\pi$ & $\pi$ & $1\,024$ & $0.149$\\
       case 3 & $64 \times 128 \times 32$ & $79.0$ & $52.7$ & $1$ & $30$ & $3\pi$ & $\pi$ & $4\,096$ & $0.195$ \\
       case 4 & $64 \times 128 \times 32$ & $79.0$ & $52.7$ & $1$ & $30$ & $3\pi$ & $\pi$ & $256$ & $0.154$\\
  \end{tabular}
  \caption{Summary of the DA runs performed for the sensitivity analysis.}
  \label{tab:summary2}
  \end{center}
\end{table}


\begin{figure}
    \centering
    \begin{tabular}{ccc}
    \includegraphics[width=0.32\linewidth]{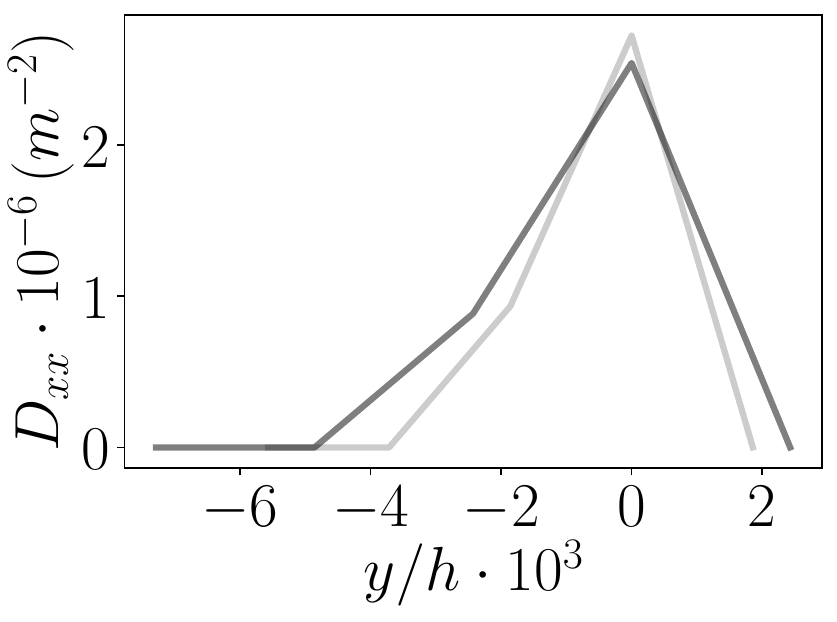} & \includegraphics[width=0.32\linewidth]{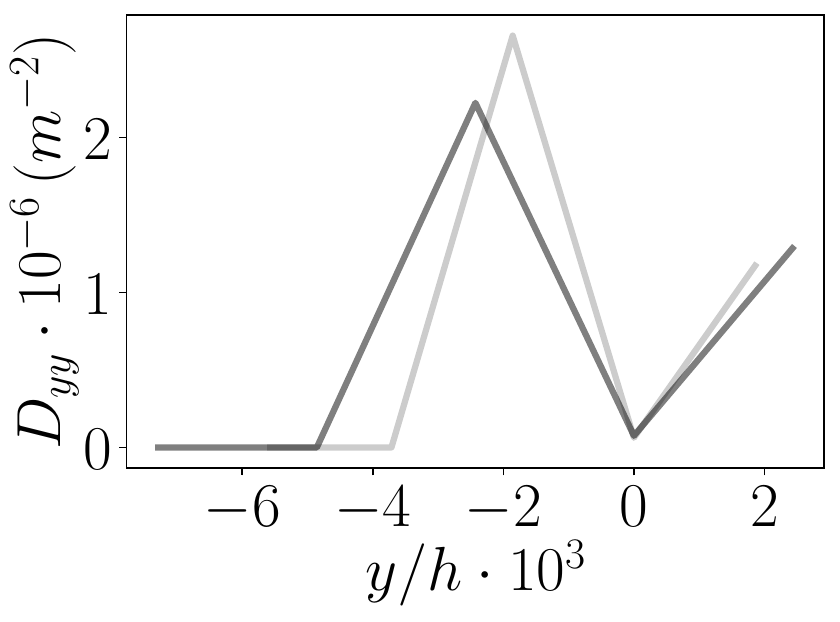} &
    \includegraphics[width=0.32\linewidth, height=0.235\linewidth]{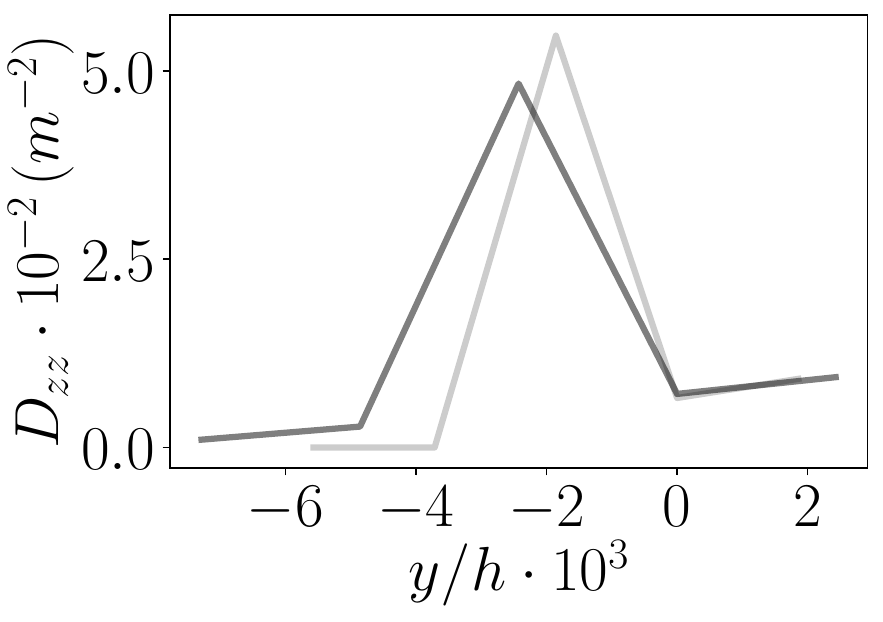} \\
    \textit{(a)} & \textit{(b)} & \textit{(c)}
    \end{tabular}
    \caption{Coefficients of tensor $\mathsfbi{D}$ for the simulations (\protect\greylinesolidstrong) case 1 and (\protect\greylinesolidsoft) case 2. Similar trends are observed in cases with identical grid refinement.}
    \label{fig:MainCoeffs}
\end{figure}

At last, the optimized values for tensor $\mathsfbi{D}$ are shown in figure \ref{fig:MainCoeffs}. Here, results are selected for the most accurate realization for each grid i.e. cases 1 and 2. First of all, the parameter distributions obtained for the two meshes are very similar. This observation confirms the robustness of the optimization procedure. In addition, for each component, the coefficients for the mesh elements of the interface $\Sigma_b$ play a predominant role when compared to the ones in the solid region $\Omega_b$. This result is in agreement with the properties of continuity and conservation that one can find in Roma's function (see equation \ref{eqn:Roma_function}) used in the discrete IBM. However, two important points need discussion:

\begin{itemize}
    \item One would expect the order of magnitude of the parameters linked to $D_{yy}$ to be similar to those characterizing $D_{zz}$ and, at the same time, to be significantly smaller than the coefficients defining $D_{xx}$. The results of the DA optimization indicate that the values for $D_{yy}$ are, on the other hand, similar to those for $D_{xx}$. This result could be associated with the improvement in the estimation of the wall shear stress.
    \item In the interface region $\Sigma_b$, the coefficients for $D_{xx}$ exhibit their maximum for $y=0$. For the coefficients for $D_{yy}$ and $D_{zz}$, a minimum is there obtained instead. The observation provided for $y=0$ is in the form of streamwise velocity, therefore it is logical to expect a local higher coefficient for $D_{xx}$. The values obtained in the neighbour mesh elements for $D_{yy}$ and $D_{zz}$ are therefore optimized to complement the constraint $u_x = 0$ with the physical condition imposed for $u_\tau$.
\end{itemize}

\subsection{Comparison of computational resources required}

Results presented in  \S\ref{sec:proposedSec4} have shown how the physics-infused procedure relying on the EnKF provides a significant improvement in the predictive capabilities of a classical penalization IBM model. In this section, the computational resources $CC$ required to perform each of the simulations of the database in tables \ref{tab:summary} and \ref{tab:summary2} are discussed. While the simulations have not been run on the same machine---therefore, an exact comparison of the computational resources used cannot be performed---the elements of discussion provided aim to obtain insights about the efficiency of the data-infused procedure i.e. gain in accuracy versus increase in costs. To this purpose, the resources required to perform the simulations are normalized over $CC_{\textrm{DNS-BF}}$, which is the total costs employed to run the simulation DNS-BF ($CC^\star = CC / CC_{\textrm{DNS-BF}}$). It is reminded that all numerical simulations are run with the same numerical schemes and grid (with the exception of the simulation R-DNS-BF, and the data-driven cases 2, 3 and 4). Therefore, differences observed in computational terms are not associated with the numerical solver. 

The two LES calculations, LES-BF and LES-BF-VD, exhibit very similar requirements when compared with the coarse-grained DNS-BF: $CC^\star = 0.995$ and $CC^\star = 0.986$, respectively. The only difference in this case is represented by the calculation of the contribution of the Smagorinsky model. Considering that this SGS closure does not require the resolution of additional equations, the variation in computational resources is negligible. Similar considerations can be performed for the simulation DNS-IBM-CF, where $CC^\star = 1.027$ since the calculation of the penalization term is straightforward in this case and does not require a significant amount of supplementary resources. The costs for these three first simulations stay in the range of a couple of percentage points from the reference simulation, which can be due to numerical uncertainty in the calculation process. On the other hand, the value of the parameter $CC^\star = 1.172$ for simulation DNS-IBM-DF is relatively higher. The reason is due to the communication and execution of the interpolation and spreading steps, which are performed by two different libraries. While this algorithmic cost can be improved with efficient coding, it still represents a non-negligible increase in the computational costs since, additionally, $0.74$ scalar hours are required when launching the simulation to load the stencils of the $16\,384$ Lagrangian markers (one Lagrangian marker for each Eulerian cell located at $y = 0, 2h$).

Now, the simulation R-DNS-BF is considered. For this simulation, one needs to take into account that the physical domain is $2$ times larger in the streamwise $x$ and spanwise $z$ directions, and the grid resolution is $4$ times more refined in every direction, which implies that the total number of mesh elements is $256$ larger when compared with the grid used for the other calculations. In addition, the time step is also five times smaller, which means that the increase of computational resources is around $10^{4}$ larger than the simulation DNS-BF. Practical estimation provides a result of $CC^\star \approx 12\,000$.    

\begin{figure}
    \centering
    \begin{tabular}{cc}
    \includegraphics[width=0.5\linewidth, trim={7cm 7cm 3cm 0}, clip]{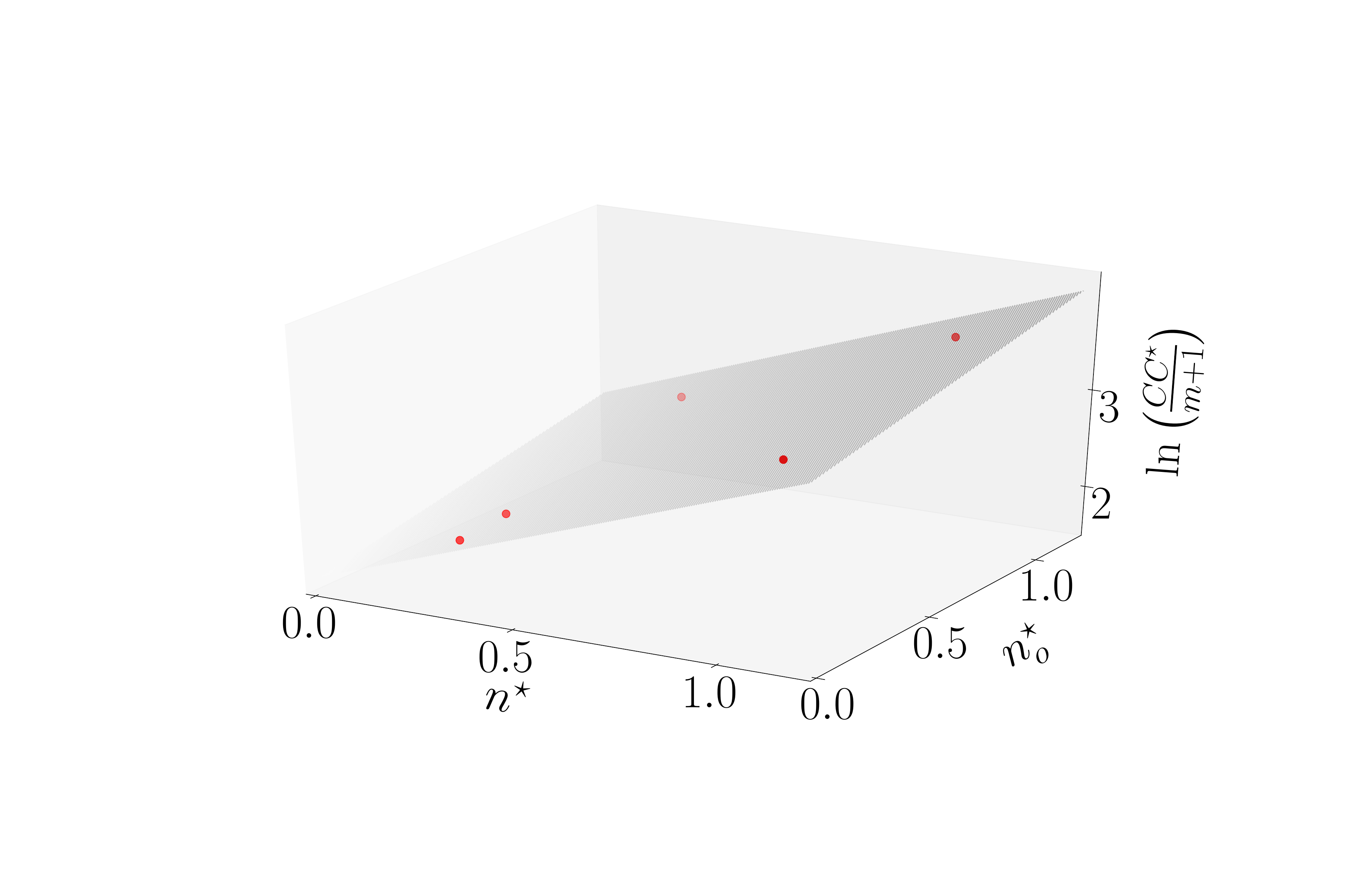} &
    \includegraphics[width=0.4\linewidth, trim={5cm 0 0cm 0}, clip]{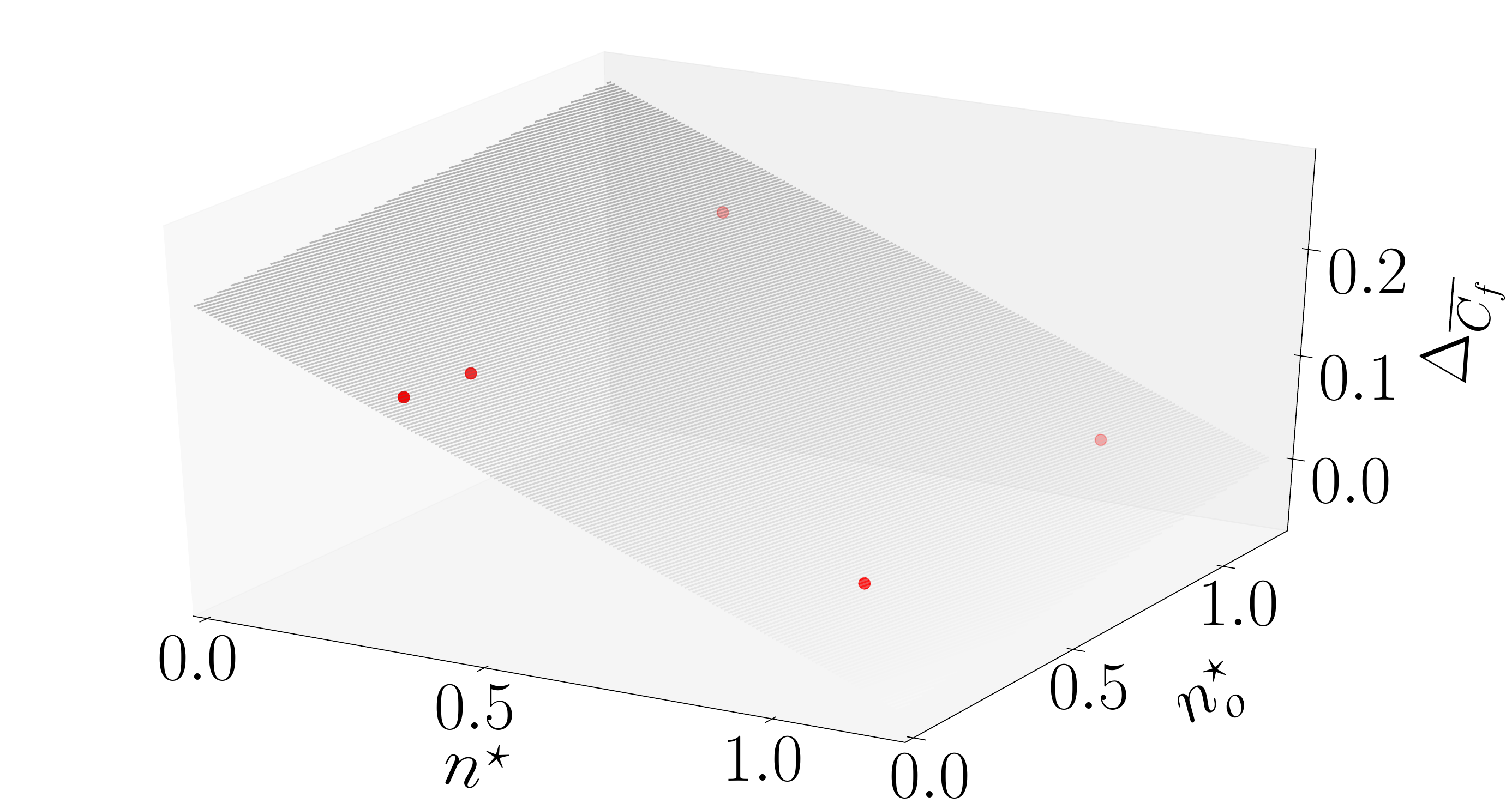} \\
    \textit{(a)} & \textit{(b)}
    \end{tabular}
    \caption{Real (red dots) and predicted (grey plane) values for \textit{(a)} computational costs $CC^\star$ and \textit{(b)} the relative error of the friction coefficient $\Delta_{\overline{C_f}}$ for the range covered in these studies.}
    \label{fig:Computational_costs}
\end{figure}

At last, the data-driven runs are considered. In this case, the computational costs are normalised by the ensemble number $m+1$ (one additional CPU core is employed to perform the analysis phases). It is noticed from the Algorithm \ref{alg:EnKF}, that the cost associated with the matrix operations depends on the number of observations employed $n_o$, the number of ensembles $m$, the number of parameters $\theta$, and the number of degrees of freedom in our system $n$. In this work, $m$ and $\theta$ are constant for every DA realisation. Therefore, the computational costs related to the data-driven procedures can be expressed as $CC^\star = f(n, n_o)$. In figure \ref{fig:Computational_costs}(\textit{a}), the costs of the five DA runs analyzed in \S\ref{sec:researchWork2} are plotted against the two variables $(n^\star, n_o^\star)$ adimensionalized over their maximum value ($n^{max} = 172\,032$ and $n_o^{max}=4\,096$). One can see that the five results for $CC^\star$ from the available DA realizations are well-fitted by a log-level regression model. This is proved by estimating the coefficient of determination $R^2$. For a dependent variable $\boldsymbol{z}$ whose predicted values are $\widehat{z}_i$, its definition is $R^2 = \sum_i (\widehat{z}_i - \overline{z})^2 / \sum_i(z_i - \overline{z})^2$, which represents the proportion of the total variation in the $z_i$ values from the average $\overline{z}$ explained by the regression equation. For the predicted $\widehat{CC}^\star$, the following equation is derived with a $R^2 = 99.902\%$: 

\begin{equation}
    \textrm{ln}\left(\frac{\widehat{CC}^\star}{m+1}\right) = 1.651 + 1.448 \,n^\star + 0.511 \,n_o^\star
    \label{eqn:CC}
\end{equation}

It is observed that, in the empirical relation expressed in equation \ref{eqn:CC}, $n^\star$ is more important than $n_o^\star$ in the range of investigation of the present study.
The data-driven runs are now sorted by increasing the computational resources required. Case 4 provides a $CC^\star/(m+1) = 8.797$, case 2 is $9.284$, case 3 is $13.708$, case 1 is $24.797$, and DNS-IBM-DA is $37.598$. This result clearly indicates that the hyperparameters driving the DA technique are critical both in terms of accuracy as well as for computational costs. Further investigations should be focused on determining these coefficients to optimize the algorithm, possibly by combining Machine Learning (ML) techniques altogether. Besides, as already expressed in \S\ref{sec:researchWork2}, one can see that more observations do not necessarily imply better accuracy. To quantify such a statement, the parameter $\Delta_{\overline{C_f}}$ from table \ref{tab:summary2} is also approximated by a log-level regression model and shown in figure \ref{fig:Computational_costs}\textit{(b)}. In this case, $R^2 =99.269\%$ for the range covered in these studies. The resulting empirical formula is:
\begin{equation}
    \widehat{\Delta_{\overline{C_f}}} = 0.219 - 0.223 \,n^\star + 0.041 \,n_o^\star
    \label{eqn:Cf}
\end{equation}
The positive coefficient of $n_o^\star$ in equation \ref{eqn:Cf} is an indication of the potential degradation of the accuracy of the DA technique due to overconstraints, such as for the DA run case 3 in the present work.


\section{Conclusions} \label{sec:proposedSec5}



A physics-infused strategy has been developed to improve the accuracy of a classical Immersed Boundary Method (IBM) tool, namely the penalization method. Physical information about the flow condition, no-slip condition and the stress at the wall, has been introduced within an online data-driven technique based on the Ensemble Kalman Filter (EnKF). This sequential tool has been used to update the physical state of the flow, as well as to optimize the parametric description of the penalization IBM integrated into the dynamic equations. The analysis of the statistical quantities of the flow, which include $u_\tau$, the mean streamwise velocity $\langle u_x \rangle$ and the components of the Reynolds stress tensor, indicate that the DA procedure is successful in accurately reproducing the flow when compared with a high-accuracy DNS. In addition, the analysis of the instantaneous features, in the form of isocontours of the Q-criterion, shows that a similar organization of the structures is obtained despite the different grid resolution. The sensitivity of the physics-infused procedures to changes in the model and the observation has then been analyzed. For the former, a grid with a different resolution in each space direction has been selected. For the latter, the position and number of sensors have been modified. The results indicate a mild sensitivity to these key elements, which is nonetheless lower than the one exhibited for classical simulations with the same mesh changes. Therefore, the procedure shows good characteristics of robustness. In addition, the current version of the online DA procedure is extremely competitive in terms of accuracy vs computational costs required, and it can be further improved using advanced EnKF strategies such as multiple local EnKF with extended localization, as performed in environmental studies \citep{Asch2016_siam}.

This study opens the perspective for including physical knowledge for numerous flow configurations within a data-driven formalism. The important point here is that the technique proposed at least partially bypasses the need to produce, store, and manipulate big databases, providing a reduction of computational resources needed for these means. Future applications currently explored by the team deal with the training of Machine Learning (ML) tools to replicate the dynamic effects of IBM models.

The present research work has been developed in the framework of the project ANR-JCJC-2021 IWP-IBM-DA. Computational resources to perform a part of the database of simulations have been obtained through the project EDARI A0122A01741 on the IRENE supercomputer (TGCC). 


\appendix
\section{}\label{appAinit}

The data-driven algorithm used to augment the IBM relies on numerical solvers based on the PISO algorithm \citep{ISSA198640}. The velocity and pressure field are iteratively updated to comply with the momentum equation and the Poisson equation. When the penalization volume force is velocity-dependent $\boldsymbol{f}_P(\boldsymbol{u}_{t,j})$, it may be solved implicitly in equation \ref{eqn:NavierStokes}. Let us consider the time $t$ and the time step advancement $\Delta t$.   Considering $j \in [1, J]$ as a single iteration from the PISO loop, the following steps are performed:

\begin{enumerate}
   \item Resolution of the momentum equation, from where $\boldsymbol{u}_{t,j}$ is obtained.

   \begin{equation}
       \mathsfbi{A} \,\boldsymbol{u}_{t,j} - \boldsymbol{b} - \boldsymbol{f}_P(\boldsymbol{u}_{t,j}) = -\nabla p_{t-\Delta t}
       \label{eqn:momentumMatrix}
   \end{equation}

   \item Estimation of the pressure $p_{t, j+1}$ through the Poisson equation. $A^\prime$ is a scalar field calculated from $\mathsfbi{A}$ and $\boldsymbol{f}_P$, whereas $\mathsfbi{T}^\prime(\boldsymbol{u}_{t,j})$ is a tensor field containing the discretized form of all the terms on the left side of equation \ref{eqn:momentumMatrix}.

   \begin{equation}
       \bnabla \bcdot \frac{1}{A^\prime} \nabla{p_{t, j+1}} = \bnabla \bcdot \left(\frac{\mathsfbi{T}^\prime(\boldsymbol{u}_{t,j})}{A^\prime}\right)
       \label{eqn:PoissonEq}
   \end{equation}

   \item Update of the velocity field $\boldsymbol{u}_{t, j+1}$ to satisfy the zero-divergence condition.

   \begin{equation}
       \boldsymbol{u}_{t,j+1} = \frac{\mathsfbi{T}^\prime(\boldsymbol{u}_{t,j})}{A^\prime} - \frac{1}{A^\prime} \nabla p_{t,j+1}
       \label{eqn:velocityCorr}
   \end{equation}
\end{enumerate}

Equations \ref{eqn:PoissonEq} and \ref{eqn:velocityCorr} are solved iteratively until reaching convergence. In the case of the presence of high-fidelity data at time $t$, when $j = J$, $\boldsymbol{u}_{t, J} = \boldsymbol{u}_k^f$, which is the forecast velocity field at the $k^{th}$ analysis phase for the EnKF. Together with the coefficients $\theta_k^f$ of the tensor $\mathsfbi{D}$, it constitutes the forecast system's state. The EnKF updates them to estimate $\left(\boldsymbol{u}_k^a, \theta_k^a \right)$. This state, however, does not necessarily respect the Navier--Stokes equations, as the velocity field is updated while the pressure field is the one obtained in the forecast. In order to obtain a consistent final solution, an additional update of the pressure is performed via an additional resolution of the Poisson equation.

\begin{enumerate}
    \item The volume force term $\left(\boldsymbol{f}_P \right)_k^a$ is updated by using the new coefficients $\theta_k^a$ from the tensor $\mathsfbi{D}$:

    \begin{equation}
        \left(\boldsymbol{f}_P \right)_k^a = -\nu \mathsfbi{D}(\theta_k^a) \,\boldsymbol{u}_k^a
    \end{equation}

    \item A Poisson equation is solved to update the pressure $p_k^a$. The complete flow field and the updated parameters $\theta_k^a$ will be used to solve the equation \ref{eqn:momentumMatrix} in $t + \Delta t$.

    \begin{equation}
        \bnabla^2 p_k^a = - \bnabla \bcdot \left(\boldsymbol{u}_k^a \bnabla \boldsymbol{u}_k^a \right) + \bnabla \bcdot \left(\boldsymbol{f}_P \right)_k^a 
    \end{equation}
\end{enumerate}

\section{}\label{appB}

The complete algorithm for the reference discrete IBM is described below. For each time $t$, the following steps are carried out:

\begin{enumerate}
\item Estimation of an initial velocity $\boldsymbol{u}_{t,j^\star}$ following the expression (\ref{eqn:momentumPredNOFORCE}) for the momentum predictor without a source term.

   \begin{equation}
       \mathsfbi{A} \,\boldsymbol{u}_{t,j^\star} - \boldsymbol{b} = -\nabla p_{t-\Delta t}
       \label{eqn:momentumPredNOFORCE}
   \end{equation}

\item Interpolation $(I)$ of the velocity $\boldsymbol{u}_{t,j^\star}$ from the element $i$ of a subspace $D_s$ (Eulerian mesh) to the Lagrangian marker $s$, with $s \in [1, N_s]$.
    
    \begin{equation}
        I[\boldsymbol{u}_{t,j^\star}]_s = \left[\,\boldsymbol{U^*}\, \right](s) = \sum_{i \in D_s} (\boldsymbol{u}_{t,j^\star})_i \,\delta_d \, (\boldsymbol{x}_i - \mathcal{X}_s) \boldsymbol{\Delta} \boldsymbol{x}
    \end{equation}

$\boldsymbol{x}_i$ and $\mathcal{X}_s$ refer to the position in the Eulerian and Lagrangian frameworks, respectively. $\boldsymbol{\Delta x}$ describes the Eulerian quadrature ($\boldsymbol{\Delta x} = \Delta x \Delta y \Delta z$ in the Cartesian frame of coordinates) and the interpolation kernel $\delta_d$ is a discretized delta function based on the Euclidean distance $r = (\boldsymbol{x}_j - \mathcal{X}_s) /d$ proposed by \citet{Roma1999_jcp}:

    \begin{equation}
\delta_d = \begin{cases} \cfrac{1}{3} \left(1 + \sqrt{-3r^2 + 1}\right) & 0 \leq r \leq 0.5 \\ \cfrac{1}{6} \left(5 - 3r - \sqrt{-3(1 - r)^2 + 1}\right) & 0.5 \leq r \leq 1.5 \\ 0 & otherwise \end{cases}
    \label{eqn:Roma_function}
    \end{equation}
    
The kernel is multiplied with a prescribed function $\mathcal{G}$ to account for the directional stretching of the mesh elements.
    
\item Calculation of $\boldsymbol{F}_P$ on the $N_s$ Lagrangian markers by employing the equation \ref{eqn:forceLagrangian}
 already presented in \S\ref{sec:introIBM}. The target values over the $N_s$ Lagrangian markers are $\boldsymbol{U_{ib}} = \boldsymbol{0}$ if the body surface is not moving.
 
\item Spreading step to obtain the Eulerian counterpart $\boldsymbol{f}_P$:
    
\begin{equation}
    \boldsymbol{f}_P\,(\boldsymbol{x}_i) = \sum_{i \in D_s} (\boldsymbol{F}_P)_k \,\delta_d \,(\boldsymbol{x}_i - \mathcal{X}_k)\, \epsilon_k
\end{equation}

The $k$-index denotes a loop over the Lagrangian markers whose support contains the Eulerian node $j$, and $\epsilon_k$ is the Lagrangian quadrature.

\begin{equation}
    \mathsfbi{C} \boldsymbol{\epsilon} = \boldsymbol{1}
\end{equation}

where the vectors $\boldsymbol{\epsilon} = (\epsilon_1, ..., \epsilon_{N_s})^T$ and $\mathbf{1} = (1, ..., 1)^T$ with size $N_s$. $\mathsfbi{C}$ is the matrix defined by the product between the $k^{th}$ and $l^{th}$ interpolation kernels:

\begin{equation}
    C_{kl} = \sum_{i \in D_s} \delta_d \,(\boldsymbol{x}_i - \mathcal{X}_k) \,\delta_d \, (\boldsymbol{x}_i - \mathcal{X}_l)
\end{equation}

\item Repetition of the momentum predictor, but with the inclusion of $\boldsymbol{f}_P$, as shown in equation \ref{eqn:momentumMatrix}, to obtain $\boldsymbol{u}_{t, j}$. However, different from the continuous IBM, in this case, $\boldsymbol{f}_P \neq \boldsymbol{f}_P (\boldsymbol{u}_{t,j})$:

\begin{equation}
    \mathsfbi{A} \,\boldsymbol{u}_{t,j} - \boldsymbol{b} - \boldsymbol{f}_P = -\nabla p_{t-\Delta t}
    \label{eqn:momentumMatrixFORCEVEL}
\end{equation}

\item Computation of the Poisson equation and the momentum corrector (equations \ref{eqn:PoissonEq2} - \ref{eqn:velocityCorr2}) iteratively until achieving convergence. $A$ is a scalar field obtained from $\mathsfbi{A}$, and $\mathsfbi{T}(\boldsymbol{u}_{t,j})$ contains the information of all the terms on the left side of equation \ref{eqn:momentumMatrixFORCEVEL}.

\begin{eqnarray}
    \bnabla \bcdot \frac{1}{A} \nabla{p_{t, j+1}} &=& \bnabla \bcdot \left(\frac{\mathsfbi{T}(\boldsymbol{u}_{t,j})}{A}\right)
    \label{eqn:PoissonEq2} \\
    \boldsymbol{u}_{t,j+1} &=& \frac{\mathsfbi{T}(\boldsymbol{u}_{t,j})}{A} - \frac{1}{A} \nabla p_{t,j+1}
    \label{eqn:velocityCorr2}
\end{eqnarray}
    
\end{enumerate}

\bibliographystyle{jfm}
\bibliography{references}

\end{document}